\pdfoutput=1

\documentclass{article}
\usepackage{arxiv}

\usepackage{amssymb}
\usepackage{amsfonts}
\usepackage{amsmath}

\usepackage[T1]{fontenc}
\usepackage{hyperref}
\usepackage[ruled,vlined,linesnumbered]{algorithm2e}
\SetKwComment{Comment}{\# }{}
\usepackage{textcomp, gensymb}

\usepackage{longtable}
\usepackage[super]{nth}
\usepackage{multirow}
\usepackage{tabularx}
\usepackage{physunits}
\usepackage{url}
\usepackage{rotating}
\usepackage{diagbox}

\usepackage{caption}
\usepackage{subcaption}
\usepackage{float}

\usepackage[utf8]{inputenc} %
\usepackage[T1]{fontenc}    %
\usepackage{hyperref}       %
\usepackage{url}            %
\usepackage{booktabs}       %
\usepackage{amsfonts}       %
\usepackage{nicefrac}       %
\usepackage{microtype}      %
\usepackage{lipsum}		%
\usepackage{graphicx}
\usepackage{doi}

\title{Combining  data from multiple  sources for urban travel mode choice modelling}

\author{
\href{https://orcid.org/0000-0002-5440-4954}{\includegraphics[scale=0.06]{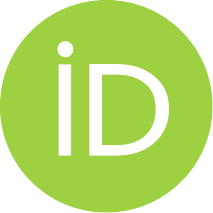}Maciej Grzenda}\\
Faculty of Mathematics and Information Science\\
Warsaw University of Technology\\
Koszykowa~75, 00-662 Warszawa, Poland\\ 
\texttt{Maciej.Grzenda@pw.edu.pl}
 \And
\href{https://orcid.org/0000-0001-7015-2956}{\includegraphics[scale=0.06]{orcid.pdf}Marcin Luckner}\\
Faculty of Mathematics and Information Science\\
Warsaw University of Technology\\
Koszykowa~75, 00-662 Warszawa, Poland\\
 \texttt{Marcin.Luckner@pw.edu.pl}
 \And
\href{https://orcid.org/0000-0002-3476-1532}{\includegraphics[scale=0.06]{orcid.pdf}Jakub Zawieska}\\
Institute of Infrastructure, Transport and Mobility\\
SGH Warsaw School of Economics\\
al.~Niepodległości~162, 02-554 Warszawa, Poland\\
 \texttt{jzawie@sgh.waw.pl}
 \And
\href{https://orcid.org/0000-0002-5479-4489}{\includegraphics[scale=0.06]{orcid.pdf}Przemysław Wrona}\\
Warsaw University of Technology\\
Koszykowa~75,00-662 Warszawa, Poland\\
 \texttt{przemyslaw.wrona.dokt@pw.edu.pl}
}

\begin{document}

\maketitle

\begin{abstract}
Demand for  sustainable mobility is particularly high in urban areas. Hence, there is a growing need to predict when people will decide to use different travel modes with an emphasis on environmentally friendly travel modes. As travel mode choice (TMC) is influenced by multiple factors, in a growing number of cases machine learning methods are used to predict travel mode choices given respondent and journey features. Typically, travel diaries are used to provide core relevant data. However,  other features such as attributes of mode alternatives including, but not limited to travel times, and, in the case of public transport (PT), also waiting times and walking distances  have a major impact on whether a person decides to use a travel mode of interest. Other factors influencing mode choices include weather conditions and built environment features. 

Hence, in this work, we propose an architecture of a software platform~\footnote{We refer to the platform as the Unified Travel Mode Choice Modelling (UTMCM) platform.} 
performing the data fusion combining data 
documenting journeys 
with the features calculated to summarise transport options available for these journeys, built environment and environmental factors such as weather conditions possibly influencing travel mode decisions.  Furthermore, we propose various novel features, many of which we show to be among the most important for TMC prediction. An important group of them 
summarises the actual functioning of the public transport system. 
We propose how stream processing engines and other Big Data systems can be used for their calculation.

The data processed by the platform is used to develop machine learning models predicting travel mode choices for individual trips.
To validate the platform, we propose ablation studies investigating the importance of individual feature subsets calculated by it and their impact on the TMC models built with them. In our experiments, we combine  survey data, GPS traces, weather and pollution time series, transport model data, and spatial data of the built environment.
The results of the development of TMC classifiers show the benefits arising from the data fusion methods and the new features calculated by them.
The  growth in the accuracy of TMC models built with the additional features 
is up to 18.2\% compared to the use of core survey data only.

\end{abstract}

\keywords{
travel mode choices \and classification \and survey data \and public transport
}

\section{Introduction}

The interest in sustainable transport modes for urban areas raises the question of how the use of individual
transport modes such as public transport, walking or cycling 
can be predicted. The problem of predicting travel mode
for a journey is known as travel mode choice modelling~\cite{hillel2021, Chang2019,tamim2022,salas2022,Grzenda2023}. Analysis of travel mode choices is a major task directly related to both predicting travel demands and transportation planning~\cite{Chang2019,cheng2019}.

The underlying process is  driven by multiple factors. These include traveller features such as lifestyle \cite{Chang2019,cheng2019}, comfort \cite{cheng2019},  age~\cite{tamim2022} and education \cite{Chang2019,cheng2019}, journey distance \cite{Chang2019, tamim2022}, and the number of cars \cite{cheng2019, tamim2022} and bicycles owned by a household \cite{Chang2019, tamim2022}. Other factors include but are not limited to the Level of Service (LOS) attributes such as cost and expected duration for each mode \cite{LondonDataset2018,hillel2021}, the attributes of the built environment \cite{cheng2019}, transport network connectivity \cite{cheng2019} and weather conditions~\cite{hagenauer2017}. Furthermore, in the case of public transport, not only planned characteristics of public transport modes, but also the actual functioning of public transport systems in different areas can influence actual mode choices. In particular, delays and frequent vehicle stops due to street congestion may discourage citizens from using public transport.

The choice of travel mode is a  complex issue and has  also been the subject of numerous studies that vary significantly in terms of the methods and approaches used. The two main approaches used to identify people's choices regarding travel modes are revealed preferences (RP)~\cite{garcia2022,batarce2015use,BROWNSTONE2000} and stated preferences (SP)~\cite{hillel2021,batarce2015use,BROWNSTONE2000}.
The main assumption of the SP approach is that SP surveys (\mbox{choice} experiments) are based on hypothetical situations where respondents face choice scenarios between mutually exclusive alternatives presented in  multiple-choice situations. 
The main advantage of the SP approach is that it enables the creation and study of  alternatives that do not yet exist, i.e. new transport modes, and allows the development of very specific settings of a choice situation where all attributes and their levels are chosen by the researcher. 
Although SP techniques are well established,
they also have certain limitations. As 
the design and choice of attributes, alternatives and the context presented for the respondent are  made exclusively by the authors of the survey,
such settings may be biased by authors' preferences,
and  not include all relevant variables. %
Moreover, in the case of travel mode choice studies, the inclusion of all factors in the hypothetical scenario such as time of  day, travel durations and distances assuming different travel modes, journey destination and destination type (such as school), weather conditions, and parking costs may make it difficult for the respondent to make decisions that the person would actually make in real life. 

Revealed preferences theory 
originally 
set out to investigate and reveal the preferences of consumers by observing their purchasing habits \cite{Wong1978FoundationsOP}. In the transportation field, this approach focuses on monitoring and collecting  data about actual trips and is usually  based on various types of travel diaries but also other data sources such as smart card data~\cite{hensher1993using, drabicki2021modelling, batarce2015use}. This provides data about respondents' past trips and the travel modes selected by them for these trips.
The core advantage of the RP approach is that the data reflect real choices and real decisions (real money, time etc. spent on given activities). %
In line with these findings, as shown in the recent survey on travel mode choice modelling with machine learning (ML) techniques~\cite{hillel2021}, travel diaries, i.e. data collected under the RP approach, are most frequently used for modelling travel mode choices. While only four of the analysed studies used SP data, the remaining 68 studies on TMC modelling relied on RP data~\cite{hillel2021}.

Importantly, travel diaries are conducted to document real transport mode choices, but they can also provide input for the models used to predict mode choices and possibly reveal what makes people select individual  modes. Hence, data from travel diaries are used as input for models relying on discrete choice methods \cite{Chang2019, garcia2022} and machine learning methods \cite{garcia2022,hillel2021,Chang2019,LondonDataset2018,Chen2023}. 
At the same time, recent studies have shown that machine learning techniques are an alternative to statistical approaches~\cite{hagenauer2017} and yield major performance gains compared to discrete choice models when applied to travel mode choice modelling~\cite{garcia2022}.
When machine learning methods are used, the travel mode choice problem is stated as a multi-class classification problem \cite{Chang2019, garcia2022}.
Random forest and XGBoost were shown to provide TMC models of particularly high accuracy \cite{Chang2019,LondonDataset2018,Chen2023}, though TMC models were also developed  with other methods including, but not limited to, decision trees~\cite{tamim2022,hagenauer2017} and support vector machines~\cite{Chang2019, garcia2022, hagenauer2017}.

An important limitation of using travel diaries as the only source for mode choice modelling is that typically  only details on the travel mode actually used are recorded in a survey. This is why details of alternative modes should be synthesised and added to the dataset used for machine learning tasks \cite{hillel2021}.
Therefore, survey data has to be combined with ground-truth data on  transport modes, such as travel times and distances when the same destination is reached by walking, cycling,  bus or  private car. 
Interestingly, while respondents can be asked about their habits such as the transport modes they tend to use on a daily basis, using such data for TMC predictions reduces the chances of discovering the impact of the level of service of different means of transport on mode choices. This is because TMC models could start to rely on the declared habits of respondents rather than e.g. travel duration when using public transport as compared to travel duration by car, which may cause these habits.

An important benefit of the RP approach is that each real trip record documents a trip made between different points in the city area at some time in the past. Hence, it can be linked to other data describing the conditions under which each trip was made, such as weather conditions, and travel duration expected on the day and time of  day given public transport schedules.
Hence, some studies extended data from travel diaries with features calculated based on other data sources. In \cite{LondonDataset2018}, inter alia straight line trip distance, duration of walking, cycling, driving and PT route
were estimated using external directions API. This provided 13 additional features for ML purposes.  Hagenauer et al. enriched survey data by including three built and natural environment features and three weather features \cite{hagenauer2017}, which provided data for a  study of classifiers used for modelling
travel mode choices. However, in their recent survey on TMC modelling with ML methods, Hillel et al. note that only 16 out of 68 analysed RP studies documented the  way attributes of mode alternatives such as duration were calculated. Importantly, as many as fourteen of these studies relied on zonal time-independent (static) models to calculate durations and/or costs~\cite{hillel2021}. Only three studies used time-dependent and/or online direction services to enrich survey data with mode choice data. Importantly, if mode choice attributes are not included, the impact of changes to the transport system on mode choices cannot be modelled~\cite{hillel2021}. Let us note that schedules and possible PT connections are typically different for different times of the day, so time-dependent features such as the duration of travel by PT are  likely to help better predict and explain travel mode choices than time-independent features averaging e.g. travel duration across the whole day.

Therefore, in this study, we propose a data fusion platform for multi-factor travel mode choice modelling. We rely on the RP approach, where travel diaries provide core data needed to model travel mode choices. The  platform automates the process of extending data from the travel diaries by calculating features quantifying circumstances possibly influencing travel model choices, such as estimated travel durations with different transport modes at a given time of  day and past real travel durations on the same route.
In this way, we aim to maximise the performance of TMC models built with ML methods. 
The Unified Travel Mode Choice Modelling platform we propose collects and preprocesses data from multiple sources. Since it enables the calculation of multiple features quantifying the factors possibly influencing travel mode choices, trip records from travel diaries can be extended not only with time-dependent features documenting alternative travel mode attributes, but also past behaviour of public transport on the same route, weather conditions, air pollution, and built environment in the area of the trip.

We validate the platform by applying it to  data collected in the years 2022-2023 for the City of Warsaw, Poland operating an extensive PT system of metro, trains, trams and buses with up to 2,000 PT vehicles used at the same time. We calculate with the platform an additional 305 features of potential impact on TMC modelling, extending the base set of features obtained from survey data. We use multiple classification methods including random forest, XGBoost, decision trees, multilayer perceptron, Support Vector Machine (SVM) and others to train and evaluate TMC models using different subsets of features arising from the data fusion.
The results of the evaluation with multiple data sets show a major improvement in the performance of TMC classifiers trained with data arising from the data fusion performed by the platform. In the case of the largest trip data set considered in this work, the best model built using additional fusion-based features yields an accuracy    14.1\% larger than that of the best model relying on the use of survey data only.

The primary contributions of this work are as follows:
\begin{itemize}
    \item We propose an architecture of a platform collecting, transforming and performing the fusion of data sets and data streams needed for urban travel mode choice modelling. We use it to develop an extensive set of features quantifying factors having an impact on travel mode choices. We describe the Unified Travel Mode Choice Modelling platform implementing this architecture.
    \item We propose different sets of features to be calculated to quantify journey features, including the level of service for different transport modes with particular attention paid to the features of expected and actually possible in the past PT connections, the built environment in the area of the journey, weather and air pollution.
    \item We provide the results of the development of travel mode choice models for the reference data sets and compare the impact of using different feature subsets on the performance of the TMC models.
\end{itemize}

 The remainder of this work is organised as follows. In Sect.~\ref{sec:RelatedWorks}, we provide an overview of related works on TMC modelling with ML methods, with an emphasis on relevant factors and studies involving data fusion. This is followed by the proposal of the Unified Travel Mode Choice Modelling platform in Sect.~\ref{sec:platform}. Next, the data fusion process enabled by the platform is discussed in Sect.~\ref{sec:fusion}. The process of training and evaluation of TMC models considering different feature sets developed with data fusion, and the results of this process for reference trip data sets are summarised in Sect.~\ref{sec:training} and Sect.~\ref{sec:Experiments}, respectively. This is followed by the conclusions in Sect.~\ref{sec:conclusions}. In addition, details on how data  were collected and processed with the platform in the case of the City of Warsaw are provided in~Appendix~\ref{appendix:Warsaw}.

\section{Related works\label{sec:RelatedWorks}}

\subsection{Travel mode choices and factors influencing them }

 Traffic congestion in urban areas is a major challenge for city councils, causing  social anxiety and negatively impacting the economy \cite{Ouallane2022}. Not surprisingly, traffic management systems are expected inter alia to reduce it by giving priority to public transport \cite{Ouallane2022} and encouraging more travellers use other means of transport than cars.
Such modal shift is an important element of any transport policy framework that aims to improve the environmental performance of transport. It is related to many  aspects of transportation planning, including e.g.  
the development of methods that are used to understand and predict urban spatiotemporal  flow~\cite{Xie2020}. 
Understanding the factors behind travel mode choices is therefore crucial in the development of measures and tools intended to both change such choices and predict transport behaviours. 
The main categories of variables used to analyse travel behaviour  are as follows: (1) socio-economic and demographic characteristics, (2) spatial development patterns, (3) policies directly or indirectly affecting travel behavior, and (4) national cultures or individual preferences~\cite{Buehler2011DeterminantsOT}. Another 
review~\cite{Witte2013LinkingMC} distinguished similar types of determinants, including travelers' socio-demographic characteristics, spatial environment, journey characteristics, and socio-psychological factors. 

A large part of the literature focused on transport mode choices   reveals different socio-economic and demographic variables influencing travel behaviour.
As an example, gender differences in travel mode availability and travel behaviour are observed
\cite{Chidambaram2021WorktripMC, Nobis2005GenderDI, Hanson2010GenderAM}.
Several studies also found a correlation between the income of the household and propensity to use a car, both for commuting and other daily activities, as well as long-distance trips~\cite{Best2004DivisionOL, Cao2007CrossSectionalAQ, Reichert2015ModeUI, Eisenmann2018AreCU}.

Another 
group of variables determining transport mode choices 
includes such aspects as 
cost of trip, travel time, and the convenience and complexity of particular transport modes and journeys~\cite{Banister2005UnsustainableTC, Racca2003FACTORSTA,LondonDataset2018}. 
Spatial aspects -- understood as land use, urban design and residential location --- are other common factors influencing transport mode choices. For instance, low density and spread-out urban areas make walking and cycling less convenient and attractive, and encourage the use of cars as the main way of travelling~\cite{Aziz2018ExploringTI, Ding2017ExploringTI, VanAcker2007WhenTG, Zhang2012HowBE}.

Cultural and behavioural aspects are more difficult to observe but are also  important elements in mode choices. These 
include environmental attitudes, perceptions, behavioral norms, subjective  beliefs, and habits~\cite{Beiro2007UnderstandingAT, Chen2011HabitualOR, Clark2015ChangesTC, Gardner2010GoingGM}.
Importantly, even between relatively similar regions and countries there can be noticed significant differences in motivations, determinants and final choices of transport mode. %
As an example,
society in the USA is much more car dependent and makes more than 70 percent of trips by car, which is the case only for the most car-oriented groups of society in Germany. 
Moreover, distance to public transport, population density, and car access have a weaker influence on car travel in the USA than in Germany~\cite{Buehler2011DeterminantsOT}.

Walking is one of the most preferred transport modes, as it is not only fully carbon neutral (environmentally friendly) but also improves the well-being and health of individuals~\cite{Marselle2013WalkingFW, Wolf2014WalkingHA, GilesCorti2003RelativeIO}. Nevertheless, it is also quite vulnerable to several barriers. In general, the concept of built environment understood as urban design factors, land use, and available public transportation in an area,
has a significant influence on peoples’ travel behaviour, in particular active mobility~\cite{Booth2000SocialcognitiveAP, Handy2002HowTB}. Such links between the built environment and walking have been shown by numerous studies~\cite{Cervero2007InfluencesOB, Ferrer2015AQS, Forsyth2009TheBE}.
Distance to the destination is  often perceived as the main obstacle to choosing walking~\cite{Goldsmith1992NATIONALBA}.
A number of studies revealed that preferences towards walking and to other active transport modes are more vulnerable to weather conditions. 
Rainy weather typically limits outdoor activities and leads to a shift towards sheltered modes of transport, including cars and public transport~\cite{saneinejad2012modelling}. 

A large number of existing studies have examined the effects of air pollution on travel behaviour, but the research results are not conclusive. Some studies have highlighted that individuals' travel activities, especially outdoor physical activity and school attendance, may be reduced during polluted periods~\cite{laffan2018every, xu2021does, chen2018air, ward2016responds}. Existing evidence also highlights the impact of air quality on both trip generation and mode choice~\cite{kim2023role}. However, it is important to note that there are studies that do not find a significant effect of air pollution on travel behaviour.
Pollution sensitivity varies between geographical regions, tending to be stronger in Western societies~\cite{liu2018severe}.

Cycling is also a popular subject of studies on travel behaviour.
Trip distance often has a significant impact on whether one chooses a bicycle as a transport mode~\cite{lovelace2017propensity}. Built environment, in particular high population density,  also has a positive impact on the prevalence of cycling~\cite{Winters2010BuiltEI}. Similarly, demand for bike sharing largely depends on land-use types \cite{Sun2023}.

Public transport (PT) is a very important sustainable transport mode, often considered the best alternative to private cars~\cite{Holmgren2007MetaAnalysisOP}. 
An extensive research review~\cite{Redman2013QualityAO} listed a number of public transport quality attributes attracting car users to public transport and divided them into two groups. Physical attributes of PT include reliability, frequency, speed, accessibility, price, but also information provision, ease of transfers/interchanges and vehicle condition. The second identified group were perceived attributes that include comfort, safety and convenience. All these attributes are also in line with the findings of  other studies~\cite{Curtis2016PlanningFP, Khan2021IncreasingPT, Dodson2011ThePO, McLeod2017UrbanPT}.
Comfortable access to the public transport network is also an important factor. A meta--analysis of data from over 100 urban areas in Mexico revealed that citizens  are less likely to use the car in areas with a better public transport network and supply~\cite{Guerra2018UrbanFT}. The walking distance and proximity  to public transport facilities were also found to be important decision factors in other studies~\cite{imekolu2015TheRO, Oa2021PublicTU, Walker2007PurposedrivenPT}. 

Among the most important identified determinants of choosing a private car as a transport mode are a feeling of freedom and independence and the connected ability to decide where to go, convenience, rapidity, comfort, and flexibility~\cite{Beiro2007UnderstandingAT}. 
The correlation between the growth of income and the probability of choosing a car was also revealed in other studies~\cite{Ko2019ExploringFA}. 
According to other studies, the car as a transport mode is less attractive in highly congested areas with limited and expensive parking spaces~\cite{carse2013factors, hensher2001parking}.
Several studies have also revealed the influence of lifestyles and lifecycle stages on preferences and choices regarding transport mode use. For instance, car use tends to increase with particular stages of life, e.g. entering the labour force, the transition from a single-person household into a two-person households, and having children
~\cite{Ardeshiri2019LifestylesRL, Kuhnimhof2012TravelTA, Prato2017LatentLA}.

\subsection{Use of machine learning for travel mode choice modelling}

There is a growing interest in using machine learning methods for TMC modelling. This is reflected by systematic reviews of ML methods for modelling travel mode choices that have been recently proposed~\cite{hillel2021, salas2022,zhao2020}. Some of these works  focus on the comparison of ML models to other models
\cite{zhao2020, salas2022}.
As observed inter alia in~\cite{Chang2019,hillel2021,garcia2022}, machine learning methods are used as an alternative to Random Utility Maximisation models. One of the key reasons for this is the better prediction capability of machine learning methods~\cite{garcia2022, zhao2020,salas2022}. As an example,~\cite{Chang2019} reported that the use of random forest improved the accuracy of predictions of travel mode choices  by 27\% compared to the multinomial logit model.
Importantly, when travel choice modelling is stated as a machine learning problem, typically it is treated as a classification task~\cite{garcia2022,hagenauer2017,zhao2020}.

As far as the methods used to build classifiers predicting travel mode are concerned, a variety of methods have been used. These include Naive Bayes~\cite{Chang2019,hagenauer2017}, decision trees~\cite{Chang2019,tamim2022,hagenauer2017}, random forest~\cite{Chang2019,zhao2020,tamim2022,garcia2022,salas2022,hagenauer2017,mohd2022}, support vector machines (SVM)~\cite{Chang2019,garcia2022,salas2022, hagenauer2017}, k Nearest Neighbours (kNN)~\cite{salas2022}, Xtreme Gradient Boosting (XGBoost)~\cite{salas2022,Chen2023,LondonDataset2018}, neural networks of different architectures, including  
multilayer perceptrons~\cite{tamim2022,salas2022,hagenauer2017}, and some methods rooted in probability theory such as logistic regression~\cite{tamim2022}. The methods reported to provide the best predictive accuracy frequently include random forest~\cite{zhao2020,hagenauer2017,garcia2022,cheng2019}. 
An important challenge related to many methods is the need to tune their hyperparameters. 

Finally, many works rely on cross-validation (CV) to evaluate different machine learning methods applied to the TMC task and possibly tune hyperparameter values. Examples include 5-fold CV, applied in~\cite{Chang2019}, and 10-fold CV used in \cite{garcia2022}, for tuning hyperparameters of different ML methods. In~\cite{LondonDataset2018}, Hillel et al. used data from the first two years of the London survey for cross-validation aimed at performance estimation and final model training, and the data of the most recent survey year as a holdout test set used for the performance evaluation of the final models. The same data split was applied in \cite{Chen2023}.

\subsection{Data used to model travel mode choices}

The data used to model travel mode choices  is most frequently based on surveys providing travel diary data. %
Dutch National Travel Survey (NTS) data have been used inter alia by Tamim et al. in their comparison of machine learning techniques used for travel mode choice modelling~\cite{tamim2022}. Garcia-Garcia et al. used the OPTIMA data set, composed of the results of a survey performed between 2009 and 2010 in Switzerland, which provided 1906 trip sequences~\cite{garcia2022}. In the same work, the NTS data were used as well. 
In~\cite{salas2022}, four different data sets were used. 
In particular, the 
‘‘Student travel at McMaster University’’ set  was collected  at McMaster University in Hamilton, Canada. It documents the travel mode choices of students travelling to school~\cite{salas2022}
and illustrates the case of  data sets documenting  behaviours of specific groups of travellers. 

It is important to note that the share of individual travel modes varies in the data sets. As an example, the NTS data were reported to contain 16.9\% of walk instances, 24.41\% of bike, 55.26\% of car and only 4.03\% of public transport instances~\cite{garcia2022}. On the other hand, the OPTIMA data set includes only 5.98\% of walk and bike instances, 65.98\% of car, motorbike and other private modes and 28.12\% of public transport instances~\cite{garcia2022}. This shows how large differences in the share of travel modes in overall transportation can be. Moreover, the number of modes and mode types vary in individual data sets.
This is because the popularity of travel modes in different regions varies. As an example, the Nanjing survey revealed that as many as 25.8\% of trips  made in this region in 2013 were by e-motorcycles~\cite{cheng2019}. 
Finally, let us note that an extensive review of works on TMC modelling and the data sets used in these works can be found in~\cite{hillel2021}.

\subsection{Role of data fusion in travel mode choice modelling}

Fusing urban data from individuals was observed to be beneficial for transport-related tasks. Examples include a recent study  combining demand data and weather data to predict mobility demand~\cite{Prado2023}.
In line with the majority of other works on travel mode choice modelling with ML techniques, we focus on proposing data fusion solutions for RP approaches.
However, RP approaches have also several limitations. The RP approach shows the actual decisions of  individuals, but it does not reveal the full choice context, including the alternatives actually 
available to the decision maker. 
Hence, in some studies survey data has been enriched by including other variables.
First of all, level of service (LOS) attributes including features, such as estimated travel duration with different travel modes for individual journeys, are calculated and added. The features in this group include travel durations, estimated distances (as e.g. walking can be made using other routes than when public transport is used), waiting time in the case of public transport and others.
As an example, in~\cite{cheng2019}, travel time was determined based on the centroids of the  pair of traffic zones for a journey and used to complement survey-based features such as traveller's age.

Another group of features considered when preparing TMC data sets based on travel diary data includes built environment attributes. In~\cite{cheng2019}, variables of the built environment were added to the travel diary data.
The data fusion performed in this way made it possible to include land utilisation, road network density, distances to the nearest metro and bus stops, and the number of bus stops and metro stations in the neighbourhood. 
Land use, as well as population density, road density, distance to rail stations and facilities such as medical centres, were the built environment features considered by Yang et al. in their study on active travel among older adults~\cite{yang2022}. 
Another group of attributes considered in some works are weather attributes. Daily precipitation, maximum temperature and average wind speed were used in~\cite{hagenauer2017}.

Nevertheless, in~\cite{LondonDataset2018}, Hllel et al.  observed that machine learning models used to predict travel mode choices are limited by their input data, which are typically the raw data from the trip diaries. To address this limitation, they proposed recreating the choice set faced by a person at the time of travel. To validate this approach and analyse the benefits arising from combining survey data with  mode alternatives data, they used the London Travel Demand Survey data. Next, requests to the Google Maps Directions API were made to calculate estimated routes and durations for walking, cycling, public transport and driving. 
 Request dates were set to two weeks before the departure times. The intention of this setting was to perform calculations for typical conditions
and  not include planned public transport disruptions or real-time trafﬁc~\cite{LondonDataset2018}. 
Moreover, travel costs 
including congestion charge costs were calculated. The performance of the models built with feature vectors extended with the aforementioned choice-set information was significantly improved compared to the use of raw survey-based data only.  Improvements were observed both during cross-validation on the training data and holdout validation of the final models with the test set~\cite{LondonDataset2018}. Importantly, weather and built environment features were not included in this study~\cite{LondonDataset2018}. In \cite{Grzenda2023}, features including both the predicted time needed to find a parking space and get from its location to the actual journey destination, and parking difficulty were proposed and used together with survey-based features and LOS features calculated with OpenTripPlanner for TMC prediction. The inclusion of the parking-related features resulted in the increased accuracy of TMC models for some data sets.

Still, as shown by the survey on TMC modelling with ML methods~\cite{hillel2021}, most analysed studies do not include attributes of mode alternatives or do not list the input features. Not including such attributes is identified as one of the major limitations related to data sets used in TMC studies~\cite{hillel2021}.

Some studies investigate the models to estimate the importance of individual features and their impact on predicted travel mode.  One of the reasons is that attracting more people to use public transport is related to an in-depth understanding of choice behaviour~\cite{Chang2019}. 
In~\cite{tamim2022},  Shapley Additive exPlanations (SHAP) values~\cite{Lundberg2017} were used to assess variable importance. The predictors  significantly influencing the travel mode decisions were found to be trip distance, travellers’ age and annual income, number of cars/bicycles owned, and trip density~\cite{tamim2022}. Salas et al. in~\cite{salas2022} analysed the importance of individual variables for four different data sets. These included variables such as travel time and parking permit. In~\cite{hagenauer2017}, Hagenauer and Helbich checked the sensitivity of the models to individual variables and found travel distance to be the most important feature for the random forest, i.e. the method which yielded the most accurate predictions among all methods tested. ~\cite{Chang2019} confirmed this finding and observed the remaining most important features to be trip duration, age, travel frequency, arrival or departure time, and whether a person has a driving license. Importantly, some variables were observed to be correlated. An obvious example is trip distance and duration~\cite{Chang2019}. 

Variable importance partly depends on whether non-survey-based attributes are additionally included.
What follows from the summary of variable importance made in~\cite{cheng2019} is that travel time (calculated based on traffic zone data), selected built environment features (calculated based on spatial data of the region) and selected survey-based attributes of individuals were among the 10 most important variables for TMC modelling. Interestingly, household attributes such as income were found to be less important. Similarly, in~\cite{hagenauer2017} some survey-based features (such as the number of cars per household), environment features (such as address density) and weather features (such as temperature) were found to be among the top 10 most important features.

\subsection{Architecture of systems used to develop travel mode choice data}

As in some studies, survey-based attributes were combined with some other attributes having a potential impact on travel mode choices, the question of the architecture of the system performing the data fusion process arises.
In~\cite{cheng2019}, ArcGIS was used to calculate built-environment features. 
However, the work is focused on the analysis of the data set and its use for TMC modelling, rather than discussing details of calculating attribute values, such as how the values of travel time were estimated and whether this included other software modules. ArcGIS was also used to calculate built environment features  in~\cite{yang2022}, where built environment variables were the only ones apart from survey-based variables.
Similarly, the details of software systems and data preprocessing used to perform the fusion of survey, environment and weather data are not discussed in~\cite{hagenauer2017}, as the study is focused on comparing different 
classifiers. %
In~\cite{LondonDataset2018}, a script was used to calculate the {\em choice-set data} extending survey data with  features quantifying duration and cost of travel made with different travel modes.
The script relied on making calls to the Google Maps Directions API. Hence, no extra storage of data such as schedule data was employed. Finally, the idea of the architecture of the Urban Traveller Preference Miner system combining both planned schedules and past real schedules developed from GPS traces for the calculation of LOS features was briefly described in \cite{grzenda2023ECML}. Other  features were not calculated by the system.

To sum up, different studies reported the importance of varied factors going beyond those collected in trip diaries for travel mode choice prediction. These included, but were not limited to choice-set data, weather data, and built-environment data. However, each of these works combined survey data with only some of the potentially relevant data sources. This raises the question of whether the inclusion of other features
calculated by a data fusion process would change the overall performance of TMC models and the ranking of the most important features in terms of their impact on TMC predictions.
Furthermore, scripting techniques aimed at linking data from the data sources selected for specific RP data sets are typically used. This suggests that an architecture of a more generic system storing and analysing relevant data and prepared to handle multiple data sets could be considered to complement these studies. Moreover, even more factors were found to have an impact on travel mode choices, including e.g. reliability of public transport, and air pollution.

Hence, in this work, we
propose the architecture and document the implementation of a software platform automating the fusion of multiple data categories  relevant to model travel mode choices for regions of interest. The platform calculates both features 
suggested by studies on TMC modelling and novel features proposed in this work.
In the case of mode-alternative attributes, it enables the calculation of time-dependent attributes reflecting travel duration variability for cars and public transport observed over the day. Furthermore, mode attributes  based on exact origin and destination trip coordinates   rather than attributes based on zonal transport models and e.g. centroids of matching zones are calculated.
The platform is prepared to store and analyse large quantities of data and use  them {\em inter alia}  to quantify the reliability of the PT system.
Moreover, the platform can be used to execute multiple experiments aimed at finding the combination of features best matching the modelling task.
Finally, through extensive tests with multiple classifiers, we analyse the contribution of each of the data categories such as weather, built-environment, choice-set, and delay data and the features of these categories to the performance of TMC models.

\section{The unified travel mode choice modelling platform \label{sec:platform}}

\subsection{Objectives of the platform}
The primary objectives of the Unified Travel Mode Choice Modelling (UTMCM) platform which we propose in this study are to a) collect all data relevant for the long-term modelling of travel mode choices, b) persist it to enable inter alia feature engineering for machine learning purposes, c) perform the calculation of relevant features and data fusion to develop TMC data sets, and d) train and evaluate TMC models with these data sets and multiple classification methods. 

By long-term modelling, we mean modelling %
including data from different waves of surveys, and enabling studies focused on both all and selected citizens of urban area(s). Importantly, whenever justified we assume the storage of relevant data such as schedule data and weather data. This is to enable iterative development of features aggregating underlying raw data such as weather time series, while analysing their impact on the performance of TMC models built with these features.  Furthermore, we aim to increase the resilience of the platform by reducing its dependency on external systems, which after some period may not offer raw data 
or not answer queries referring to past periods. The platform extends some of the ideas and solutions introduced in \cite{grzenda2023ECML} and builds upon the idea of calculating features present inter alia in \cite{LondonDataset2018,cheng2019,yang2019,Grzenda2023}.

\subsection{Identification of relevant data}
The development of TMC models typically relies on trip records with known travel mode choices.  Hence,  the UTMCM platform supports the processing of survey data including traveller features, household features and trip features collected using the revealed preferences approach. We refer to all data collected during surveys as {\em survey data} and use the term of {\em survey features} to refer to the features placed e.g. in trip data sets, which directly reflect answers provided by respondents to survey questions. Let us note that by survey data we mean  data from both surveys such as Computer Assisted Web Interview (CAWI) surveys, and  e.g. a dedicated mobile application used by city inhabitants to record and share their trip data.

In line with the past studies, we selected the most promising data categories to be additionally included in TMC modelling.
The other categories of data that the platform is capable of collecting and transforming into features used to develop machine learning models fall into two categories. First of all, data describing circumstances under which travel is undertaken, common for all modes, are collected. These include  built environment data used to calculate built-environment features, and weather and pollution time series data, as these have been shown to have an impact on mode choices. The second group of data sources are data sources related to some means of transport only.
As cities aim to increase the use of public transport rather than private cars,  data on public transport services are particularly important. Hence, the UTMCM platform collects public transport timetables to use them to calculate candidate PT connections. However, planned timetables do not reveal the scale and spatial distribution of potential disturbances such as delays and missed transfers. Hence, the location of public transport vehicles is also collected in the form of time series of geocoordinates. Other relevant data sources are travel time and demand matrices from a transport model, which can  be used to reveal hourly variation in travel time by car. Let us note that we focus in this section on the architecture of the travel mode choice modelling platform. Details on both data sources used to perform data fusion for the reference survey data for the City of Warsaw and platform settings influencing the calculation of features are provided in Appendix~\ref{appendix:Warsaw}. 

\subsection{High-level overview of the platform \label{sec:HighLevelArchitecture}}
\begin{figure}[ht]
\centering
\includegraphics[width=\textwidth]{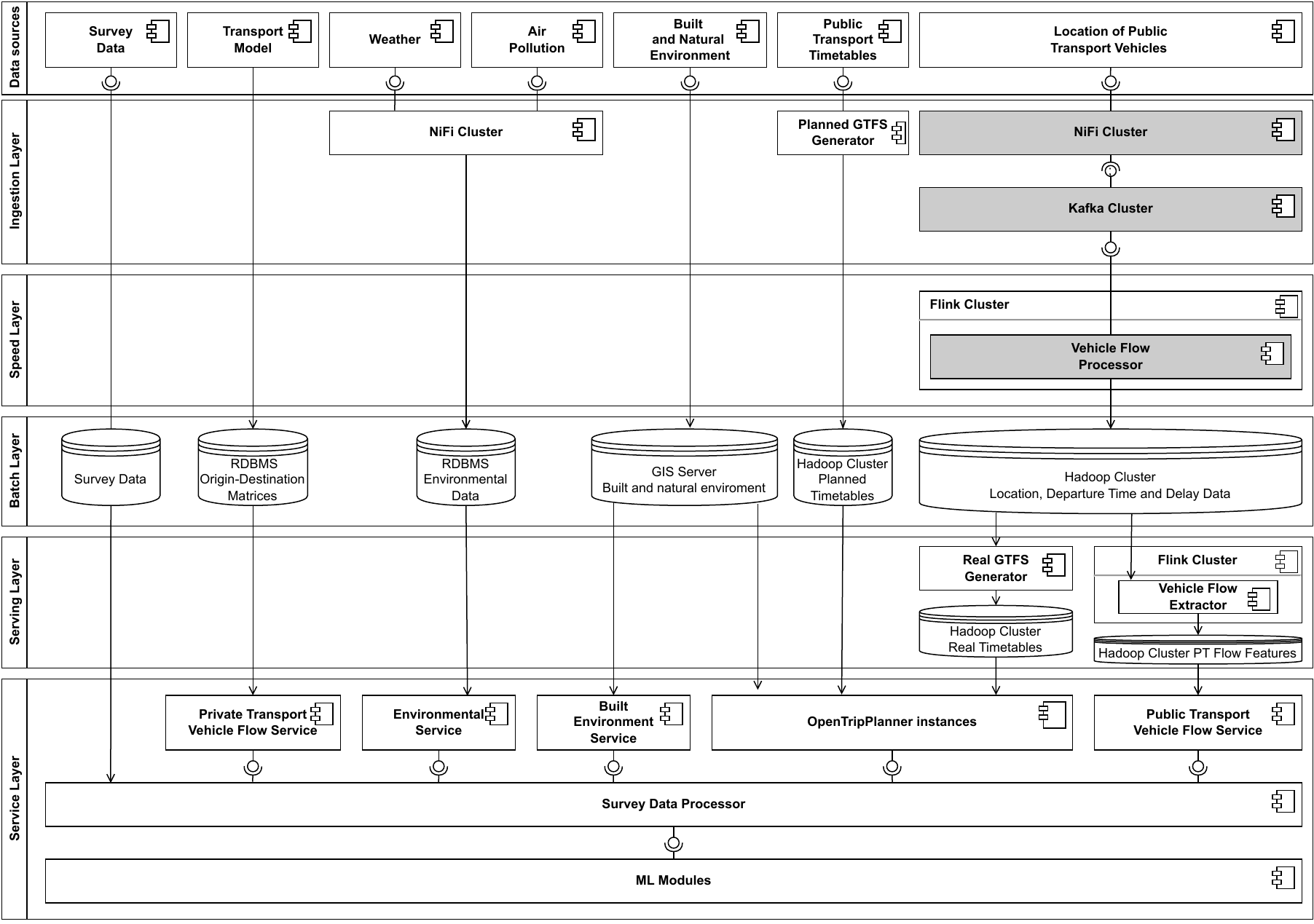}
\caption{\label{fig:main_architecture}The architecture of the Unified Travel Mode Choice Modelling (UTMCM) platform. Some secondary modules and interactions have been omitted for clarity. Online stream processing modules shown in grey.}
\end{figure}
The architecture of UTMCM platform is shown in Fig.~\ref{fig:main_architecture}. It extends the Lambda pattern~\cite{Marz2015}, i.e. the design pattern frequently used for developing platforms combining batch and online processing of data with storing raw data referred to as the 
{\em master data set}. As the UTMCM architecture is planned to provide generic solutions applicable to different cities, we extend the Lambda architecture with the Ingestion layer, the role of which is to contain components used to collect data from data sources unique for a given urban area and convert them to open standardised formats.
All the processing performed in the remaining layers relies on standardised formats, irrespective of the formats used by individual city systems. Moreover, we rely on Big Data frameworks (Apache NiFi, Apache Hadoop, Apache Kafka, and Apache Flink) whenever large quantities of data are possible. This is to make the platform scalable, especially when processing sensor data from possibly thousands of PT vehicles with a time resolution of a few seconds.

As far as Lambda layers are concerned, the UTMCM architecture includes the Batch layer for storing an immutable constantly growing master data set $\Omega$ including raw data of the aforementioned categories complementing the survey data.
The Speed layer is used to process data arriving as data streams, which in the analysed case means online processing of the GPS traces of public transport vehicles. In line with the Lambda pattern, the role of the Serving Layer is to perform batch processing of raw data. In the case of the UTMCM platform, we propose to develop in this way two categories of data. First of all, real timetables including real departure times of public transport vehicles are periodically developed based on the GPS traces stored in the Batch Layer. In this way, the real functioning of the PT system is documented. Another form of using the same data is the development of features quantifying delays, vehicle speed and the number of vehicle stopping events, which may occur due to traffic jams and negatively affect the perception of PT services.

Finally, to develop and evaluate TMC models, we designed a number of services calculating individual features comprising the data used to develop classifiers. These services and the modules calling them are placed in the Service Layer and can be used on an as-needed basis.

\subsection{Ingestion layer\label{sec:IngestionLayer}}
The primary objective of the data ingestion layer is to collect data relevant to TMC modelling. In the case of  timetables and time series data such as weather and pollution time series, and GPS traces, data have to be regularly received from data sources to be stored in the UTMCM platform. This is because not all data sources publish historical data or clearly declare which data resources will be published and for how long in the future. Below we focus on these data sources which require dedicated modules to collect data from them and put into the batch layer of the platform.

\subsubsection{Time series data}
In the case of time series data, we propose a unified approach relying on Apache NiFi data flows to collect all the series including vehicle location traces. Depending on the envisaged volume of data, the data are stored in a Relational Database Management System (RDBMS) or processed as a data stream to be persisted in Apache Hadoop. As in the case of weather and pollution data, data collected every 10 minutes can be considered sufficient for modelling purposes and the number of monitoring points is substantially lower than the number of PT vehicles in use, a relational database is recommended for the storage of this category of data. 

Let us note that in the case of GPS traces of PT vehicles, to capture exact departure times for individual stops and calculate how frequently vehicles stop because of traffic conditions, which may negatively affect the comfort of travellers, multiple location readouts per minute are needed. For this reason, the UTMCM platform relies on storing location traces of vehicles in the Apache Hadoop cluster and its Hadoop Distributed File System (HDFS). 
In this way, the scalability of the solution is ensured. In particular, the location of all vehicles potentially used by multiple travellers, including PT vehicles, but also taxis, e-scooters or cars available through car-sharing models, can be recorded to be used for future analysis. 

\subsubsection{Public transport schedules}
Apart from Apache NiFi-based data collection, we propose 
\texttt{Planned GTFS Generator} in the Ingestion Layer. The role of this module is to collect PT schedule data
and convert it if needed to one open format. The platform assumes storing schedule data in General Transit Feed Specification (GTFS) format~\cite{gtfs} due to its simple structure and popularity.  In our approach, the \texttt{Planned GTFS Generator} is activated daily to collect data of frequently changing schedules for the next $k$ consecutive days. The timetables created for the day $d+n,n>1$ are overwritten with their newer version based on downloads made on days $d+i,i<n$. This solution reduces the risk of not downloading some of the timetables due to 
data source failures.

\subsubsection{Travel time and demand matrices \label{sec:BatchTravelTime}}
If a city runs a transport model, in order to estimate the average time of travel between two points by car, we propose to make use of Travel Time By Car (TTBC) matrices from the transport model. In this way, differences in travel times between different times of the day can be accounted for.
Each TTBC matrix can be seen as a graph defined as $G = (V, E)$, where $V$ represents the set of zones, which the city area is divided into, and $E$ represents the edges, $e_i=(v_j,v_k),e_i\in E, v_j,v_k\in V$. Each edge is described by the travel time by car  between the two zones connected by the edge.

The TTBC matrices can be present in the transport model in two forms: containing estimated travel times under free vehicle flow conditions and under street congestion.
Moreover, for each hour of the day, a separate matrix can be present in the transport model. The TTBC matrices refer to a vector layer defining the spatial location of each transport zone. The boundary of each zone is defined by a polygon object. 
The TTBC matrices can be used to estimate travel time under street congestion between any origin and destination in the city area. In addition, we propose the use of travel demand (TD) matrices documenting the estimated demand for transportation between transport zones. Similarly to the TTBC matrices, these are also square matrices.

We propose to upload both the vector layer defining zone boundaries and TTBC and TD matrices to an RDBMS and use PostgreSQL in the reference implementation of the platform. 
We will use $M_\mathrm{TR}(o,d,h)$ to refer to TTBC matrices under traffic conditions, i.e. containing travel duration by car for individual zone pairs $(o,d)$ at different hours of a day $h$ and $M_\mathrm{NTR}(o,d,h)$ to refer to TTBC matrices under free vehicle flow conditions, i.e. when no street congestion is observed. The latter matrices  can 
reveal regular disturbances related to the time of the day.

\subsection{Speed layer}
The objective of the Speed Layer is to process GPS traces of PT vehicles. Based on planned GTFS schedules and GPS traces forwarded as a data stream, the \texttt{Vehicle Flow Processor} streaming application extends each newly received location record with additional features. These features are: destination stop, previously visited stop, next stop to be visited and delay (hereafter understood as a possibly negative difference between true and expected departure time) calculated by matching location record with the PT schedules. Notably, location records are not aggregated. Hence, multiple records at and between PT stops are typically present both in the input and the extended output location stream.
The application relies on a distributed Apache Flink processing engine.
As shown in~Fig.~\ref{fig:main_architecture}, the extended data stream is persisted in the Apache Hadoop cluster for future use. 
Hence, it complements the master data set by documenting the real functioning of a PT system at a micro scale of PT location traces with multiple readouts per minute for the same vehicle.

\subsection{Batch layer  \label{sec:BatchLayer}}
\begin{table}[tbh!]
\begin{center}
\caption{The summary of data resources proposed to be included in the master data set.
\label{tab:master_data_set}
}

\begin{tabular}
{@{}p{1.7cm}p{2cm}p{1.6cm}p{6.5cm}}
\hline
Data source&Data collected from the source&Suggested storage&Remarks\\
\hline
Survey data&Entire survey files&Flat files or RDBMS, depending on raw input format&In the case of input data decomposed into tables of e.g. respondents and trips, they can be reflected as tables in RDBMS; otherwise, flat files can be used, as these data are processed sequentially and random access is not needed. Moderate volume of data.\\
\hline
Transport model&TTBC and TD matrices&RDBMS&These data will be queried by zones between which trips were made. Hence, indexed storage is recommended. Moderate volume of data.\\
\hline
Weather and air pollution&Geolocated weather and air pollution time series & RDBMS & These data will be queried by locations and time. Hence, indexed storage is recommended. Moderate volume of data.\\

\hline
Built and natural environment&Spatial layers of streets, bike routes, green areas, walking paths and other related&
Geographical Information System (GIS) Server&Spatial data require  storage of multiple layers. Spatial queries have to be supported to identify the features of area around e.g. trip origin. Hence, a GIS server rather than a collection of layer files is recommended. Online use of external services such as OpenStreetMap is also possible.\\
\hline
Location of PT vehicles& GPS traces, extended in the Speed layer with delays and departure times from PT stops&Apache Hadoop&As the volume of data is large and sequences of readouts may have to be collected in 24x7 mode and periodically processed to  e.g. estimate departure times and deal with missing data, cluster-based scalable storage supporting batch processing of large volumes of data is recommended. Apache Hadoop and its HDFS can be used to address these objectives.\\
\hline
Public transport timetables& Schedules  including coordinates of PT stops and stations&Files in GTFS format and possibly tables in RDBMS&Schedules should be collected on a daily basis to reflect schedule changes. GTFS files can be directly used by engines such as OpenTripPlanner. As Apache Hadoop is used to host a distributed file system, it can be used to store also daily schedule files.\\
\hline

\end{tabular}

\end{center}
\end{table}

The primary role of the master data set is to contain data for future use when calculating features describing individual trips such as mode attributes, which depend on built environment data and in the case of public transport on schedules. The summary of data categories we propose to store in the master data set and suggested storage options are provided in Table~\ref{tab:master_data_set}.

\subsection{Serving layer \label{sec:ServingLayer}}

\underline{Real timetables.\label{sec:RealTimetables}}
Based on raw data present in the master data set, data needed to develop TMC models are calculated. 
One of the core categories of transformed data developed in the Serving Layer is the real timetables. These timetables are generated by \texttt{Real GTFS Generator} using vehicle location data persisted in HDFS and follow the solutions briefly discussed in \cite{grzenda2023ECML}. One real timetable per day is developed. Raw location records provide a basis for developing schedules not only containing  real departure times but also possibly reflecting the fact that some planned connections were not made at all e.g. due to vehicle failures.
We propose that the real timetables be placed in GTFS files in HDFS in the same way as the planned timetables. 
Let us note that the real timetables can  be developed  also when the planned timetables define the frequency of service such as metro service rather than exact departure times.

\underline{Public transport vehicle flow features.\label{sec:ServingPTFlowFeatures}}
The  role of \texttt{Vehicle Flow Extractor} \texttt{(VFE)} module is to calculate features quantifying the movement of PT vehicles. Similarly to \texttt{Real GTFS Generator}, the \texttt{VFE} module relies on tabular vehicle location data stored in HDFS as an input.
The output data are also tabular data, but this time aggregating vehicle location data for different pairs of consecutive PT stops. These data are stored in Apache Parquet format, i.e. an open-source, column-oriented format dedicated to storing big data \footnote{See \url{https://parquet.apache.org/docs/} for format details}. This is to enable queries referring to individual stops and periods of time.
Every row represents aggregated data between pair stops for a single vehicle and a single travel of this vehicle between the two stops.

Each Parquet file developed by the \texttt{VFE} module contains three groups of columns. The first group identifies a vehicle, the second group represents a pair of stops and the time of the travel of the vehicle between the stops, and the last group contains aggregated attributes of vehicle travel between the stops, including but not limited to average speed between the stops, departure time for each of the two stops, and delay at each of the two stops.
Moreover, we calculate two types of metrics related to  short and long  stops of a PT vehicle.
Whenever the vehicle's velocity drops below a threshold such as 5$km/h$, we identify this as a stopping event. Next, we measure the duration of a period when the speed is below the indicated speed.
We distinguish long and short stops by the duration of the stopping event. 
As an example, when processing the data for the City of Warsaw, we have introduced 4 thresholds: {30s, 60s, 90s, and 120s} which finally gave us 8 metrics quantifying the number of stopping events below and above each threshold for each pair of vehicle and transport system edge given by two consecutive stops. These features were used to quantify the perception of a traveller, i.e. how smooth the movement of a vehicle was.

\subsection{Service layer \label{sec:ServiceLayer}}

\begin{table}[tbh!]
\begin{center}
\caption{Feature engineering services present in the Service Layer of the UTMCM platform. 
\label{tab:service_layer_feature_services}
}

\begin{tabular}
{@{}p{2.3cm}p{3cm}p{3.5cm}p{3cm}}

\hline
Service name & Data calculated by the service & Input data & Data returned by service\\
\hline
\texttt{Private Transport Vehicle Flow Service} & Travel time by car  between $o(\mathcal{T})$ and $d(\mathcal{T})$ at  time $t_\mathrm{D}(\mathcal{T})$ considering possible street congestion and estimated parking difficulty in the area around trip destination at this time & $o(\mathcal{T}),d(\mathcal{T})$ and time of departure $t_\mathrm{D}(\mathcal{T})$ & The values of the following features: travel time by car in [sec] between $o(\mathcal{T})$ and $d(\mathcal{T})$, and the features quantifying estimated parking difficulty\\
\hline
\texttt{Environmental Service} & 
Average weather and air pollution parameters
& For trip data, $o(\mathcal{T})$ and the period before departure $[t_\mathrm{D}(\mathcal{T})-\Delta t,t_\mathrm{D}(\mathcal{T})]$. Calculation for other locations and time periods is also possible.
& For instance, average ambient temperature 
and average PM10 concentration in the period $[t_\mathrm{D}(\mathcal{T})-24h,t_\mathrm{D}(\mathcal{T})]$ 
around $o(\mathcal{T})$\\
\hline
\texttt{Built Environment Service} & Aggregated features of the area around trip origin and home of the traveller& For trip data, typically $o(\mathcal{T})$. Calculation for any location possible. & For instance, road density around trip origin $o(\mathcal{T})$, and distance to the nearest metro station from $o(\mathcal{T})$\\
\hline
\texttt{Public Transport Vehicle Flow Service}&Aggregated features of the movement of PT vehicles such as delays and stop events & 
Coordinates of public transport stops on the route $s_1,\ldots,s_n$, PT line $l$ and  the period before departure $[t_\mathrm{min},t_\mathrm{max}]$ e.g. the period of one of $k$ preceding days at the same time of the day as the time of the trip $t_\mathrm{D}(\mathcal{T})$ & For instance, average delay at the PT stops relevant for a trip $\mathcal{T}$\\
\hline
\end{tabular}
\end{center}
\end{table}

Once all relevant data are stored in the master data of the Batch Layer, 
they can be used for developing the data needed for the training and evaluation of TMC models. To make this possible, the Service layer includes novel software modules, designed and developed as a part of the UTMCM platform and  instances of open-source OpenTripPlanner \footnote{\url{https://www.opentripplanner.org}} service used to calculate routes and PT connections in urban areas. 

As far as the newly developed modules are concerned, the Service layer includes {\em feature engineering services} calculating the values of individual features to be used in the data fusion process.  The values of these features depend on the location of a trip and the home address of a person, but also the time of the trip to reflect the conditions prior to the trip in the relevant area. Hence, let $o(\mathcal{T})$, $d(\mathcal{T})$
denote geocoordinates of the origin and destination of a trip $\mathcal{T}$, respectively, and $t_\mathrm{D}(\mathcal{T})$ and $t_\mathrm{A}(\mathcal{T})$ departure and arrival time.
A summary of the feature engineering  services is provided in Table~\ref{tab:service_layer_feature_services}. 

Moreover, the layer includes \texttt{Survey Data Processor (SDP)}. The role of this module is to extract trip records from survey data and perform data fusion by calling the feature engineering services to obtain values of additional features such as estimated travel time by car under street congestion. Hence, 
the feature engineering services are called by \texttt{Survey Data Processor} to 
quantify conditions known before a trip and having a possible impact on the travel mode choices made for this trip.
Importantly, as raw data are stored in the master data, multiple experiments aimed at finding the best way of transforming raw data by the feature engineering services into features used by ML methods can be performed.
All the feature engineering services services 
hide the underlying complexity of the calculation of individual features that will be appended to survey-based features. 
Finally, let us note that not only the \texttt{Survey Data Processor} but also third-party applications can call the aforementioned services e.g. to request the calculation of built environment features values in the area of interest.

\subsubsection{Feature engineering services
\label{sec:FeatureEngineeringServices}}

\underline{Private Transport Vehicle Flow Service}.
This service
consists of two subservices. The first one estimates driving time by car at different times of the day to calculate the feature of a trip $F_\mathrm{d\_traffic}(\mathcal{T})$ (\texttt{DurationInTraffic\_CAR}) considering possible street congestion and related features such as the difference between this duration and the duration under free flow conditions $F_\mathrm{d\_free\_flow}(\mathcal{T})$ (\texttt{Duration\_CAR}). 
It estimates the  duration of trip $\mathcal{T}$ made by car in traffic as
$
F_\mathrm{d\_traffic}(\mathcal{T})=
F_\mathrm{d\_free\_flow}(\mathcal{T}) * \frac{M_\mathrm{TR}(O(\mathcal{T}),D(\mathcal{T}),h(\mathcal{T}))}{M_\mathrm{NTR}(O(\mathcal{T}),D(\mathcal{T}),h(\mathcal{T}))}$.
Hence, it multiplies the estimated travel duration under free flow conditions $F_\mathrm{d\_free\_flow}(\mathcal{T})$ while considering the exact $o(\mathcal{T})$ and $d(\mathcal{T})$ coordinates by the proportional increase 
in the travel duration for the zone pair $(O(\mathcal{T}),D(\mathcal{T}))$ between which the trip $\mathcal{T}$ is made at the hour $h$ of the day due to street congestion.
Hence, it considers both the real origin and destination of a trip and travel time following from these coordinates, and time-dependent transport model data to go beyond static zonal approaches, typically used in such cases as shown in~\cite{hillel2021}.

The second subservice estimates how difficult it is to find a parking place for different destinations.
It calculates  parking difficulty factors based on the number of arrivals to  the area around~$d(\mathcal{T})$.
The estimated number of arrivals from TD matrices is used.
Three methods are implemented to estimate the difficulty of finding a parking place, as described in~\cite{Grzenda2023}. This provides the values of three features to be used for training TMC models. All estimations are based on the processing of the data from the transport model on the number of arrivals to the zone in which the destination of the journey is and to the neighbouring zones. The intuition behind the features of a journey calculated in this way is that travellers may be reluctant to travel by car to the destinations many travel to, and finding a parking place may be more difficult in such zones.

\underline{Environmental Service}. The role of this service is to calculate averaged weather conditions and air quality features at the time preceding the travel.
When a request is made to the service, the raw time series data are first filtered to find measurements of the parameter of interest such as wind speed coming from the measurement point most approximate to the requested location.

\underline{Built Environment Service\label{sec:ServiceBuiltInfrast}}.
We define this service to calculate several urban features in the area around trip origin $o(\mathcal{T})$ and the neighbourhood of  the respondent's home.  It aims to provide additional data on the 
neighbourhood which may influence the habits of the person and TMC choices in turn. Data necessary to calculate the features are taken from the open spatial data shared by OpenStreetMap (OSM)\footnote{ \url{https://www.openstreetmap.org}}. This makes the service easily applicable to different cities. Let us note that this service calculates mostly the features of the built environment, but also natural environment features. However, for simplicity, we use the name reflecting its primary use.
The  features calculated by the service include inter alia the density of road network, address points, population,  and green area, as suggested in \cite{hagenauer2017}. Moreover, distances to PT stops of different categories e.g. distance to the nearest tram stop are calculated.

\underline{Public Transport Vehicle Flow Service.  \label{sec:ServicePTFlow}}
This service is 
responsible for 
the exposition of PT flow feature values. It provides service clients with the data on delays, vehicle stopping events and other features calculated by the \texttt{Vehicle Flow Extractor} application by querying  these data from the Parquet files.
Importantly, the requests refer to individual edges of the public transport graph i.e. the edges linking consecutive public transport stops e.g. bus stops. The service is used by the \texttt{Survey Data Processor} to calculate features aggregating the experience of a person travelling with PT on a certain route and e.g. suffering from delays or reduced comfort due to buses frequently stopping in a traffic jam when travelling this route. The features calculated in this way make it possible to take into account the varied scale of disturbances in different parts of the city, possibly influencing mode choices of travellers. As an example, one of the features calculated for a trip $\mathcal{T}$ by the \texttt{Survey Data Processor} and obtained by calling this service is \texttt{maxCurrentStopDelay\_LOW\_TRANSIT} i.e. the feature containing maximum delay in seconds observed in the period preceding the trip $\mathcal{T}$ at one of the stops of the connection from the origin to the destination of the trip $\mathcal{T}$. The feature aggregates past data for the same PT line(s) as the one(s) found suitable for the trip $\mathcal{T}$. Hence, whether a person possibly using the same line(s) for a similar trip suffered from delays in the preceding days is quantified by one of  the features added to a trip instance $\mathcal{T}$. Unlike the remaining feature engineering services, this service can be called by \texttt{Survey Data Processor} many times per trip, as there can be many PT connections suitable for a trip. Hence, the ultimate aggregation of its responses, which is necessary to develop feature values used for ML purposes, is performed by \texttt{Survey Data Processor}.

\subsubsection{OpenTripPlanner  \label{sec:ServiceOTP}}

The OpenTripPlanner (OTP)  engine can calculate for each pair $o(\mathcal{T}),d(\mathcal{T})$  both travel routes under different travel modes and their features such as distance and duration, and in the case of PT also candidate public transport connections for a trip. 
The OTP API  is called for every trip to calculate the  trip LOS. 
Two types of OTP instances are used in the UTMCM platform. The first one is based on official schedules for a given day, i.e. planned timetables. The second one is based on real timetables also provided as GTFS files. Both instances answer  queries from \texttt{Survey Data Processor} to enable the calculation of features of PT connections under planned and real trip circumstances.

\subsubsection{Survey Data Processor and ML modules}
The core module of the architecture combining data of different categories is \texttt{Survey Data Processor}. The module performs data fusion to extend trip records collected in surveys with the data documenting the conditions faced by travellers and possibly influencing their choices. 
The way the data fusion process is performed is described in the next section.

The data sets developed by the \texttt{Survey Data Processor} can be used as input for calculating aggregate statistics, such as empirical cumulative distribution functions of feature values such as travel duration assuming the use of different means of transport. Another aspect, which we focus on in this study is the use of data fusion performed by SDP to perform travel mode choice modelling with machine learning techniques. A batch learning module  was implemented and used in this work.

\section{Data fusion for travel mode choice modelling \label{sec:fusion}}

\subsection{Core solutions}

\begin{algorithm}[ht!]
\caption{Development of fusion-based trip data  \label{alg:main_method}}
\SetNoFillComment
\KwIn{
$D$ - a set of survey records, $\Omega$ - master data set collected by the platform, $L$ - the list of travel modes other than PT, by default $L=\{\mathrm{WALK},\mathrm{CYCLE},\mathrm{CAR}\}$, $\delta_\mathrm{S},\delta_\mathrm{F}$ - parameters of PT mode choice time window
}
\KwData{
$a(r)$ - the number of journeys present for the $r-th$ respondent, 
$\mathcal{A}=\mathcal{T}_1,\mathcal{T}_2,...$ - the sequence of trips extracted from a data set $D$ 
}
\KwResult{$\mathcal{S}_1,\mathcal{S}_2,\ldots$ - stream of labelled instances $(\mathbf{x},y)$, each describing one trip with the vector of features $\mathbf{x}$ and transport mode $y$ used for it}
\Begin{
$\mathcal{A}=\phi$\;
\tcc{For every respondent}
\For{$r=1,\ldots$}
{
 \tcc{Extract trip records from $r$-th survey record.} 
$\{\mathcal{T}_1,\mathcal{T}_2,\ldots,\mathcal{T}_{a(r)}\}=DevelopTrips(D^r)$\;
$\mathcal{A}=\mathcal{A}\cup \{\mathcal{T}_1,\mathcal{T}_2,\ldots,\mathcal{T}_{a(r)}\}$\;
}
$\mathcal{T}_1,\mathcal{T}_2,\ldots,\mathcal{T}_{card(\mathcal{A})}=SortByDateAndTime(\mathcal{A})$\;}
\For{$k=1,\ldots,card(\mathcal{A})$}
{
\tcc{Develop  vector with the features based on survey data}
  $\mathbf{x}[\mathrm{SURVEY}]=FeatureVector(\mathcal{T}_k)$; $\mathbf{x}=\mathbf{x}[\mathrm{SURVEY}]$\label{alg:survey_line}\;
\tcc{Calculate and append mode-choice features}
  \For{$\mathcal{M} \in L$\label{alg:fusion_line}}
{
$\mathbf{x}=[\mathbf{x},DevelopTransportFeatures(\mathcal{T}_k,\mathcal{M})]$\; 
}
  
   \tcc{Append travel time by car considering road congestion and parking difficulty features}
  $\mathbf{x}=
  [\mathbf{x}%
,DevelopTransportFeatures(\mathcal{T}_k,\mathrm{E\_CAR\_LOS},\Omega)]$\label{alg:car_e_line}\;
  \tcc{Build  vector containing the values of aggregate features of candidate planned PT connections}
  $\mathbf{x}[\mathrm{PLAN\_PT\_LOS}]=DevelopPTFeatures(\mathcal{T}_k,\mathrm{PLAN\_PT\_LOS},\Omega,\delta_\mathrm{S},\delta_\mathrm{F})$ \label{alg:planned_pt_los_line}\;
  \tcc{Build  vector containing the values of aggregate features of past real PT connections}
  $\mathbf{x}[\mathrm{REAL\_PT\_LOS}]=DevelopPTFeatures(\mathcal{T}_k,\mathrm{REAL\_PT\_LOS},\Omega,\delta_\mathrm{S},\delta_\mathrm{F})$\label{alg:real_pt_los_line}\;
    \tcc{Build feature vector describing past movement of PT vehicles including delays}
  $\mathbf{x}[\mathrm{PT\_EXPERIENCE}]=DevelopTransportFeatures(\mathcal{T}_k,\mathrm{PT\_EXPERIENCE},\Omega)$\;
$\label{alg:env_line} \mathbf{x}[\mathrm{ENV}]=[CalcFeatures(\mathcal{T}_k,\mathrm{WEATHER},\Omega),CalcFeatures(\mathcal{T}_k,\mathrm{POLLUTION},\Omega)]$\;

$\label{alg:built_line}\mathbf{x}[\mathrm{BUILT\_INF}]=CalcFeatures(\mathcal{T}_k,\mathrm{BUILT\_INF},\Omega)$\;

$\mathbf{x}=
[\mathbf{x},
\mathbf{x}[\mathrm{PLAN\_PT\_LOS}],\mathbf{x}[\mathrm{REAL\_PT\_LOS}]]$\;
$\mathbf{x}[\mathrm{DIFF}]=DevelopDifferentialFeatures(\mathbf{x})$\;
$\mathbf{x}=
[\mathbf{x},\mathbf{x}[\mathrm{PT\_EXPERIENCE}],\mathbf{x}[\mathrm{ENV}], \mathbf{x}[\mathrm{BUILT\_INF}],\mathbf{x}[\mathrm{DIFF}]]\label{env:label}\label{alg:final-concatentation-line}$\;
  
  $\mathcal{S}_k=(\mathbf{x},y)$\; 
}
\end{algorithm}
The process of data fusion which we propose in this work and which is implemented by \texttt{SDP} module is described in Alg.~\ref{alg:main_method}. 
The core input for the algorithm is a trip data set $D$ collected under the RP approach, i.e. documenting real trips made in the urban area of interest.  The algorithm also relies  on the data previously collected by the platform in the master data set $\Omega$ for the period during which real trips were made and the period preceding it. Furthermore, $\delta_\mathrm{S},\delta_\mathrm{F}$ define the period $[t_\mathrm{D}(\mathcal{T})-\delta_\mathrm{S},t_\mathrm{D}(\mathcal{T})+\delta_\mathrm{F}]$ to be considered when identifying PT connections considered suitable for a trip starting at time $t_\mathrm{D}(\mathcal{T})$. As an example, if $\delta_\mathrm{S}+\delta_\mathrm{F}=15min$ then PT connections within a 15-minute {\em mode choice window} are considered relevant for the trip. 

Let us note that travel mode choices $y$ under the same conditions $\mathbf{x}$ may change with time. Moreover, the probability of observing different travel conditions $p(\mathbf{x})$ may also change, i.e. both virtual concept drift \cite{ditzler2015} (a change in $p(\mathbf{x})$) and real concept drift \cite{ditzler2015} (a change in $p(y|\mathbf{x})$)  may occur. 
Hence, trips provided by respondents are extracted from input trip data $D$ and sorted based on travel start dates and times to form a sequence of trips $\mathcal{A}$. This is to produce a stream of labelled instances $(\mathbf{x},y)$ as an outcome of Alg.~\ref{alg:main_method} revealing trips made over time and providing a basis for the use of incremental machine learning methods, possibly responding to concept drift by updating travel mode choice models. When batch machine learning methods are used, all instances produced by Alg.~\ref{alg:main_method} are placed in the data set to be used for the training and evaluation of a TMC model. Importantly, instances $\mathbf{x}$ include both survey features and fusion-based features calculated by SDP. Moreover, the ultimate dimensionality $dim(\mathbf{x})$ is a sum of the number of survey features and the number of fusion-based features. These depend on survey data used as an input of Alg.~\ref{alg:main_method} and the content of master data set $\Omega$ for the region of the survey, respectively. As an example, the number of features quantifying air pollution depends on the availability of air pollution data. Hence,  features calculated for the reference data used in this work will be discussed in Sect.~\ref{sec:Experiments}.
In addition, the data sources influencing the content of $\Omega$ and the settings of the platform applied for the City of Warsaw are provided in Appendix~\ref{appendix:Warsaw}.

In line \ref{alg:survey_line} of Alg.~\ref{alg:main_method}, the answers provided by respondents for individual questions related to a trip, and possibly to a person and their household are converted into feature values.
The data fusion process starts in line~\ref{alg:fusion_line} of the algorithm. For each travel mode other than the use of public transport, such as walking and cycling, LOS feature values are calculated. 
This is done by submitting requests to OpenTripPlanner, which estimates e.g. walking duration while considering the spatial data for the city. 
The LOS features calculated with OpenTripPlanner include inter alia distance and duration for walking and cycling, but also elevation changes based on the elevation model.
For car travel, LOS features include estimated travel time, average vehicle speed under free flow conditions, distance, walking distance to and from the parked car, where the parking place is estimated as the nearest point on a public road from the travel origin and destination. 
Importantly, the values of LOS features for these travel modes do not depend on PT schedules. Next, in line \ref{alg:car_e_line}, additional LOS features for a trip assuming it was made by a car including parking difficulty features and travel time under road congestion conditions are calculated. This is done by calling the \texttt{Private Transport Vehicle Flow Service}.

In lines \ref{alg:planned_pt_los_line} and \ref{alg:real_pt_los_line}, LOS features assuming the use of public transport, including the use of multimodal connections, are calculated. We propose calculating two groups of features. The first of these groups are PLAN\_PT\_LOS features, i.e. the LOS features aggregating the features of PT connections starting during the mode choice window of a trip $\mathcal{T}_k$ and available under planned PT system behaviour as defined by planned schedules for the day of the trip $\mathcal{T}_k$.
Importantly, this relies on schedules
downloaded into the master data set before the actual time of  travel.
In this way, candidate planned connections and their features, such as travel duration and the number of transfers that might have been known to a traveller prior to each journey are determined.
However, travellers' decisions can also be influenced  by the actual prior system behaviour. Hence, another OTP instance configured with the real timetables produced by \texttt{Real GTFS Generator} is called to determine the actually feasible connections on the day(s) preceding each journey. In this way, the actual travel duration, possibly including delays and other features such as waiting times related to the use of actual connections of which the traveller might have been aware, are determined. 
Importantly, both PLAN\_PT\_LOS and REAL\_PT\_LOS features are calculated in two stages. First, candidate PT connections within the mode choice time window given by $\delta_\mathrm{S},\delta_\mathrm{F}$ are determined using an OpenTripPlanner instance. Next, the features of these connections are aggregated into the features placed in $\mathbf{x}[\mathrm{PLAN\_PT\_LOS}]$ and $\mathbf{x}[\mathrm{REAL\_PT\_LOS}]$ vectors, respectively. An example of such an ultimate feature is \texttt{minDuration\_TRANSIT}, which contains the minimum duration of a trip from $o(\mathcal{T}_k)$ to $d(\mathcal{T}_k)$ relying on PT, i.e. assuming the use of the fastest PT connection starting during the mode choice window $[t_\mathrm{D}(\mathcal{T}_k)-\delta_\mathrm{S},t_\mathrm{D}(\mathcal{T}_k)+\delta_\mathrm{F}]$ according to planned schedules available for the day of the trip.

As a consequence, two groups of features, i.e. the features $\mathbf{x}[\mathrm{PLAN\_PT\_LOS}]$ documenting planned PT system behaviour for the period of travel and the features $\mathbf{x}[\mathrm{REAL\_PT\_LOS}]$ reflecting real PT system behaviour on the day(s) preceding the journey during the same time of the day are determined. In both cases, these features are calculated for the connections linking the origin and the destination of a reported trip $\mathcal{T}_k$.  However, PT features other than travel duration  may also affect travellers' decisions. These include delays and frequent stops of buses during congestion periods, which reduce the comfort of travellers. Hence, we propose the calculation of PT\_EXPERIENCE features summarising the real movement of PT vehicles in the days preceding travel. This is done by calling \texttt{Public Transport Vehicle Flow Service}, which extracts and aggregates vehicle trajectory data for the same   sequences of PT stops and lines as those which could potentially be used for the trip. 
The aggregation is performed on data from the day(s) preceding the trip.

Next, in line~\ref{alg:env_line}, weather features and air pollution features are calculated. This is done by calling \texttt{Environmental Service}, which is requested to return aggregate weather and pollution values. For each parameter such as  wind speed, data from the measurement point closest to the origin of the trip $o(\mathcal{T}_k)$ for which data for the period preceding a trip exist, i.e.  were successfully collected and stored in the master data set, are used. This provides for the increased resilience of the platform when some sensors fail to capture relevant data.
In line~\ref{alg:built_line}, this is followed by calling 
\texttt{Built Environment Service} to calculate the values of features summarising the area around the trip origin and the home of the traveller, as both groups of features may influence mode choices.

Let us note that to calculate all PT-related feature groups, schedules have to be collected both for the days for which trips were reported in the surveys (PLAN\_PT\_LOS features) and for the days preceding these trips 
(REAL\_PT\_LOS and PT\_EXPERIENCE features). Similarly, GPS traces of PT vehicles, weather and pollution data for the periods preceding the trips have to be collected. This confirms the need for collecting master data set $\Omega$.

Next, we propose calculating {\em differential features}. By these these features we mean features explicitly showing e.g. the difference in travel duration between the trip made by car and the trip to the same destination made with PT and the ratio of these durations. First of all, such features are likely to help build better TMC models, as for some of the models such as tree-based models, capturing dependencies based on the quotient of two variables may be  difficult when relying on each feature separately. Secondly, such features may result in simpler models, which in consequence are easier to interpret.

Finally, in line \ref{alg:final-concatentation-line} the values for all features are placed in one vector. Hence, for each trip documented in a trip diary, the values of LOS features under different travel modes, weather, pollution and built-environment features are collected.

\subsection{Feature groups}

\begin{table}[tbh!]
\begin{center}
\caption{Feature sets calculated in Alg.~\ref{alg:main_method}. All feature sets are calculated by the UTMCM platform \label{tab:fsets}. Data categories: SR - survey data, SP - spatial data, SC - schedules, TS - time series, TD and TTBC - matrices from transport model
}
\begin{tabular}
{@{}p{2.8cm}p{7.8cm}p{2cm}}
\hline
Feature set  
& 
Features in the set & Input data\\
\hline

SURVEY 
& Survey-based features, i.e.  features containing  values provided in the survey 
& SR\\
\hline
WALKING\_LOS,
CYCLING\_LOS, 
CAR\_LOS,
...  & LOS attribute values calculated for travel modes other than PT, based on spatial data of inter alia streets, walking paths and cycling paths &SP \\
\hline
E\_CAR\_LOS & Additional car-related features including parking difficulty features and estimated travel time by car, taking into account street congestion& SP,  TD, TTBC \\
\hline
PLAN\_PT\_LOS & LOS attribute values calculated for PT based on  spatial data and planned PT schedules for the day of the trip &  SP, SC \\
\hline
REAL\_PT\_LOS & LOS attribute values calculated for PT based on both spatial data and real schedules recreated from GPS traces 
for the day(s) preceding the trip & SP, SC, TS (GPS traces) \\
\hline
DIFF & Differences and ratios  based on LOS attributes calculated for different travel modes & LOS features for different travel modes \\
\hline
PT\_EXPERIENCE & Features quantifying  actual PT vehicle trajectories, such as average speed and number of short and long stops of vehicles, calculated based on GPS traces of PT vehicles for the day(s) preceding the trip & TS (GPS traces)\\
\hline
WEATHER & Weather features, e.g. average ambient temperature in the area of trip origin in 24 hours before the trip& TS (weather data)\\
\hline
POLLUTION & Air pollution features, e.g. average PM2.5 concentration in the area of trip origin in 2 hours before the trip& TS (pollution data)\\
\hline
BUILT\_ENV & Built environment features such as the prevalence of green areas around trip origin
& SP\\
\hline
\end{tabular}
\end{center}
\end{table}

The summary of feature sets which we propose in this work to quantify the circumstances possibly influencing travel mode choices for individual trips is provided in Table~\ref{tab:fsets}.
These features are calculated by Alg.~\ref{alg:main_method} and its reference implementation in the UTMCM platform. The calculation of different feature sets necessitates the availability of different input data. 

\section{Training and evaluation of TMC models \label{sec:training}}

\begin{algorithm}[ht!]
\caption{The training and  evaluation of TMC models using feature set scenarios
\label{alg:evaluation}}
\KwIn{$D$ - matrix of $n$ feature vectors containing the values of $d$ features,
$P \in M^n$ - vector of corresponding travel modes $m_i \in M$, 
 $S$ - modelling scenario defining which features to use to develop TMC models,
 $\mathcal{M}$ - set of machine learning methods to use,
$K$ - the number of CV folders, 
 $\mathcal{P}=\{\mathcal{P}_i,i=1,\ldots,N_\mathrm{P}\}$ - performance measures to be calculated, $Q\in \{1,\ldots,N_\mathrm{P}\}$ - the performance measure used to select the best ML method and its settings,
 $\alpha\in (0,1)$ - the share of most recent trip instances to be used for final evaluation
}
 \KwResult{
 $P_S^{\mathcal{P}_j,\mathrm{F}}$
- performance of the best TMC model calculated on final trips set and 
$P_\mathrm{S}^{\mathcal{P}_j,\mathrm{CV}}(\mathbb{M})$ - average performance of the best models under ML method $\mathbb{M}$ calculated on CV test folds, both quantified with $j$-th performance measure
}
\Begin{
$(D,P)$ = SortByTripDateAndTime$(D,P)$\;
 \tcc{Leave in $D$  only features present in scenario $S$}
$D$ = SelectFeatures$(D,\mathcal{F}(S)) \label{alg:evaluation:featuresel_line}$\;
\tcc{Remove at least $\alpha\times n$ last (most recent) trip instances from $(D,P)$ to put them in $(D_\mathrm{F},P_\mathrm{F})$ and use them for final tests}
$((D,P),(D_\mathrm{F},P_\mathrm{F}))$ = Split$(D,\alpha)\label{alg:evaluation:data_split}$\;
	$\{D_j,P_j\}_{j=1,\ldots,K}$ = DivideSet($D$,$P$,$K$)\;
\For{$\mathbb{M} \in \mathcal{M}$ \label{alg:evaluation:ML_methods}}
{
 \For{$k=1,\ldots, K$}{

	   $D_\mathrm{T}$ = $D_k$; $D_\mathrm{V}$ = $D_{(k+1) \; mod \; K}$\;	   
	   $D_\mathrm{L}$ = $\cup_{j\in\{1,\ldots,K\}-\{k\}-\{(k+1) \; mod \; K\}} D_j$\;	   
\For{$\mathbf{h} \in hyperVals(\mathbb{M})$}{
	   $M=train(\mathbb{M},\mathbf{h},D_\mathrm{L},P_\mathrm{L},D_\mathrm{V},P_\mathrm{V})$\;
    
    \For{$j=1,\ldots, N_\mathrm{P}$}{
 
	   $P^{\mathcal{P}_j,\mathrm{CV}}
    (\mathbb{M},\mathbf{h}) = P^{\mathcal{P}_j,\mathrm{CV}}
    (\mathbb{M},\mathbf{h}).append(\mathcal{P}^{j}(M(D_\mathrm{T}),P_\mathrm{T}))$\;
    }
    }
	  }
   }
 \For{$j=1,\ldots, N_\mathrm{P} \label{alg:evaluation:mean_HP}$}{
 \For{$\mathbb{M} \in \mathcal{M}$}{
 \For{$\mathbf{h} \in hyperVals(\mathbb{M})$}{
 	$P_\mathrm{S}^
  {\mathcal{P}_j
    ,\mathrm{CV}}(\mathbb{M},\mathbf{h}) = mean(P^{\mathcal{P}_j,\mathrm{CV}}(\mathbb{M},\mathbf{h}))$ \;
 }
 	$P_\mathrm{S}^
  {\mathcal{P}_j
    ,\mathrm{CV}}(\mathbb{M}) = PerfUnderBestHP\Big(\{P_\mathrm{S}^{\mathcal{P}_j,\mathrm{CV}}(\mathbb{M},\mathbf{h}),\mathbf{h}\in hyperVals(\mathbb{M})\},D\Big)$ \;
 }
 }
 \tcc{Find the best ML method and its settings on $(D,P)$ data}
$(\mathbb{M}_\mathrm{B},\mathbf{h}_\mathrm{B})=FindBestMethodAndHPValues(P_\mathrm{S}^{\mathcal{P}_Q,\mathrm{CV}})$\;
\tcc{Train the final TMC model with the best ML and HP settings}

$M^\mathrm{F}=train(\mathbb{M}_\mathrm{B},\mathbf{h}_\mathrm{B},D,P)$\;
 \tcc{Evaluate the final TMC model using $D_\mathrm{F}$ data}
 \For{$j=1,\ldots, N_\mathrm{P}$}{
 $P^
 {\mathcal{P}_j
 ,F}_\mathrm{S} = \mathcal{P}^{j}(M^\mathrm{F}(D_\mathrm{F}),P_\mathrm{F})$\;
 }
}
\end{algorithm}

What follows from Table~\ref{tab:fsets} is that to calculate different feature sets, different raw data have to be obtained. 
As in different cities, the distributions of feature values and their impact on mode choices are likely to differ, the question of the importance of different feature subsets in terms of their impact on TMC models can be asked. By understanding this impact we could better decide which feature sets are particularly important and could even possibly be extended with additional features. To answer the question of the importance of different feature sets, TMC models can be developed  with machine learning methods under different scenarios, i.e. with different subsets of all possible features comprising the input data used to develop TMC models. For each scenario, the best modelling technique and the values of its hyperparameters should be determined. Next, the TMC model developed with these settings for a given scenario should be evaluated.

Hence, when machine learning methods are to be used to develop TMC classifiers, the question arises of how to select the best machine learning method and select its hyperparameter values to train and evaluate TMC models. 
An important aspect is to avoid the risk of data leakage~\cite{Chen2023}. This could happen if some trip records from the same respondent were used for the training of a model while the remaining ones were used for the testing of this model~\cite{hillel2021}. All these decisions  are directly related to the evaluation of the impact of using fusion-based trip data and different subsets of features on the performance of TMC models. Hence, let us summarise the process of  evaluation of the impact of the use of different feature sets arising from data fusion on the performance of TMC models. This process is proposed in Alg.~\ref{alg:evaluation}. It extends  methods used to evaluate TMC models inter alia in \cite{garcia2022,Chang2019,LondonDataset2018} by enabling  analysis of the impact of different feature subsets on the performance of TMC models developed with them.

The key input for the algorithm is trip data set $D$ and the modelling scenario $S$, defining which features out of all present in the set $D$ should actually be  used for the development and evaluation of models.
First, all $n$ trip instances present in $D$ are sorted by trip dates and times.  Next, in line~\ref{alg:evaluation:featuresel_line}, only features indicated by scenario $S$ are left in the data. In line \ref{alg:evaluation:data_split}, the most recent trip instances are left for evaluation purposes and placed in the holdout data $(D_\mathrm{F},P_\mathrm{F})$, which is in line with the approach used for the London data in \cite{LondonDataset2018,Chen2023}. 
The holdout data includes both trip data and corresponding true labels, i.e. travel modes. 

In line \ref{alg:evaluation:ML_methods}, the main loop of the model development starts. For different classification methods~$\mathcal{M}$, $K$-fold cross-validation is performed, within which models with different hyperparameter values are developed and their performance on testing folds $D_\mathrm{T}$ is calculated. This provides the basis for the calculation of the averaged performance of the models trained with each method  $\mathbb{M}\in\mathcal{M}$ under its different hyperparameter settings. 

Finally, the best machine learning method $\mathbb{M}_\mathrm{B}$ and its best hyperparameter settings $\mathbf{h}_\mathrm{B}$ are determined. This relies on the results obtained on the testing folds of cross-validation performed on the subset  $D$ of the original trip data. The same subset $D$ is used to train the final model $M^\mathrm{F}$. Finally, the performance of this model on the holdout data set of final trips $(D_\mathrm{F},P_\mathrm{F})$ is quantified with different performance measures $\mathcal{P}^{j}$, such as accuracy and $\kappa$.
Importantly, this performance is influenced by the set of features given by modelling scenario $S$, i.e. the set of features used for the training  of the models and providing the basis for determining the best ML method  under scenario $S$. In the remainder of this work, we will refer to the $(D_\mathrm{F},P_\mathrm{F})$ data as {\em final testing data}.

\section{Experiments\label{sec:Experiments}}
\subsection{Testbed environment}
To  run all systems, services and applications of the UTMCM platform and collect and preprocess all data used in experiments, a server with 1.5TB RAM and 96 cores was used. In particular, seven virtual machines with 578GB RAM and 22 cores were used to perform 24x7 data collection, aggregate data with feature engineering services, and run six OpenTripPlanner instances to estimate connections under both planned and real schedules for different survey periods. Moreover, a dedicated VM with 12 cores and 96GB RAM was used to perform data fusion with \texttt{Survey Data Processor}. 
The remaining resources of the server were used for experiments with  ML techniques and data preprocessing methods.

\subsection{Reference trip data sets}

To validate the platform proposed in this work and evaluate the impact of data fusion on TMC models, we  selected survey data sets collected under the RP approach in the City of Warsaw, Poland in 2022-2023. The data sets include travel diaries documenting travel modes selected by travellers for  trips made in Warsaw.

\begin{table}[ht]
    \centering
    \caption{Summary of reference survey data}
    \label{tab:survey:info}
   \begin{tabular}{lllrr}
   \hline
Survey	 &
Started	&
Ended	&
No of. respondents &
No of. described trips \\ \hline   
PARENTSW1 & 2022-03-15 & 2022-04-28 & 523&	1861\\
CITIZENSW1  & 2022-05-10 & 2022-06-23 & 1170&	2961\\
PARENTSW2 & 2022-11-22 & 2022-12-15 & 316&	798\\
CITIZENSW2  & 2023-05-02 & 2023-06-15 & 1157&	3316\\
\hline
    \end{tabular}
\end{table}

The survey data were collected for representative samples of different respondent groups, i.e.  a sample of parents of primary school pupils from three selected schools (PARENTSW1), a sample of parents of primary  school children from all schools (PARENTSW2), and a sample of Warsaw city citizens (CITIZENSW1 and CITIZENSW2). The PARENTS* data sets can  reveal the impact of one of the stages of life on travel mode choices. Such  impact has been shown in some studies, as follows from the preliminary analysis in Sect.~\ref{sec:RelatedWorks}.

All data sets include travel diaries documenting trips made by a person during one working day, including their origins, destinations and transport modes selected for these trips. The survey data also contain social and economic information about a respondent and their household. This includes features influencing travel mode choices such as income, whether the person has a driving licence and the number of household members. 
The data sets enable the comparison of the results of TMC modelling and the role of different feature sets across populations with potentially different behaviours. 
The summary of survey data sets is given in Table~\ref{tab:survey:info}. While  PARENTSW2 can be classified as a small data set according to the taxonomy of TMC data sets proposed in the survey on TMC modelling~\cite{hillel2021}, the remaining three data sets used in our study can be classified as medium data sets (1,000-10,000 entries) according to the same survey. Hence, they fall into the category of data sets most frequently used for TMC modelling with ML techniques, according to the same work~\cite{hillel2021}. Finally, let us add that some of these survey data sets were used in the study proposing  parking-related features for TMC modelling~\cite{Grzenda2023}.

\subsection{Development of fusion-based trip data}

\begin{table}[ht]
    \centering
    \caption{Summary of fusion-based trip data sets used for the training and evaluation of TMC models. Columns C(ar), P(public) T(transport), W(alk), and B(ike) present the percentage of trips with a given mean of transport in the dataset. Other modes were skipped due to insufficient number of trips. CITIZENS represents merged surveys CITIZENSW1 and CITIZENSW2.}
    \label{tab:MLdataset:info}
  \begin{tabular}{llrrrrrrrr}
  \hline
   \multicolumn{6}{c}{}& \multicolumn{4}{c}{Transport modes [\%]}\\ 
\hline   
Data set  & Surveys	 & \#Resp. & $|D|$ & $|D_\mathrm{F}|$ & $dim(D)$
&C&PT&W&B\\ \hline   
PAR\_W1& PARENTSW1 & 489&1697&555&317&47.4&19.4&28.4&4.8\\ 

PAR\_W2& PARENTSW2 & 274&660&152&320&38.1&30.1&30.9&0.9\\

CIT\_W1&CITIZENSW1  & 1096&2675&653&320&33.4&43.0&19.6&4.0\\
CIT\_W2&CITIZENSW2  & 1094&2996&767&320&30.7&42.8&21.7&4.8\\
CIT\_W1\_2&CITIZENS  & 2190&5671&1190&320&31.9&42.9&20.7&4.4\\
\hline
\end{tabular}
    
\end{table}

We used raw survey data sets listed in Table~\ref{tab:survey:info} to develop data sets used for travel mode choice modelling. First of all, only trips made in the city area were selected out of all those reported in travel diaries. Moreover, survey data sets were processed with the UTMCM platform using Alg.~\ref{alg:main_method}. As a consequence, every trip instance originally including only survey-based features was extended by appending the values of features of all types listed in Table~\ref{tab:fsets}.  In this way, fusion-based trip data sets documented in Table~\ref{tab:MLdataset:info} were obtained. The impact of life stage on mode choices can be clearly observed in the case of PARENTS* data sets, as the most frequently used travel mode in these sets is car, which is unlike in the case of CITIZENS* data sets.
Let us note that the set of survey questions was extended after performing the PARENTSW1 survey, which provided an increased number of survey features in the remaining survey data sets.

Furthermore, as CITIZENSW1 and CITIZENSW2 survey data sets document trips made by  representative samples of citizens of Warsaw and are the outcomes of two waves of the survey, one more data set CIT\_W1\_2, i.e. the set of trips combining data from both waves of the Warsaw survey was included. 
The set illustrates the impact of using an increased number of trip records from different periods of time on model development. The summary of TMC data sets used for TMC model development and evaluation, including the total number of all instances $|D|$ and the number of holdout instances selected for the final evaluation of TMC models $|D_\mathrm{F}|$ out of instances available in $D$, is provided in Table~\ref{tab:MLdataset:info}. Let us note that the dimensionality of data $dim(D)$, i.e. the number of features  present in each data set, is equivalent to the number of all features present in instances $\mathcal{S}_k$ produced by Alg.~\ref{alg:main_method}.

\begin{figure}[th]
\centering
         \includegraphics[width=\textwidth]{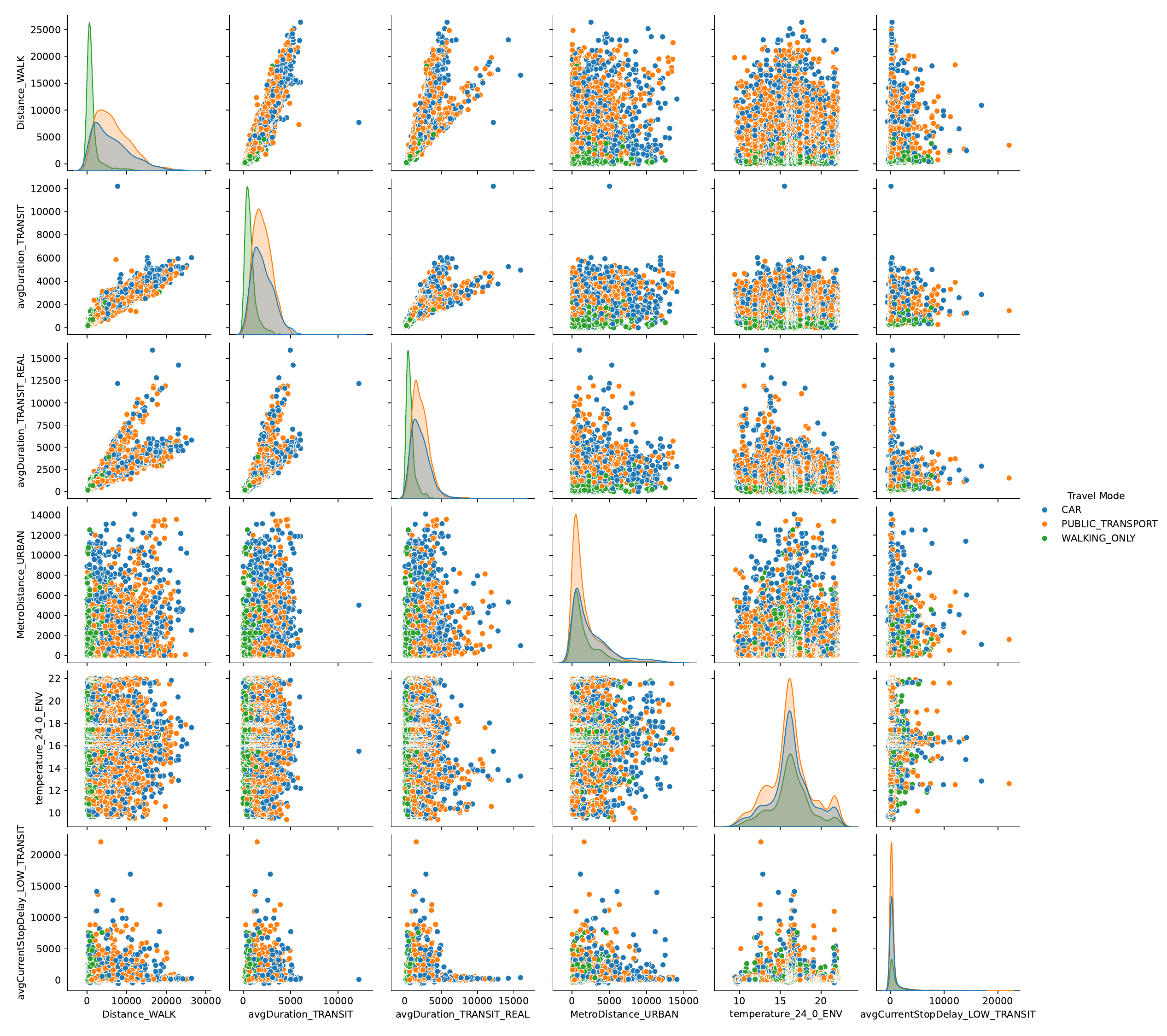}
         \caption{The distributions of selected fusion-based features. CIT\_W1\_2 data.}
         \label{fig:sample_feature_plots}
     \end{figure}

 The distributions of selected feature values obtained in this way are provided in Fig.~\ref{fig:sample_feature_plots}. On the main diagonal of a matrix, the distributions of the values of a feature of interest for trips performed with different transport modes are shown.
As an illustration, we have selected various features for the plot, inter alia the distance in meters to the trip destination, assuming a person walks to it (\texttt{DISTANCE\_WALK} feature, an example of the WALKING\_LOS feature set).  It can be observed that low values of this feature are largely related to trips actually walked by a person.
The next feature, \texttt{avgDuration\_TRANSIT}, is 
the average duration of the trip in seconds across different connections possible with public transport assuming planned schedules, including both the time of walking needed to use these connections and waiting times. This is one of many features from the PLAN\_PT\_LOS group. The \texttt{avgDuration\_TRANSIT\_REAL}, unlike the former feature, was calculated in the same way, but using real schedules, i.e. it reflects connections feasible in the past based on GPS traces of public transport vehicles. It can be observed that the duration of some of these connections is likely to be much longer due to the longer tail of the distribution. This reflects the impact of delays on overall travel duration, which may also cause missed transfers.
This feature illustrates one of the features from the REAL\_PT\_LOS feature set. The other features shown in the plot are \texttt{MetroDistance\_URBAN}, i.e. the distance to the metro station nearest the trip origin. It can be observed on the main diagonal of the matrix that low values of this feature from the BUILT\_ENV feature set contribute to the high use of public transport.

Finally, the  last two features are average ambient temperature during the 24 hours preceding the time of departure (sample feature from the WEATHER group) and average delay observed at public transport stops present in the PT connections relevant for the trip. The distribution of  the  \texttt{avgCurrentStopDelay\_LOW\_TRANSIT} feature reveals that most connections are not affected by significant delays, but occasionally major delays occur. The latter feature is an example of PT\_EXPERIENCE features quantifying the movement of PT vehicles documented by GPS traces.

 \subsection{Training and evaluation of TMC models}
Let us investigate the impact of the use of various feature sets on the performance of TMC models.
 To train and evaluate the TMC models relying on different modelling scenarios, we  used trip data sets $D$ documented in Table~\ref{tab:MLdataset:info}.
For each combination of a TMC data set $D$ and modelling scenario $S$, multiple travel mode choice models were developed as defined in Alg.~\ref{alg:evaluation}.

While the objective of this work was not to search for the best machine learning methods for TMC modelling, by evaluating multiple candidate machine learning methods typically used to train classifiers, we aimed to reflect the usual process of selecting the most promising models to estimate the expected performance of classifiers given training data and features present in it.
Hence, the following machine learning methods $\mathcal{M}$ were used to develop the models: Naive Bayes, CART decision trees (DT) (rpart implementation \cite{rpart}), random forest (RF), ranger~\cite{ranger}, support vector machine with linear and radial kernel, multi layer perception (MLP), $k$ nearest neighbour (kNN), and XGBoost~\cite{Chen2016} (XGBTree).
These included the methods found to yield the best TMC models in prior works, i.e. random forest \cite{Breiman2001} and XGBoost~\cite{Chen2016,LondonDataset2018}.
These methods were used in the first stage to train and evaluate multiple candidate models in the cross-validation loop of  Alg.~\ref{alg:evaluation}. In this way, for each data set and the set of features selected to be placed in this data set, the most promising method $\mathbb{M}_\mathrm{B}$ and the values of its hyperparameters  $\mathbf{h}_\mathrm{B}$ were found. The hyperparameters were adjusted according to the machine learning method used. For each method, ten sets of parameters were used and the best set was chosen.
The criterion for selecting such a combination was maximising the value of Cohen's $\kappa$~\cite{japkowicz2011}, which was used as the performance measure $\mathcal{P}_Q$ in Alg.~\ref{alg:evaluation} to better match imbalanced data. 
In this way,  the selection of models was driven by the proportion of correctly predicted labels above chance agreement each model provided, rather than the value of its accuracy measure.
However, accuracy was the other calculated measure. Furthermore, $K=10$ and $\alpha=0.2$ were used. Moreover, trip records from entire days were placed in the final testing data.

\subsection{Evaluation of importance of feature groups}

\begin{table}[htbp!]
    \centering
    \caption{Feature set scenarios used in the evaluation. Each scenario includes only feature sets marked with (+) and possibly differential features based on these feature sets.}

    \label{tab:evaluation_scenarios}
    \begin{tabular}{|l|c|c|c|c|c|c|c|c|c|c|c|r|r|}
        \hline
        \diagbox{Scenario}{Feature set}& \begin{sideways}SURVEY\end{sideways} & 
    \begin{sideways}WALKING\_LOS\end{sideways} & 
\begin{sideways}CYCLING\_LOS\end{sideways} & 
\begin{sideways}CAR\_LOS\end{sideways} & 
 \begin{sideways}E\_CAR\_LOS \;\end{sideways} & 
\begin{sideways}PLAN\_PT\_LOS\end{sideways} & \begin{sideways}REAL\_PT\_LOS\end{sideways} & 

\begin{sideways}WEATHER\end{sideways} & 
\begin{sideways}POLLUTION\end{sideways}  &
\begin{sideways}BUILT\_ENV\end{sideways} & 
        \begin{sideways}PT\_EXPERIENCE~~\end{sideways} & \begin{sideways}Feature count\end{sideways}\\
        \hline
        S\_ONLY & + & - & - & - & - & - & -  & - & - & - & - & 14 \\
        S\_P\_LOS & + & + & + & + & - & + & - &  - & - & - & - & 126 \\
        S\_P\_LOS\_TR & + & + & + & +&+&+ & - & - & - &  - & - & 151\\
        S\_R\_LOS & + & + & + & + & - & - &  + & - & - &-&-&126\\
        S\_R\_LOS\_TR & + & + & + & + & +&- & +  & - & - & - &-& 151\\
        S\_BE & + & - & - & - & - & - & -  & - & - & + & - & 28\\
        S\_ENV & + & - & - & - & - & - & - & +  & + & - & - & 44 \\
        S\_ALL & + & + & + & + & + & + & + & +  & + & + & + & 319 \\
        \hline
    \end{tabular}
\end{table}

The primary objective of the work is to propose a way additional features of trips can be calculated and combined with survey-based features. Hence, it is interesting to analyse the impact of the inclusion of different feature sets on the observed performance of TMC models.  Thus, let us analyse the impact of the use of different modelling scenarios on the change of performance measures of TMC models compared to the use of scenario \texttt{S\_ONLY}, i.e. when only survey-based features are used. 
To do this, first we distinguish different modelling scenarios based on both survey features and other sets of features developed by \texttt{Survey Data Processor}. The summary of modelling scenarios we consider is provided in Table~\ref{tab:evaluation_scenarios}. The baseline scenario \texttt{S\_ONLY} refers to the use of survey  features only. The remaining scenarios assume the data fusion of survey data with other data, including both aggregated features of public transport connections estimated based on planned schedules (\texttt{S\_P\_LOS}) and past system behaviour determined based on GPS traces (\texttt{S\_R\_LOS}). 
Both these scenarios can be extended to include travel duration by car, estimated using travel time matrices from the transport model and parking difficulty features. Importantly, these scenarios are feasible only when transport model travel TTBC and TD matrices are available. The remaining scenarios rely on the use of other feature sets, i.e. built environment (BE), weather and pollution data. Moreover, the \texttt{S\_ALL} scenario includes all features present in the remaining scenarios. 

Differential features are present in the feature set of a scenario if and only if LOS features of different transport modes are included in the scenario. As a rule, the differential features placed in the set of features in a scenario are calculated based only on the LOS features included in the scenario. As an example,   the \texttt{minDurationRatioCarToTransitReal\_DIFF} feature is included in the \texttt{S\_R\_LOS\_TR} scenario. It is calculated as the duration of travel by car  divided by the minimum duration of travel by public transport, given real schedules of public transport. This is because public transport LOS features in this scenario rely on real schedules.

Finally, let us observe that each scenario $S$ other than the \texttt{S\_ONLY} scenario includes some data fusion-based features and is used to perform a separate ablation study used to verify whether the inclusion of additional features suggested by feature set $\mathcal{F}(S)$ yields TMC models of better performance than the models developed with survey data only.

\begin{table}[th!]
\caption{The performance of travel mode choice models 
for a) all classification methods calculated on testing folds of cross-validation and b) the best classifier applied to the final testing data.\label{tab:acc:all}}
\centering
\begin{tabularx}{\textwidth}{X>{\raggedright\arraybackslash}X>{\raggedleft\arraybackslash}
X>{\raggedleft\arraybackslash}
X>{\raggedright\arraybackslash}
X>{\raggedleft\arraybackslash}
X>{\raggedleft\arraybackslash}
X
}
  \hline
\multicolumn{2}{c}{}&\multicolumn{2}{c}{(a) All methods - OOB data} &\multicolumn{3}{c}{(b) Best method - final testing data} \\
Data set &Scenario~$S$   &mean~ACC[\%]& mean $\kappa$  &  $\mathbb{M}_\mathrm{B}$&  $P_S^{\mathrm{ACC},\mathrm{F}}$[\%] & $P_S^{\kappa,\mathrm{F}}$  \\ 
  \hline

\multirow{8}{*}{PAR\_W1} & S\_ONLY  &56.9 & 0.265 & SVMRadial & 51.3 & 0.225 \\ 
  
      & S\_P\_LOS  & 66.8 & 0.444 & SVMRadial & 67.5 & 0.506 \\  
        & S\_P\_LOS\_TR  & 67.5 & 0.458 & SVMRadial & \textbf{69.5} & \textbf{0.536} \\   
          & S\_R\_LOS  & 67.3 & 0.456 & ranger & 66.5 & 0.480 \\   
            & S\_R\_LOS\_TR  & \textbf{68.1} & 0.465 & SVMRadial & 67.5 & 0.493 \\    
              & S\_BE  & 61.4 & 0.349 & XGBTree & 60.1 & 0.382 \\\  
                & S\_ENV  & 58.4 & 0.308 & SVMLinear & 53.5 & 0.287 \\   
                  & S\_ALL  & 67.8 & \textbf{0.471} & XGBTree & 67.9 & 0.521 \\   
                \hline
\multirow{8}{*}{PAR\_W2} & S\_ONLY  & 55.3 & 0.328 & XGBTree & 58.5 & 0.359 \\  
  
      & S\_P\_LOS  & 63.7 & 0.458 & RF & 66.2 & 0.489 \\
        & S\_P\_LOS\_TR  & 62.5 & 0.437 & XGBTree & 69.7 & 0.538 \\   
          & S\_R\_LOS  & 63.6 & 0.453 & ranger & 68.3 & 0.511 \\ 
            & S\_R\_LOS\_TR  & 63.0 & 0.445 & ranger & 67.6 & 0.504 \\  
              & S\_BE  & 59.2 & 0.387 & ranger & 57.7 & 0.358 \\   
                & S\_ENV  & 53.5 & 0.298 & ranger & 52.8 & 0.286 \\ 
                  & S\_ALL  & \textbf{64.3} & \textbf{0.465} & DT & \textbf{70.4} & \textbf{0.553} \\
                \hline
\multirow{8}{*}{CIT\_W1} & S\_ONLY  & 56.3 & 0.292 & Naive~Bayes & 60.0 & 0.127 \\
  
      & S\_P\_LOS  & 61.0 & 0.387 & ranger & 66.2 & 0.456 \\ 
        & S\_P\_LOS\_TR   & 60.9 & 0.387 & ranger & 66.0 & 0.456 \\ 
          & S\_R\_LOS  & \textbf{62.4} & 0.413 & ranger & 68.2 & 0.452 \\ 
            & S\_R\_LOS\_TR  & 61.9 & \textbf{0.421} & ranger & \textbf{69.5} & \textbf{0.501} \\    
              & S\_BE  & 56.8 & 0.299 & RF & 64.4 & 0.351 \\  
                & S\_ENV  & 55.8 & 0.294 & DT & 60.6 & 0.275 \\  
                  & S\_ALL  & 59.7 & 0.369 & ranger & 65.8 & 0.454 \\   
                 \hline
\multirow{8}{*}{CIT\_W2} & S\_ONLY  & 55.0 & 0.279 & SVMLinear & 53.7 & 0.295 \\
  
      & S\_P\_LOS  & 63.1 & 0.413 & ranger & \textbf{65.5} & \textbf{0.491} \\
        & S\_P\_LOS\_TR  & 62.8 & 0.407 & ranger & 65.0 & 0.483 \\  
          & S\_R\_LOS  & \textbf{63.3} & \textbf{0.418} & ranger & 64.3 & 0.471 \\   
            & S\_R\_LOS\_TR  & 63.0 & 0.414 & ranger & 64.5 & 0.472 \\   
              & S\_BE  & 55.1 & 0.266 & RF & 54.6 & 0.308 \\ 
                & S\_ENV  & 54.0 & 0.260 & ranger & 49.7 & 0.231 \\    
                  & S\_ALL  & 62.9 & 0.414 & ranger & 64.3 & 0.466 \\  
                 \hline
\multirow{8}{*}{CIT\_W1\_2} & S\_ONLY  & 56.4 & 0.317 & SVMRadial & 52.6 & 0.266 \\ 
  
      & S\_P\_LOS  & 62.4 & 0.411 & ranger & \textbf{67.2} & \textbf{0.505} \\ 
        & S\_P\_LOS\_TR   & 64.9 & \textbf{0.464} & ranger & 67.0 & 0.501 \\  
          & S\_R\_LOS  & 62.7 & 0.416 & ranger & 65.7 & 0.484 \\  
            & S\_R\_LOS\_TR  & \textbf{65.0} & 0.463 & ranger & 67.1 & 0.502 \\  
              & S\_BE  & 57.3 & 0.326 & ranger & 57.6 & 0.325 \\  
                & S\_ENV  & 54.4 & 0.279 & ranger & 53.4 & 0.270 \\  
                  & S\_ALL  & 62.7 & 0.418 & ranger & 66.7 & 0.494 \\  
                \hline
\end{tabularx}
\end{table}

The results of the evaluation using all the scenarios listed in Table~\ref{tab:evaluation_scenarios} are provided in Table~\ref{tab:acc:all}. All the results were obtained by running Alg.~\ref{alg:evaluation} for individual data sets listed in Table~\ref{tab:MLdataset:info}. 
The results provided in Table~\ref{tab:acc:all} include  average accuracy and $\kappa$ values across testing folds of $K$-fold cross-validation and multiple ML methods, i.e. average accuracy $avg(P_\mathrm{S}^
  {ACC
    ,\mathrm{CV}}(\mathbb{M})), \mathbb{M}\in{\mathcal{M}}$ and average Cohen's $\kappa$
    $avg(P_\mathrm{S}^
  {\kappa
    ,\mathrm{CV}}(\mathbb{M})), \mathbb{M}\in{\mathcal{M}}$ respectively. In this way, the impact of the use of both survey features only (the \texttt{S\_ONLY} scenario) and the inclusion of different fusion-based features on different ML models is quantified.
In practical terms, the performance of the best ML model relying on  the most promising hyperparameter values and applied to holdout data is of particular importance.
Hence, the values of performance measures $P_S^{\mathrm{ACC},\mathrm{F}}$ and $P_S^{\kappa,\mathrm{F}}$ calculated on the most recent trip instances not used for method selection, hyperparameter tuning and model development are provided. These reflect the use of the best model trained for each data set and scenario to predict travel mode choices for the instances developed out of most recent trips.

First of all, let us note that in the case of all data sets, the inclusion of all considered features (the \texttt{S\_ALL} scenario) in the data used to train TMC models extending the feature set of the baseline scenario \texttt{S\_ONLY}, yields major performance gains both on the CV test folds and final testing data. For example, the inclusion of all features extending survey data results in an increase in accuracy on final data from 51.3\% to 69.5\% for the PAR\_W1 dataset. In the case of the largest data set \texttt{CIT\_W1\_W2}, a similar accuracy growth from 52.6\% to 67.2\% is observed. This confirms the importance of features calculated by the UTMCM platform in travel mode choice modelling. The best overall results are observed for the set documenting the trips of a representative sample of parents of primary school children, i.e. the \texttt{PAR\_W2} data set, which is shown not just by the highest accuracy but also by the highest $\kappa=0.553$ value. It is important to note that the use of the \texttt{S\_ALL} features substantially increases the value of the $\kappa$ measure, helping to make model predictions significantly higher than those arising from the \texttt{S\_ONLY} scenario.

Interestingly, the best results for individual data sets are attained under the \texttt{S\_ALL} scenario (the \texttt{PAR\_W2} data set), but also other scenarios, namely \texttt{S\_P\_LOS}/\texttt{S\_R\_LOS} and its \texttt{TR} extensions. The fact that the results for the \texttt{S\_P\_LOS}* and \texttt{S\_R\_LOS}* scenarios are quite similar is worth emphasising. This shows that features aggregating PT connections under planned future schedules can  potentially be replaced with  features aggregating PT connections feasible in the past, as determined based on recent GPS traces. This can be explained by the fact that  connections from the days preceding the trip are likely to be similar to those possible according to schedules for the day of the trip. Moreover, the travel mode choices of respondents are likely to be affected both by the characteristics of the future connections they consider using, as given by trip planners, and by their recent experience of using PT on the same route. In practical terms, this means that when schedules are difficult to obtain for automated processing or do not specify exact departure times but just the frequency of a service, GPS traces can be used as their substitute. The GPS traces of PT vehicles once processed with the UTMCM platform can be used to estimate real--though past--departure times and schedules, and can also used for the calculation of PT features by the platform, here represented by the \texttt{S\_R\_LOS}* scenarios. 

Moreover, let us note that the inclusion of travel duration by car under street congestion features and parking difficulty features, given by the *\texttt{\_TR} scenarios, contributed to performance improvement for the \texttt{PAR\_W*} and \texttt{CIT\_W1} data sets. This shows the need to evaluate different candidate feature sets depending on the population for which travel mode choices are analysed. Interestingly, in the case of the CIT\_W1 data set, the best performance was attained under the \texttt{S\_R\_LOS\_TR} scenario, i.e. when relying on features quantifying the past behaviour of the PT system rather than its planned behaviour, and including additional car-related features based on the transport model. This shows that  mode choice data quantifying both planned and past PT connections are vital for predicting travel mode choices, as the planned and real behaviour of PT systems may be substantially different.

\subsection{Performance of TMC models for different trip data sets}

\begin{figure}[th]
\centering
     \begin{subfigure}[b]{0.32\textwidth}
         \centering
         \includegraphics[width=\textwidth]{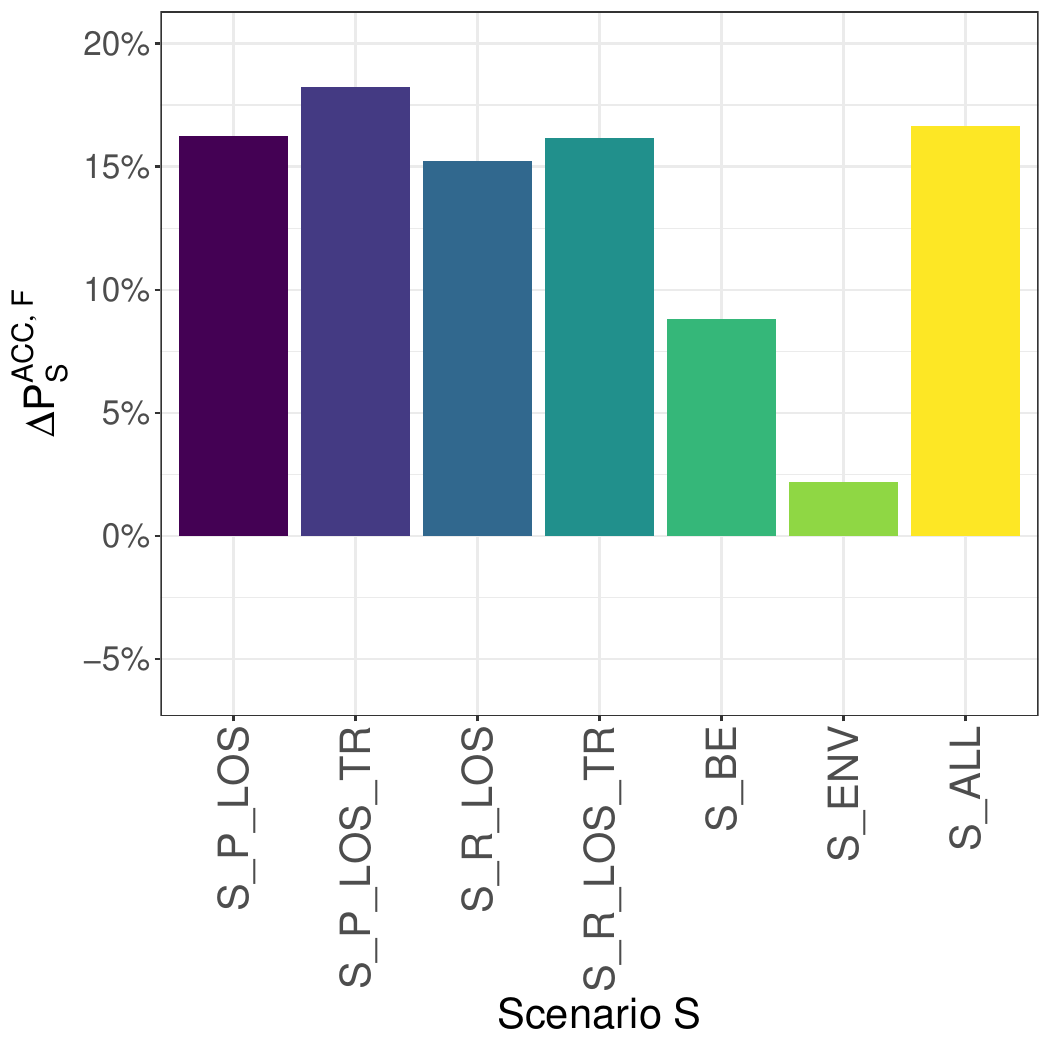}
         \caption{PAR\_W1}
         \label{fig:inc:PARW1}
     \end{subfigure}
     \begin{subfigure}[b]{0.32\textwidth}
         \centering
         \includegraphics[width=\textwidth]{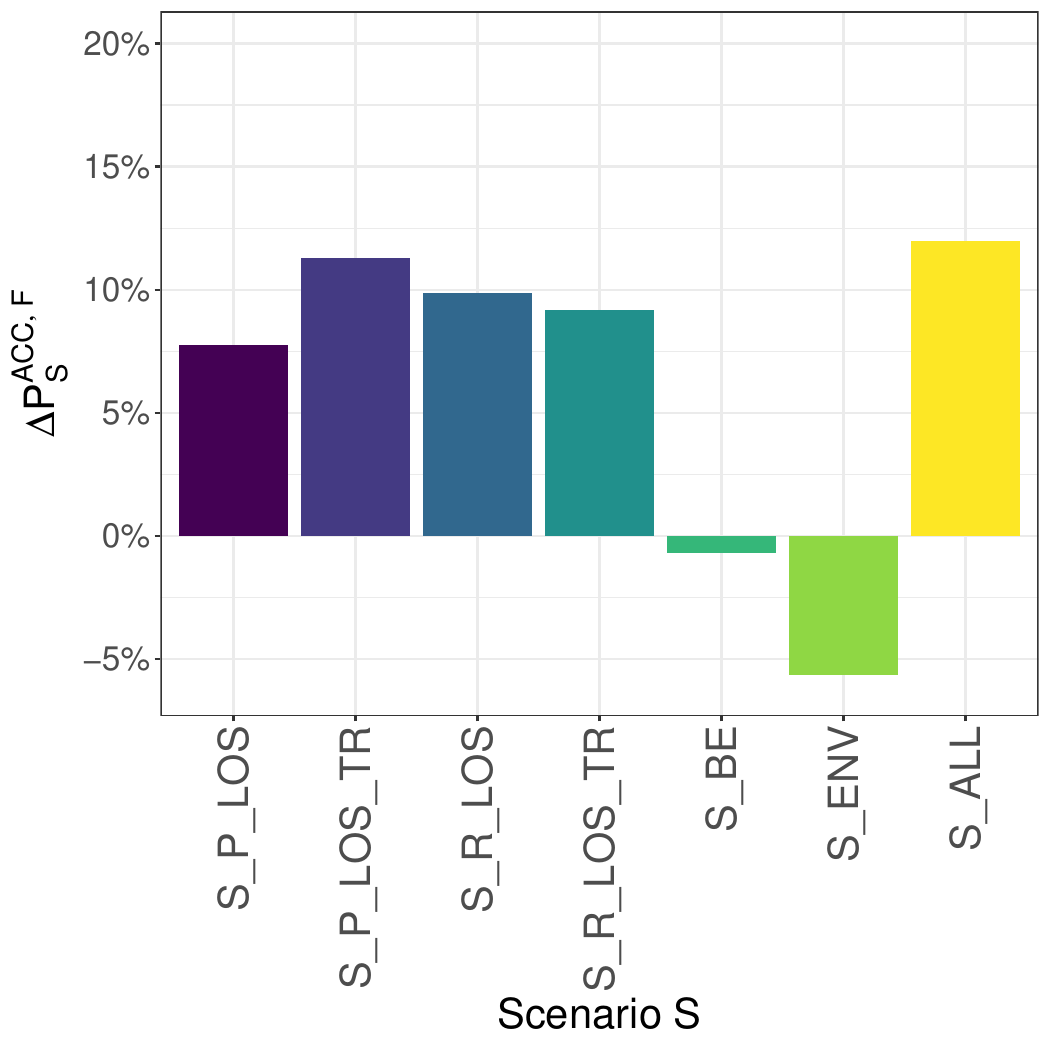}
         \caption{PAR\_W2}
         \label{fig:inc:PARW2}
     \end{subfigure}
     \hfill
     
     \begin{subfigure}[b]{0.32\textwidth}
         \centering
         \includegraphics[width=\textwidth]{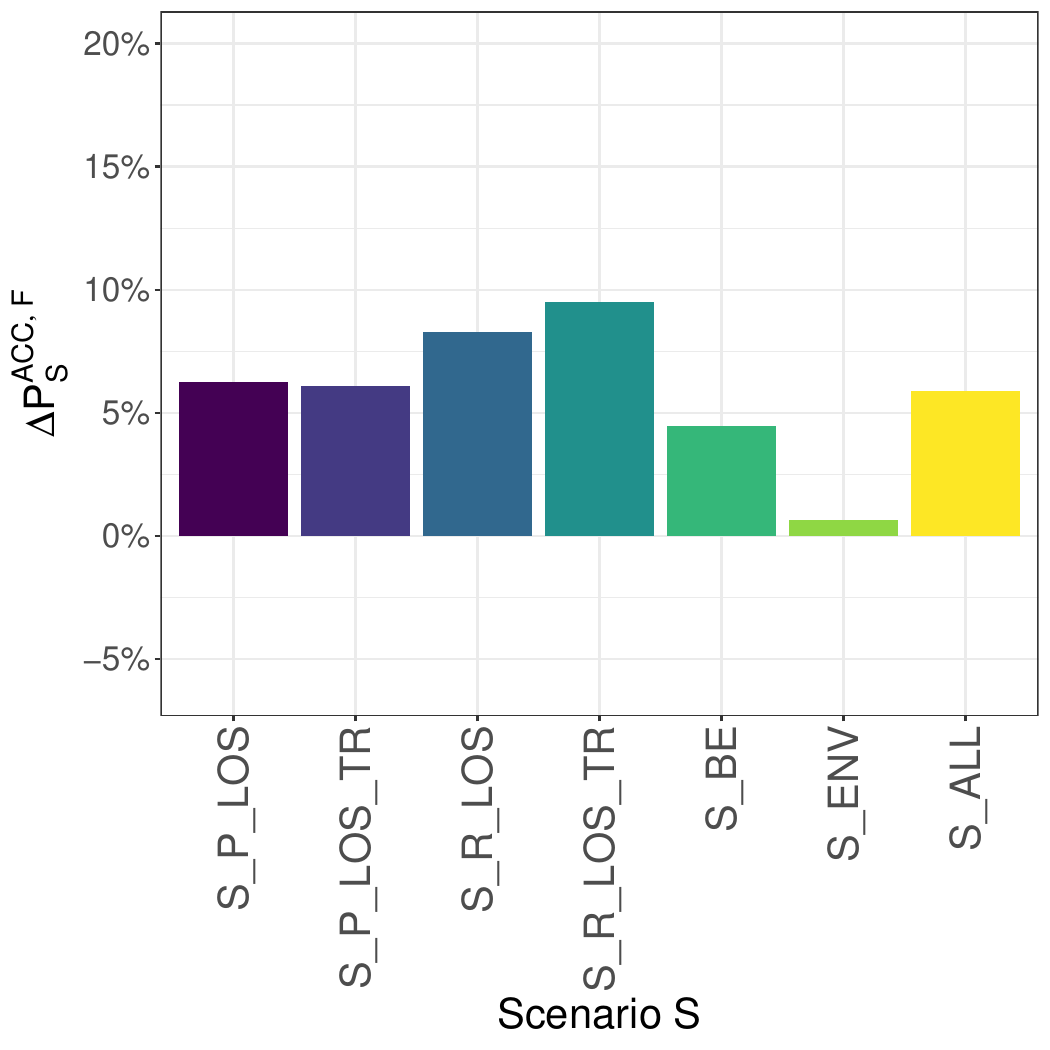}
         \caption{CIT\_W1}
         \label{fig:inc:CITW1}
     \end{subfigure}
     \begin{subfigure}[b]{0.32\textwidth}
         \centering
         \includegraphics[width=\textwidth]{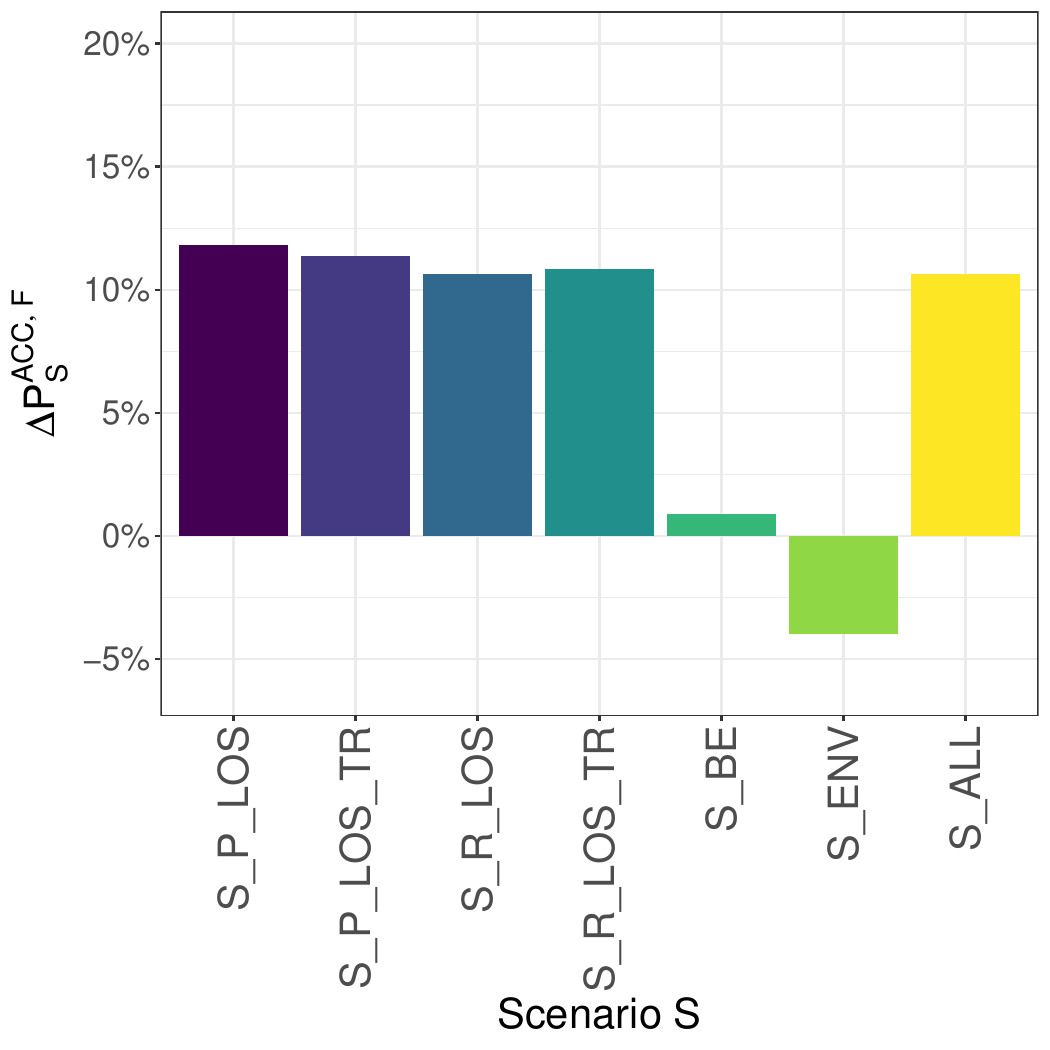}
         \caption{CIT\_W2}
         \label{fig:inc:CITW2}
     \end{subfigure}
     \begin{subfigure}[b]{0.32\textwidth}
         \centering
         \includegraphics[width=\textwidth]{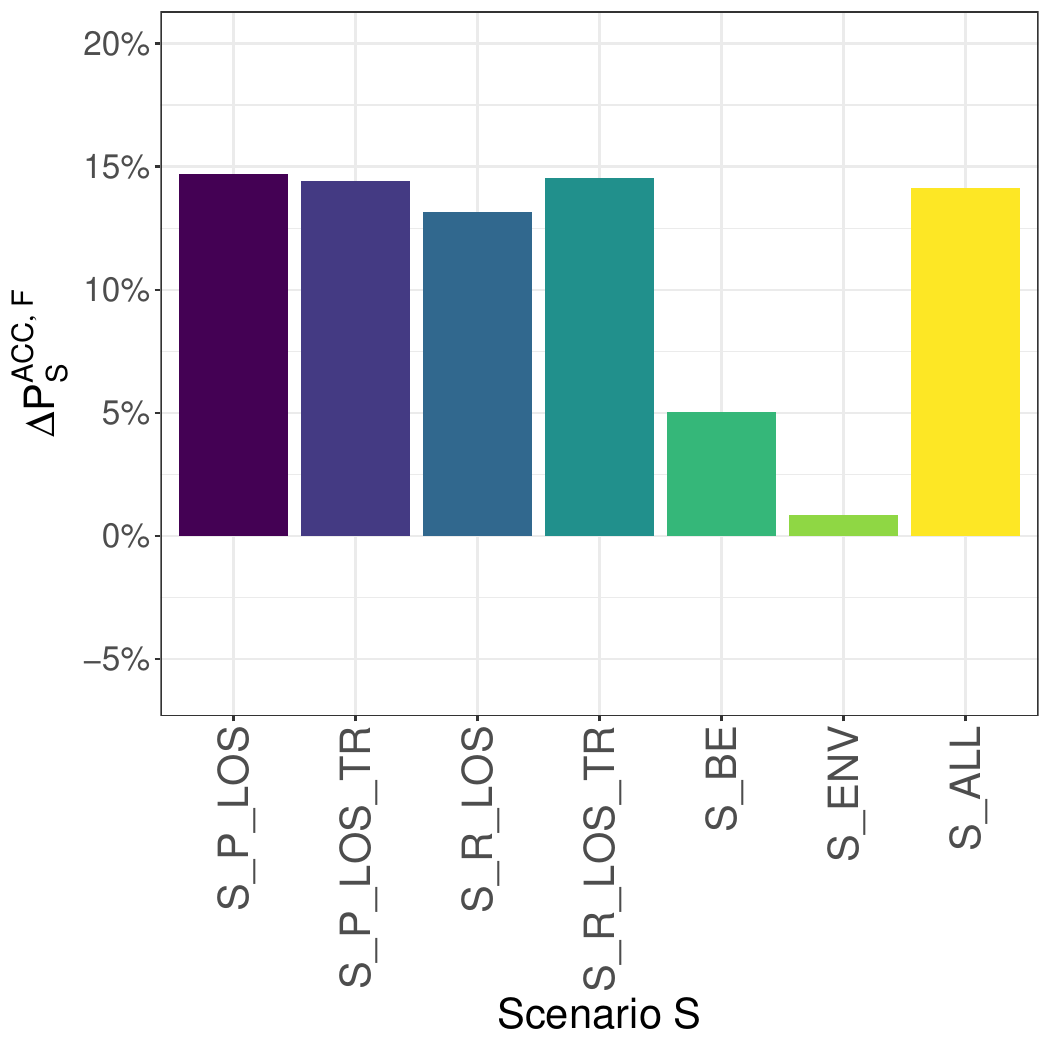}
         \caption{CIT\_W1\_2}
         \label{fig:inc:CITW12}
     \end{subfigure}     
        \caption{Change of accuracy of travel mode choice models for different data sets in comparison to the use of survey data only, i.e. the \texttt{S\_ONLY} scenario.}
        
        \label{fig:accuracyPerSetIncrease}

\end{figure}

\begin{figure}[th!]
 \centering
     \begin{subfigure}[b]{0.25\textwidth}
         \centering
         \includegraphics[width=\textwidth]{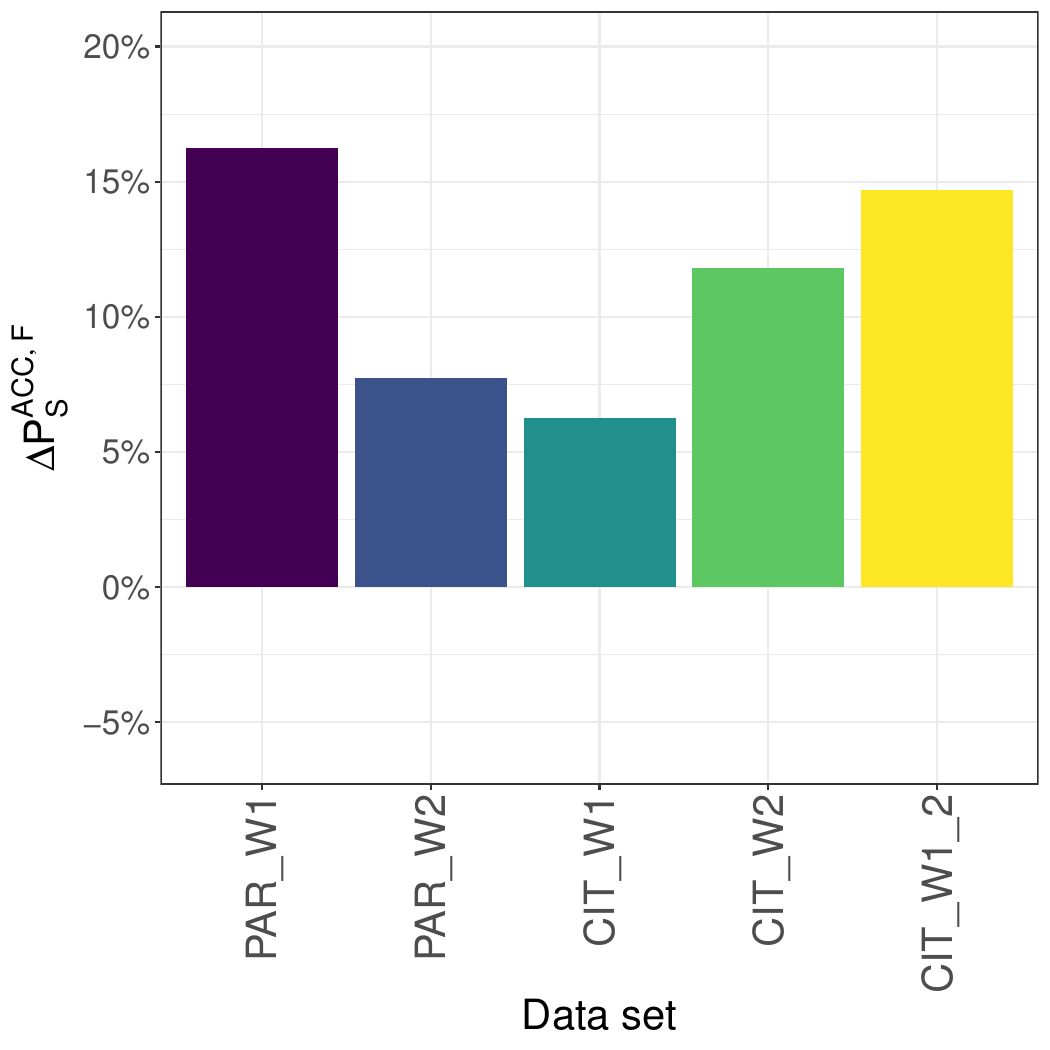}
         \caption{S\_P\_LOS}
         \label{fig:inc:IF_S_P_LOS}
     \end{subfigure}
     \hfill
     \begin{subfigure}[b]{0.24\textwidth}
         \centering
         \includegraphics[width=\textwidth]{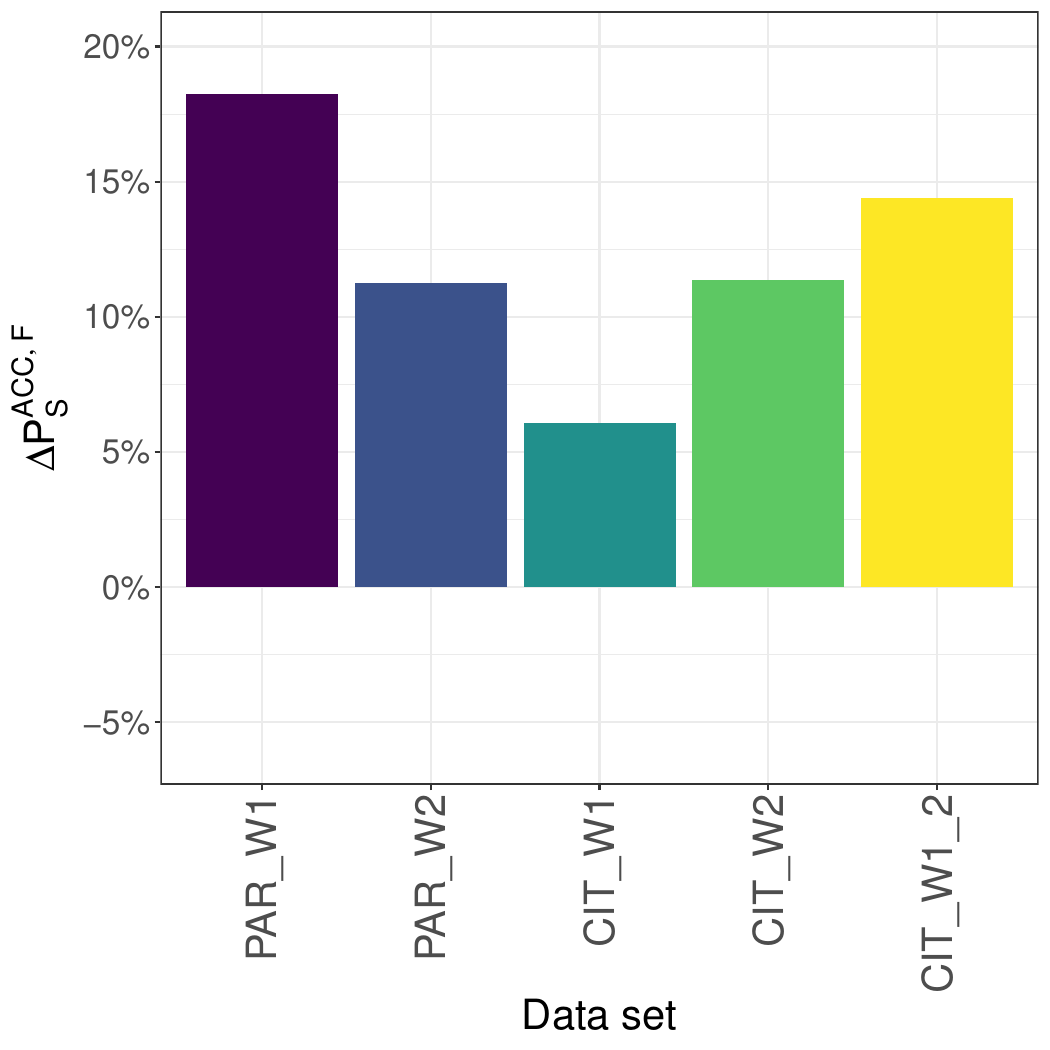}
         \caption{S\_P\_LOS\_TR}
         \label{fig:inc:IF_S_P_LOS_TRAFFIC}
     \end{subfigure}
     \hfill
     \begin{subfigure}[b]{0.24\textwidth}
         \centering
         \includegraphics[width=\textwidth]{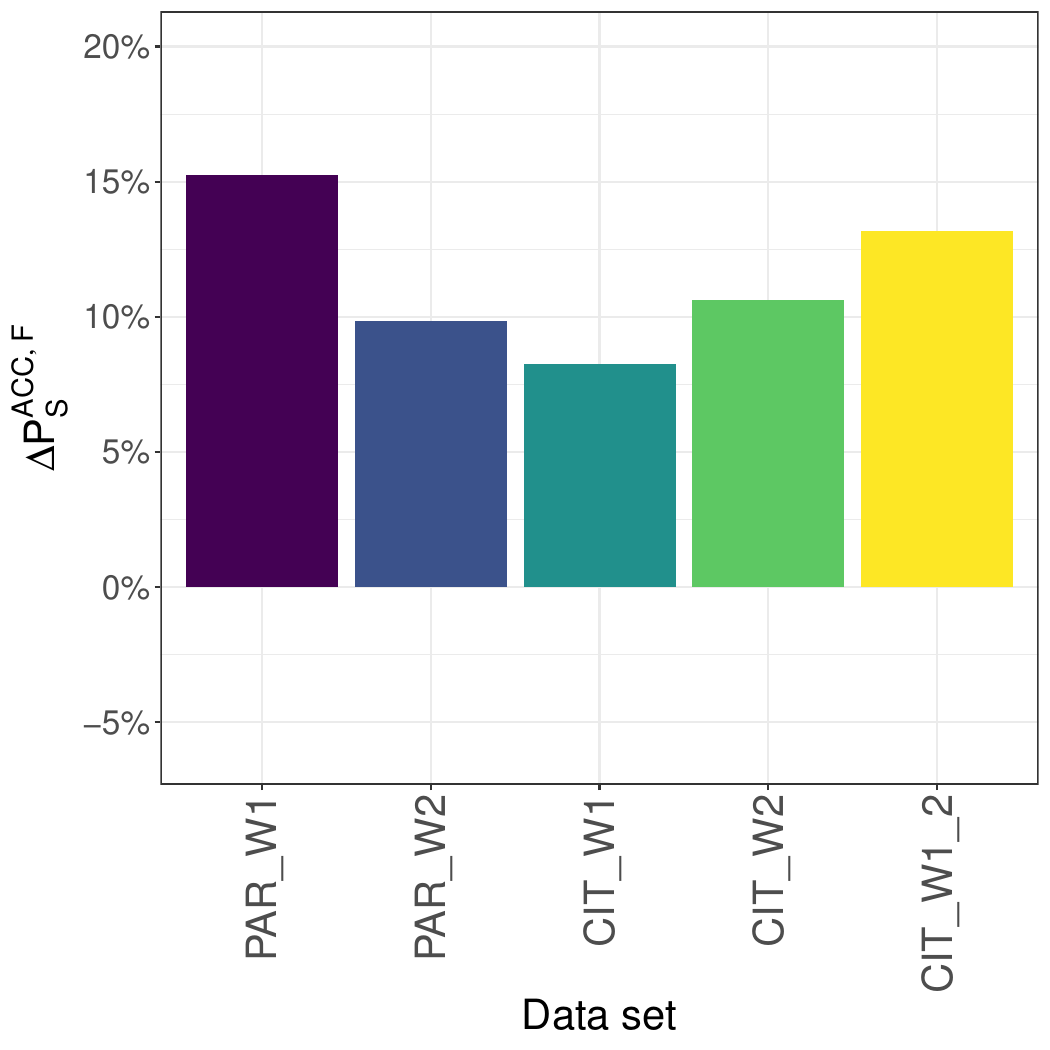}
         \caption{S\_R\_LOS}
         \label{fig:inc:IF_S_R_LOS}
     \end{subfigure}
     \begin{subfigure}[b]{0.24\textwidth}
         \centering
         \includegraphics[width=\textwidth]{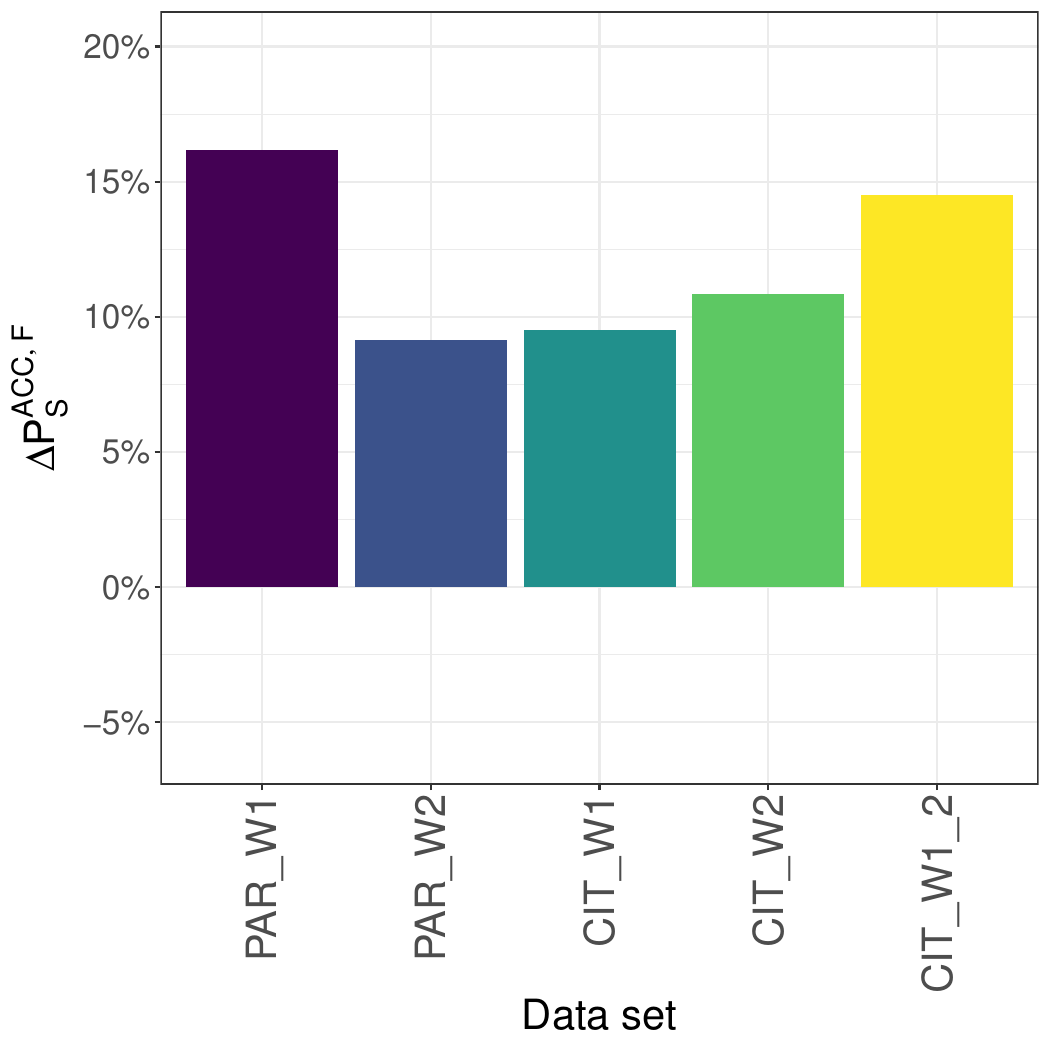}
         \caption{S\_R\_LOS\_TR}
         \label{fig:inc:IF_S_R_LOS_TRAFFIC}
     \end{subfigure}
     \begin{subfigure}[b]{0.24\textwidth}
         \centering
         \includegraphics[width=\textwidth]{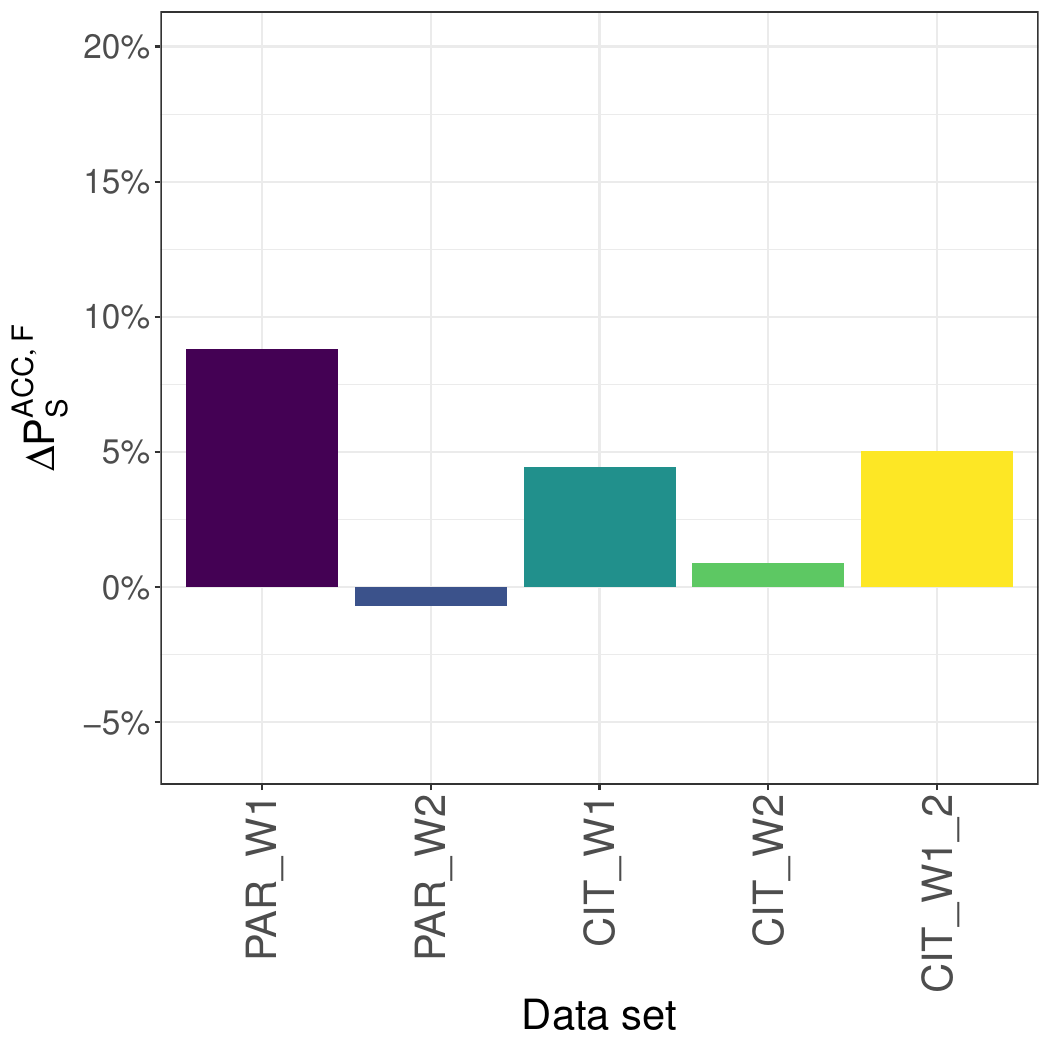}
         \caption{S\_BE}
         \label{fig:inc:IF_S_BE}
     \end{subfigure}
     \begin{subfigure}[b]{0.24\textwidth}
         \centering
         \includegraphics[width=\textwidth]{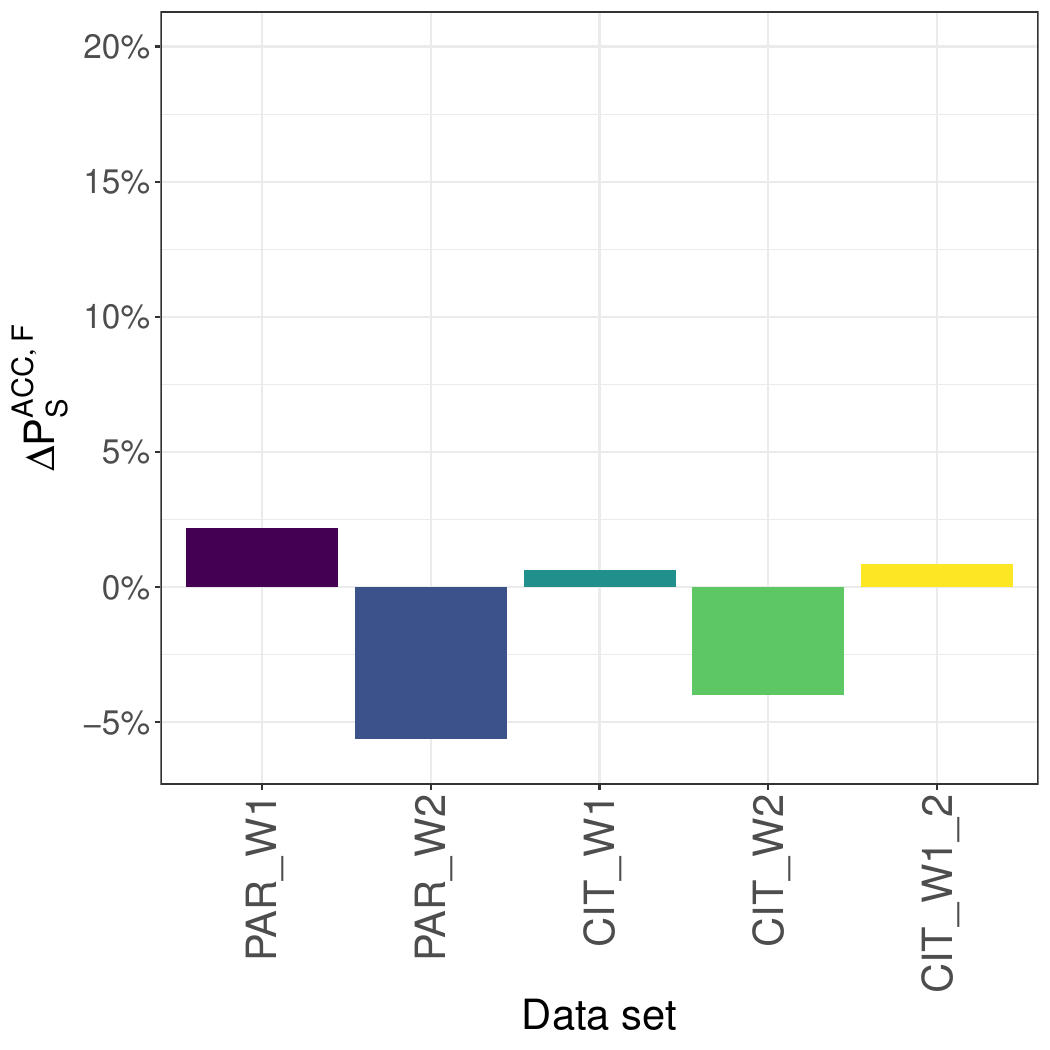}
         \caption{S\_ENV}
         \label{fig:inc:IF_S_ENV}
     \end{subfigure}     
     \begin{subfigure}[b]{0.24\textwidth}
         \centering
         \includegraphics[width=\textwidth]{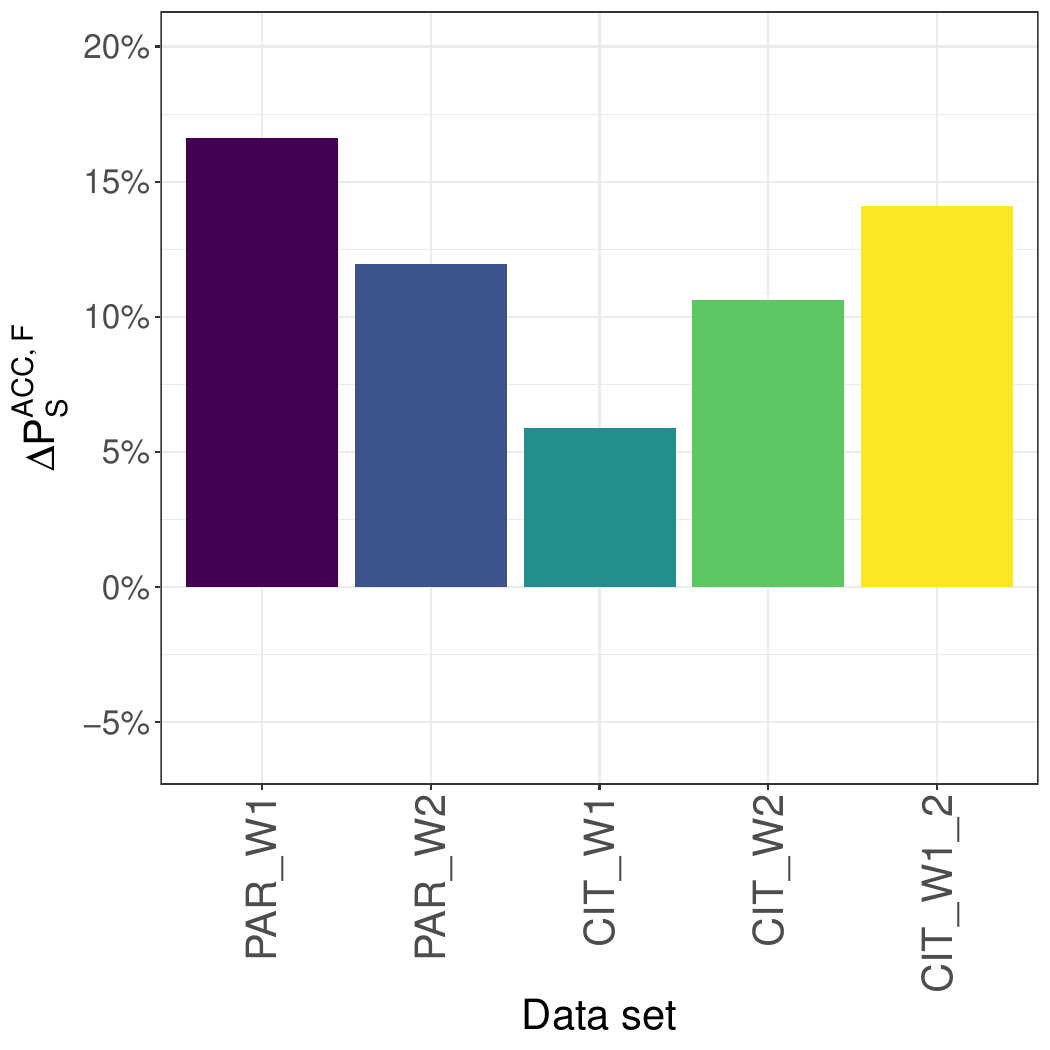}
         \caption{S\_ALL}
         \label{fig:inc:IF_ALL}
     \end{subfigure}          
        \caption{Change of accuracy of travel mode choice models for different scenarios in comparison to the use of survey data only, i.e. the \texttt{S\_ONLY} scenario.}

        \label{fig:accuracyPerScenarioIncrease}
\end{figure}

It is interesting to analyse the improvement in the  performance
of TMC models compared to the \texttt{S\_ONLY} scenario under other modelling scenarios. Fig.~\ref{fig:accuracyPerSetIncrease} shows the difference in  accuracy $\Delta P_S^{\mathrm{ACC},\mathrm{F}}$ between a model developed with an extended set of features as defined by scenario $S$ and the \texttt{S\_ONLY} set for individual trip data sets. This difference is defined as $\Delta P_S^{\mathrm{ACC},\mathrm{F}}=P_S^{\mathrm{ACC},\mathrm{F}}-P_\mathrm{S\_ONLY}^{\mathrm{ACC},\mathrm{F}}$.
It follows from the plots that accuracy gains under the \texttt{S\_}*\texttt{\_LOS}* and \texttt{S\_ALL} scenarios may  even exceed 15 per cent points for the PAR\_W1 data set.  This can be explained by the fact that the PAR\_W1 data set documents the trips of parents of  pupils from three primary schools. Hence, the behaviour of this group including the areas in which they frequently travel is likely to be more homogeneous than the behaviours and trips of respondents represented in all the other trip data sets.
Moreover, when larger data sets are collected,  even higher performance gains than in the case of smaller data sets can be expected. This is illustrated by the comparison of results under the \texttt{S\_ALL} scenario for the \texttt{CIT\_W1\_2} data set with  the results for the \texttt{CIT\_W1} and \texttt{CIT\_W2} data. $\Delta P_S^{\mathrm{ACC},\mathrm{F}}$ for TMC models built with all features is by far larger when data from both survey waves are used than in the case of building models for each wave separately.
This can be explained by the fact that to model complex dependencies between  features present in instances and travel mode choices, larger data sets may be necessary.

The findings for the \texttt{S\_BE} scenario are more complex. In the case of \texttt{PAR\_W1}, \texttt{CIT\_W1} and \texttt{CIT\_W1\_2}, substantial performance gains arising from the use of built-environment features can be observed. While the scale of these improvements is lower than for the mode-choice scenarios \texttt{S\_}*\texttt{\_LOS}*,
it is important to note that built-environment features do not require schedule data and/or GPS traces. As they rely on spatial data only, they can be easily calculated by  \texttt{Built Environment Service} for any city.
They are also likely to partly yield information related to mode-choice data. This is because the planning of transport services takes into account spatial data such as population density. Moreover, some BE features such as distance to PT stop of any type are correlated with the features calculated based on candidate PT connections such as walking distances needed to use these connections.
Hence, built environment features typically help better explain some mode choices. The findings for environmental features (the \texttt{S\_ENV} scenario) are less conclusive. It follows from the experiments that their inclusion may even reduce the performance of the models compared to the \texttt{S\_ONLY} scenario. This is likely due to overfitting training data, i.e. the fact that models may learn spurious correlations present in the data used for cross-validation and development of final models. Such correlations between weather and pollution and mode choices may  not be valid for the period from which trips are present in the testing data. This risk is likely to be mitigated by larger data sets, as shown by the results for the \texttt{CIT\_W1\_2} data. This confirms that the larger cardinality of data sets is likely to increase model performance gains arising from the use of additional features and the data fusion process.

Moreover, results for the \texttt{S\_ALL} scenario show that for all data sets the inclusion of all considered features improves the accuracy of travel mode choice models compared to models created with the \texttt{S\_ONLY} features. 
Results for the \texttt{CIT}* data sets again confirm  that larger data sets make modelling easier. This is not obvious in the analysed case, as by combining data from CITIZENSW1 and CITIZENSW2 we train models mostly on CITIZENSW1 data and evaluate them only on trips from CITIZENSW2 data, i.e. from a different period with partly different weather conditions compared to those documented in the travel data from the CITIZENSW1 survey. All the results show that not only can travel mode choice models be developed, but also successfully used to predict future mode choices of other persons.

Finally, let us briefly summarise the performance of TMC models developed under each scenario. As before, let us use $\Delta P_S^{\mathrm{ACC},\mathrm{F}}$ to analyse performance changes arising from the inclusion of additional data fusion-based features. Fig.~\ref{fig:accuracyPerScenarioIncrease} shows the summary of changes under each modelling scenario $S$. It follows from the figure that for every trip data set, the use of \texttt{S\_P\_LOS} and \texttt{S\_R\_LOS} feature sets substantially increased the performance of TMC models compared to the use of \texttt{S\_ONLY} features. The same applies to the \texttt{S\_P\_LOS\_TR} and \texttt{S\_R\_LOS\_TR} scenarios, i.e. scenarios additionally  including features relying on transport model data and the \texttt{S\_ALL} scenario. The inclusion of only weather and pollution data (\texttt{S\_ENV} scenario) or built environment data (\texttt{S\_BE} scenario) can result in TMC models of improved performance (\texttt{PAR\_W1} and \texttt{CIT\_W1\_2} data), but can also negatively influence model development. This may be due to overfitting. Hence, future studies can address feature selection techniques for TMC modelling.

\subsection{Identification of  most important features}

\begin{figure}[tbh!]
\centering
     \begin{subfigure}[b]{0.49\textwidth}
         \centering
         \includegraphics[width=\textwidth]{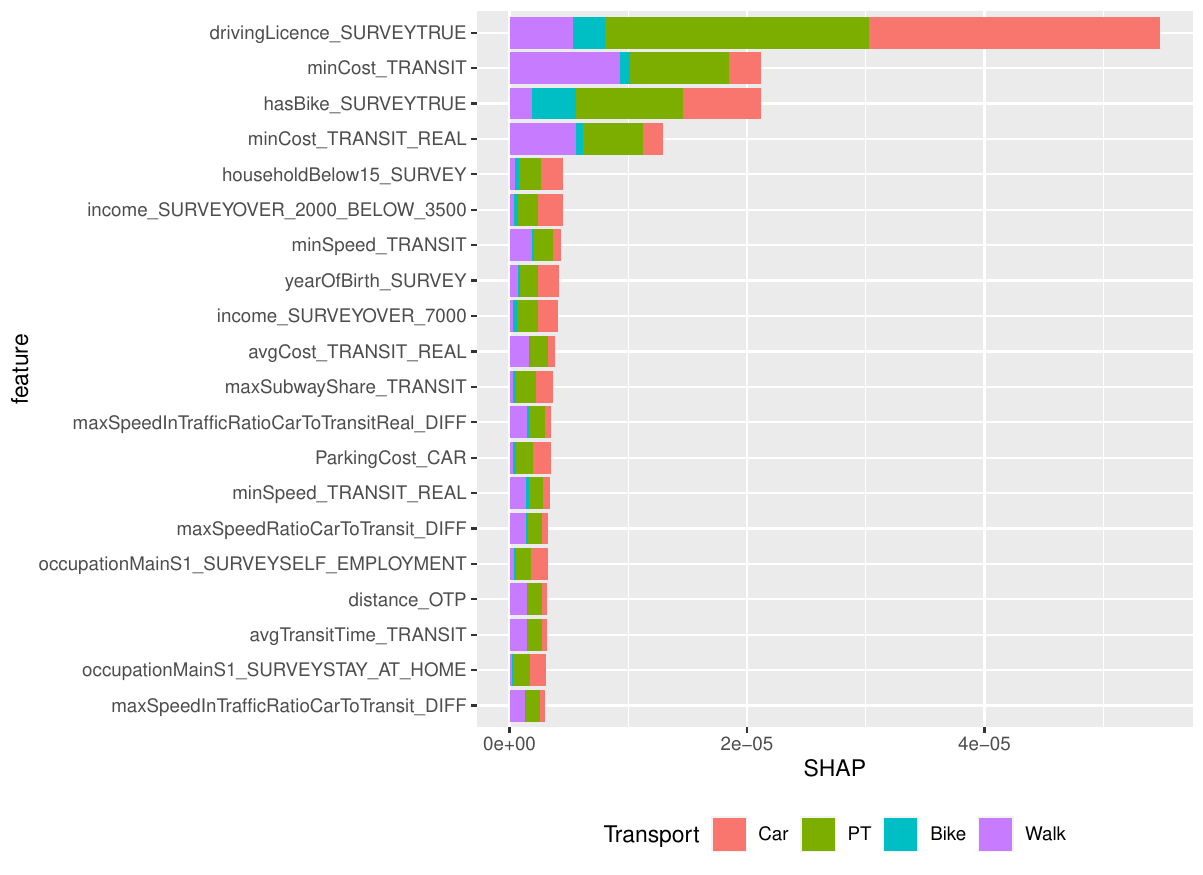}
         \caption{CIT\_W1 data set}
         \label{fig:SHAP:ALL:CIT_W1}
     \end{subfigure}
     \hfill
     \begin{subfigure}[b]{0.49\textwidth}
         \centering
         \includegraphics[width=\textwidth]{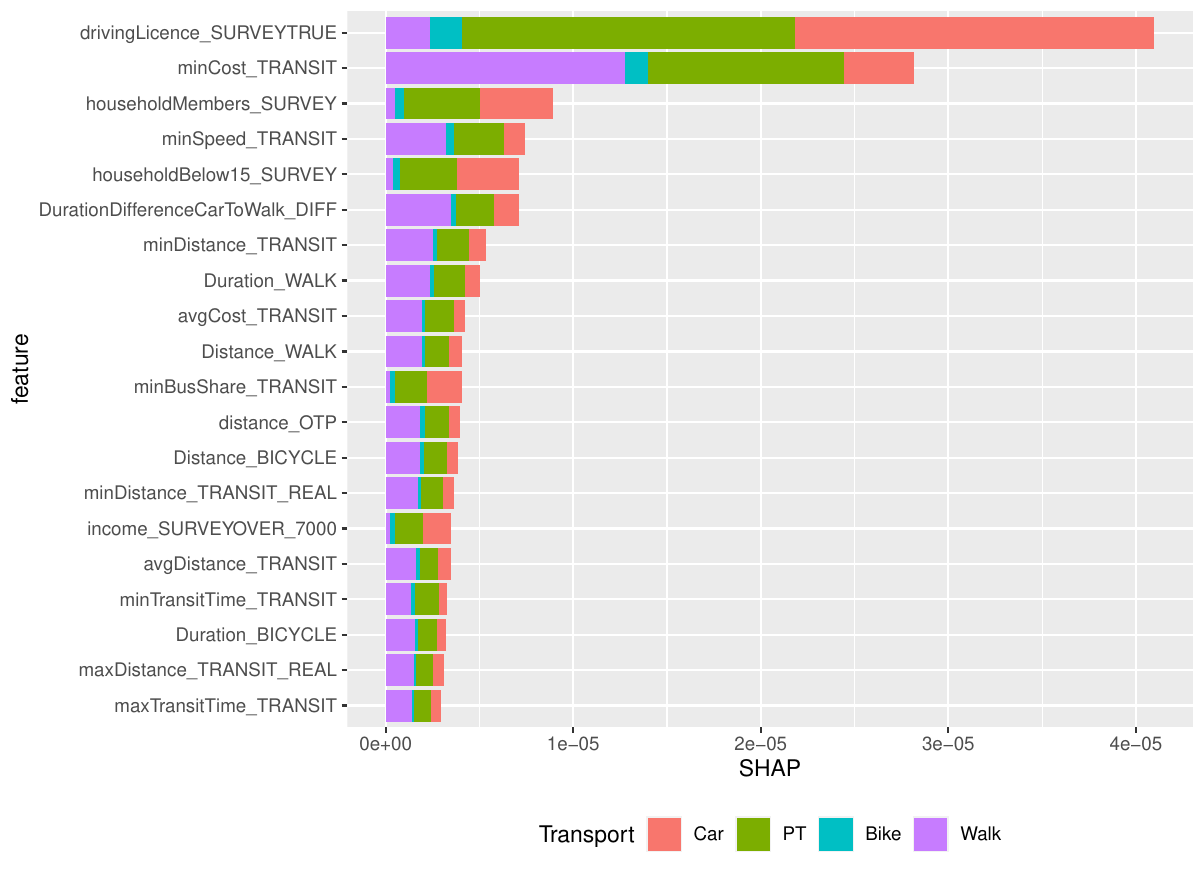}
         \caption{CIT\_W2 data set}
         \label{fig:SHAP:ALL:CIT_W2}
     \end{subfigure}
     \hfill
     \begin{subfigure}[b]{0.49\textwidth}
         \centering
         \includegraphics[width=\textwidth]{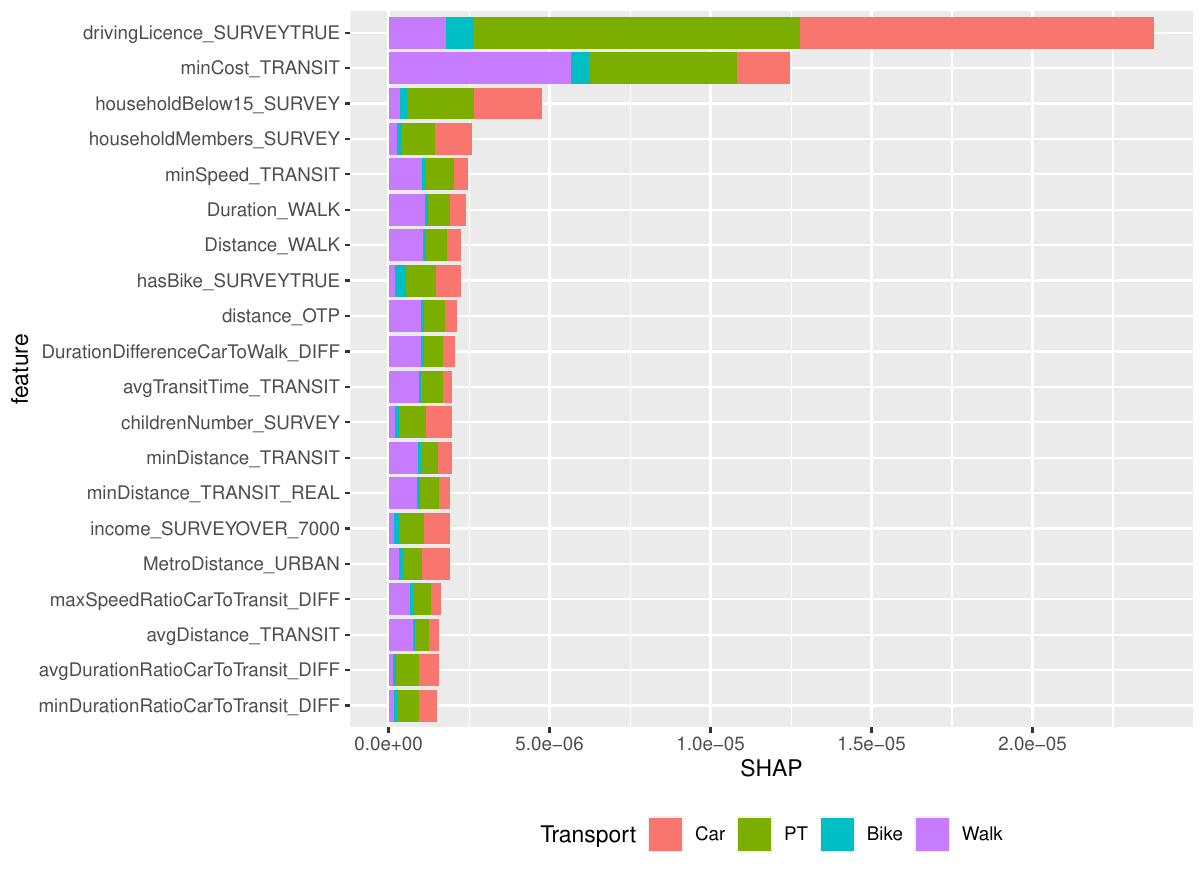}
         \caption{CIT\_W1\_2 data set}
         \label{fig:SHAP:ALL:CIT_W1_2}
     \end{subfigure}
        \caption{SHAP values for TMC models developed under the same scenario \texttt{S\_ALL}, classification method (ranger) and different data sets from the CIT* group.}
        \label{fig:SHAP:S_ALL_scenario}
\end{figure}     

Let us investigate feature importance quantified by SHAP values. It follows from Table~\ref{tab:acc:all} that most frequently the ranger method provides the best performing TMC models. Hence, to avoid varied preferences of different ML methods for features of different types such as categorical and numerical features, let us analyse feature importance for sample models developed with the ranger method and assuming the use of the \texttt{S\_ALL} scenario in all cases. Fig.~\ref{fig:SHAP:S_ALL_scenario} provides SHAP values for three models developed for individual \texttt{CIT\_W*} data sets. In all of the cases, survey features  are among the features having the largest impact on model predictions. Notably,  the most influential feature in this group is the \texttt{drivingLicence\_SURVEYtrue} feature. Not surprisingly, whether a person has a driving license has a major impact on predicting the use of both public transport and cars, as indicated by the length of the car and PT bars.

Other survey features which are among those having the greatest impact on travel mode choice predictions of all models include \texttt{householdBelow15\_SURVEY}, i.e. the number of persons under the age of 15 in the household of a respondent. Let us note that the second most important feature is a representative of the \texttt{PLAN\_PT\_LOS} feature group, i.e. the \texttt{minCost\_TRANSIT} feature. This feature represents the cost of a single ticket for the fastest PT connection available for a trip. Similarly to other features from this group, the calculation of the value of this feature necessitates the use of the PT schedules and spatial data of the city to determine feasible PT connections.
The value of the feature is calculated according to Warsaw PT pricing strategy and takes as an input the duration of the whole public transport route. The feature takes values corresponding to estimated travel duration lower than 20min, 20-75 min, 75-90min and above 90min.
Interestingly, this feature also has an impact  on walking class predictions. This is because, when there is no PT connection available with a shorter overall walking distance to/between/from the final PT stop involved than the distance of a direct walk from the origin to the destination, the value of the feature is set to 0.  

As far as the remaining most important features are concerned, in the case of all data sets there are just a few features of outstanding importance and multiple features of similar importance. While the importance rankings for individual data sets are different, the ranking for the largest data set, i.e. \texttt{CIT\_W1\_2} data set is of particular importance. The features present in it include, apart from survey features, the features from \texttt{PLAN\_PT\_LOS}, \texttt{REAL\_PT\_LOS}, \texttt{WALKING\_LOS},  but also \texttt{BUILT\_ENV} and \texttt{DIFF} features. This shows that calculating all fusion-based features extending survey features is vital to both properly  model TMC choices and quantify the importance of all factors influencing these choices. These include, but are not limited to the minimum speed of PT connections (\texttt{minSpeed\_TRANSIT}) and duration and distance of the trip if walked.
One more interesting example of an important feature is the minimum distance of the route if travelled by PT under the real functioning of the PT system (\texttt{minDistance\_TRANSIT\_REAL}). This feature reflects the fact that  connections possible in reality and possibly influencing travellers' experience may require longer routes than the ones which should be available according to planned schedules.
Another related feature is \texttt{maxDistance\_TRANSIT\_REAL}. This shows that minimum values over candidate PT connections are not the only aggregate values summarising these connections that are found important by ML methods. 

Interestingly, which types of PT vehicles can be used for a trip is also important. Maximum share of subway line(s) in the PT connections (\texttt{maxSubwayShare\_TRANSIT}) and 
 the minimum share of bus(es) in PT connections (\texttt{minBusShare\_TRANSIT}) are among the most important features for CIT\_W1 and CIT\_W2 models. Another related feature is the \texttt{MetroDistance\_URBAN} feature. This feature, which represents the \texttt{BUILT\_ENV} group, shows that the distance to the nearest metro station also has a major impact on predicting travel mode choices, especially the use of cars and PT.

Many features present in the rankings of the most important features are differential features. The most important of them is the \texttt{DurationDifferenceCarToWalk\_DIFF} feature revealing the difference between the estimated durations of driving to the trip destination and walking to it, assuming free-flow driving conditions. This is just one example of four differential features present in the ranking for the CIT\_W1\_2 model. Another important feature in this group is \texttt{avgDurationRatioCarToTransit\_DIFF}. This shows that the ranger method finds it useful to rely on the explicit ratio of travel durations with different modes to predict mode choices. 

To sum up, the investigation of the features having the largest impact on the predictions of TMC models let us observe that both survey features and data fusion-based features of different categories are among the most important features. This and the overall summary of the performance of TMC models provided in Table~\ref{tab:acc:all} confirms that the different feature groups implemented in this work are vital for TMC predictions. However, not all of the features are of equal importance. Some of them, such as weather and pollution features, were not found to substantially influence model predictions for the reference data sets. In particular, they are not present among the most important features.
This confirms the need for efficient calculation of multiple feature sets and analysis of which of them should be considered in an urban region of interest. The platform proposed in this work can be used to collect all relevant data, perform the fusion of preprocessed data, and enable the evaluation of different feature sets and individual features in terms of their impact on TMC predictions in the region of interest. Importantly, the most important features for the reference models include features from many different groups. This shows that neglecting some of the feature groups is likely to negatively influence both prediction and understanding of travel mode choice decisions.

\section{Conclusions and future works}
\label{sec:conclusions}
Understanding travel mode choices is vital for transportation planning and promoting sustainable mobility. These choices are affected by multiple factors. Some of them are traveller's features such as whether a person has a driving licence, the number of children in the traveller's household or the purpose of a trip, such as taking a child to school. Different studies successfully integrated such survey-based features with some of the other features such as mode-choice features, weather features or built-environment features by using dedicated scripts. This raises the question of how all feature types could be calculated with an integrated software platform and whether other features such as the differential features could be used instead of or apart from those calculated so far. A related question is how design patterns such as the Lambda pattern developed for Big Data platforms could be adapted to develop a platform enabling the storage of related data and the fusion of data needed to develop travel mode choice data sets.

In this work, we have proposed the architecture of a platform collecting, transforming and integrating the data needed to develop travel mode choice models. We show that the use of the platform provides  increased performance of travel mode choice models developed with extended sets of features rather than survey-based features only. The platform enables the ranking of different features and feature groups in terms of their contribution to travel mode choice modelling for the trips of a population of interest. Experiments performed with sample trip data sets show that major accuracy improvement was attained thanks to the use of data developed with the platform. This highlights how important data fusion can be for both classifier development and an in-depth understanding of complex human choices.
Furthermore, we analysed the relative importance of different feature groups typically considered separately and for different survey data. The analysis performed in this work illustrates the use of the platform to quantify the impact of different factors on choices made by urban travellers.

Importantly, the platform relies on open standards and formats. Thanks to its ingestion layer it can be adapted to different urban settings. In the future, we will address inter alia the issue of feature selection respecting the composition of feature groups. Furthermore, 
we plan the development of a semi-supervised method in which trip data can be extended with unlabelled instances illustrating  origin-destination trips made at different periods of the day.

\section*{CRediT authorship contribution statement}
{\bf M. Grzenda}: Conceptualization, Formal analysis, Funding acquisition, Investigation, Methodology, Project administration, Resources, Software, Supervision, Validation, Writing - original draft,  Writing - review \& editing. {\bf M. Luckner}: Conceptualization; Data curation; Investigation, Methodology,  Software,  Validation, Visualization, Writing - original draft. {\bf J. Zawieska}: Conceptualization,  Data curation, Formal analysis, Funding acquisition,  Writing - original draft
{\bf P. Wrona}:  Conceptualization, Data curation, Methodology, Software, Visualization, Writing - original draft.

\section*{Acknowledgements}
Funding: The research leading to these results has received funding from the EEA / Norway Grants 2014-2021 through the National Centre for Research and Development. CoMobility benefits from a 2.05 million € grant from Iceland, Liechtenstein and Norway through the EEA Grants. The aim of the project is to provide a package of tools and methods for the co-creation of sustainable mobility in urban spaces. 

Data made public by the City of Warsaw including schedules and GPS traces from the City Open Data portal (\url{http://api.um.warszawa.pl}) acquired  and processed in the years 2022-2023 were used to develop public transport schedules. Further details on these data can be found on the Open Data portal. Open map data copyrighted by OpenStreetMap contributors available from \url{https://www.openstreetmap.org} were also used as input data.

\appendix

\section{The data sources and the configuration of the platform applied for the City of Warsaw \label{appendix:Warsaw}}

This appendix describes details of the data used to apply data fusion  and perform experiments documented in Sect.~\ref{sec:Experiments}. This includes both survey data sets and data   collected in the Ingestion layer of the UTMCM platform for the City of Warsaw and used as input for data fusion. Moreover, platform settings influencing the number of and the values of features placed in the data sets used for travel mode choice modelling are discussed.

\subsection{Survey data}

The survey data were provided in CSV files, one for each survey data set listed in Table~\ref{tab:survey:info}. One line of a CSV file contains all answers provided by a single respondent, including a person's entire travel diary, i.e. a sequence of trips made by the person on the described day. Every person was asked to describe one most recent typical working day between Tuesday and Thursday. The survey files were used as an input for \texttt{Survey Data Processor}.
As observed in \cite{hillel2021}, an important benefit of including attributes of mode alternatives in the modelling is the ability to analyse the impact of changes to the transport network on mode choices. To fully understand the impact of these attributes, i.e.  the LOS features on mode choices, we decided to include core trip, respondent and household features in the modelling rather than e.g. habits such as travelling on the same route typically by car. This is because the values of such features are more than likely to be influenced by the LOS attributes such as lack of direct PT connections and long transfer times for the route. As a consequence, the inclusion of such survey answers  would reduce the ability to understand to what extent LOS attributes influence travel mode choices, possibly via habits and opinions they result in.

\subsection{Built and natural environment}

The built and natural environment features were calculated by \texttt{Built Environment Service} using OpenStreetMap services available at \url{ https://api.openstreetmap.org/}. 
Additionally, demographic features were computed using a local government register of citizens. In this way, the density of address points and population density were calculated. 

\subsection{Public transport timetables\label{sec:appendix_schedules}}
The timetables for individual lines of buses, trams, and trains of Fast City Railway Lines (SKM) operated by the city were downloaded by the \texttt{Planned GTFS Generator} by polling
the City of Warsaw APIs three times a day to update schedules with the most recent updates applicable for the next day. The data were saved by the module in the form of GTFS files.
The underground system timetables were generated separately %
based on the frequency of metro trains in a given hour made public by the metro operator at \url{https://metro.waw.pl}. Furthermore, publicly available schedules from railway companies   operating short and medium-distance trains, which can be used for travel within the City of Warsaw, i.e. Mazovian Railways (Koleje Mazowieckie, KM)  and Warsaw Commuter Railway (Warszawska Kolej Dojazdowa, WKD) were also obtained. All these schedules were used to configure public transport data in OpenTripPlanner instances present in the Service layer of the platform. 

\subsection{Location traces of public transport vehicles}
The real-time position of PT buses and trams is accessible through a public open data API at \url{http://www.api.um.warszawa.pl/api/action/busestrams_get}.
The public API updates the values approximately every 10 seconds, providing the current location of vehicles in Warsaw. Importantly, the historical location is not provided.
The values returned by the service are: 
the position of the vehicles provided in the WGS84 coordinate system, the time of reading the location from a GPS device, and line number and brigade values that precisely identify the vehicle.
These data were polled by a dedicated Apache NiFi data flow every 5 seconds from the open data API and forwarded via Apache Kafka for further processing with \texttt{Vehicle Flow Processor}. The downloaded data alone does not contain an explicit link to the schedule.
However, every time a vehicle's presence at a stop is recognised, the corresponding timetable can be assigned and possible differences between expected and actual departure times can be calculated.  \texttt{Vehicle Flow Processor} was performing this operation. Importantly, the input for the latter module is the sequence of location records. 
Hence, all the aspects unique for capturing data from the open data portal of the City of Warsaw are encapsulated by the Apache NiFi data flow. 

 Approximately 8.2 up to 9.2 mln deduplicated location records from up to 2,000 PT buses and trams operated at the same time by the City of Warsaw were persisted in Apache Hadoop per working day.

\subsection{Transport model\label{sec:ODModel}}

\subsubsection{Travel area zones}

In 2015, the City of Warsaw conducted a traffic study named WBR2015 to determine the flow of people and vehicles in the Warsaw Agglomeration Area. The interviews of 17000 residents of Warsaw and more than 1100 passengers were collected~\cite{KosteleckaSzarataJacyna2015}.  In 2019, the study was updated and published as the MRAW2019 model. 

The collected data was used to determine homogeneous travel demand zones, i.e. areas that share spatial planning and transport behaviours of their residents. The division consists of  801  municipal travel demand zones and 25 suburban zones. 
The zone boundaries were obtained as a polygon layer (ESRI Shape file format) and uploaded into RDBMS for further use by the platform. To enable spatial queries, the zone layer was saved into the database using the PostGIS extension.

\subsubsection{Travel times and travel demand matrices}

The Travel Time by Car Matrices (TTBC) and Travel Demand (TD) matrices for the City of Warsaw present in the transport model were calculated using PTV VISUM software\footnote{\url{https://www.myptv.com/en/mobility-software/ptv-visum}} and saved in the form of MTX files. The calculations were performed for the peak hours of working days considered together and the times between 7:30 am and 8:30 am and from 4 pm to 5 pm. The results for the peak hours were next extrapolated based on the transport model on the one-hour time slots between 6 am and 9 pm. A separate set of matrices documenting transport system behaviour under free flow conditions was used for the remaining hours.

The hourly TTBC matrices include matrices containing estimated private car travel time calculated with and without considering possible street congestion. These matrices were used to estimate the influence of street congestion in the E\_CAR\_LOS feature set.
The TD matrices consist of two groups. The first series of hourly matrices describes the number of trips between zones using private transport and the number of cars per hour. The second series contains information about the number of passengers on public transport per hour.  Both types of these matrices were used to estimate parking difficulties in the E\_CAR\_LOS set.

The TTBC and TD  matrices in the form of the files in the MTX format obtained from the MRAW2019 model  were uploaded into the RDBMS of the platform by a dedicated software module we developed. 
This provided for time-dependent calculation of both travel time by car and parking difficulty by \texttt{Private Transport Vehicle Flow Service}. Hence, the calculation of these features considers time-dependent travel durations and time-dependent number of arrivals to zones.

\subsection{Weather and air pollution data}
The weather data for Warsaw was collected from public archives of the local Institute of Meteorology and Water Management through its webpage\footnote{\url{https://danepubliczne.imgw.pl/data/dane_pomiarowo_obserwacyjne/dane_meteorologiczne/terminowe/synop/}}. The data were downloaded by dedicated Apache NiFi flow on monthly basis. Each monthly package contains hourly readings from a weather station located in Warsaw. The following weather data were collected: the air temperature [\degree C], the total rainfall in the preceding six hours [mm], the wind speed [m/s], and the cloudiness.

The air pollution data for Warsaw was collected using an endpoint provided by the Chief Inspectorate of Environmental Protection. Recent data is available via API, and web archive can be used to access historical data. The API address is \url{https://powietrze.gios.gov.pl/pjp/content/api}. The data was collected daily from six Warsaw stations. They provide the measurements of benzene ($C_6H_6$), ozone ($O_3$), carbon oxide (CO), nitrogen dioxide ($NO_2$), PM2.5, and PM10 [$\micro g/m^3$] concentration. All the weather and pollution data were persisted in the RDBMS. In this way, data for the periods of the days documented in trip diaries and preceding them were collected. This enabled the calculation of weather and pollution features by the \texttt{Environmental Service}. Both in the case of weather and pollution data, the features containing average values of each of the weather and air pollution parameters calculated above were calculated. For each parameter, two features,  containing the average value of this parameter during 24h and 2h period preceding the start of a trip, were calculated and placed by \texttt{Survey Data Processor} in a trip data set.

\subsection{Data fusion performed with Survey Data Processor}
\subsubsection{Calculation of LOS features for public transport}

For each trip $\mathcal{T}$, public transport connections were determined for $\delta_\mathrm{S}=5 min$ and $\delta_\mathrm{F}=10min$ settings. Hence, the PLAN\_PT\_LOS features were obtained from OpenTripPlanner assuming the use of PT connections available during the mode choice window $[t_\mathrm{D}(\mathcal{T})-5min,t_\mathrm{D}(\mathcal{T})+10min]$ according to planned schedules available for the day of the trip. Similarly, the REAL\_PT\_LOS features were calculated using real schedules recreated from GPS traces for the same mode choice window hours of the day  as the PLAN\_PT\_LOS features, but for the period of the preceding regular working day, i.e. Tuesday, Wednesday or Thursday. In both cases, PT connections assuming the use of one or many means of transport out of metro, buses, trams and trains of different operators listed in~\ref{sec:appendix_schedules} were determined. Next, LOS features aggregating the features of these connections were calculated.

Elevation data for active forms of travelling (walking and cycling) was obtained from the Digital Elevation Model (DEM)
made public by the Head Office of Geodesy and Cartography 
in the form of a web service \url{https://mapy.geoportal.gov.pl/wss/service/PZGIK/NMT/GRID1/WCS/DigitalTerrainModelFormatTIFF}. The bounding box for the model was calculated from the Warsaw city polygon provided by OSM.

\subsubsection{Calculation of PT\_EXPERIENCE features}

The features of the PT\_EXPERIENCE group were calculated for the same PT line(s) as the one(s) present in candidate PT connections determined by \texttt{Survey Data Processor} to calculate LOS features for public transport. The requests to \texttt{Public Transport Vehicle Flow Service} were made to aggregate data for individual stop sequences relevant to the trip and related PT lines. The requests to the service were made
for the data from the preceding working days, from Tuesday to Thursday, i.e. the same working days as the ones for which trips were reported in the travel diaries. The data for the periods determined by travel time and rush hours were requested from the service. 

For Warsaw, the rush hours are from 6 am to 9 am and  from 3 pm to 7 pm. The data for PT vehicle flow estimation were computed for all  peak periods and/or periods between them that overlapped with the actual trip period. In this way, the possible experience of travellers travelling on the same route on the preceding days at a similar time of the day was quantified and PT\_EXPERIENCE features were calculated.

\bibliographystyle{unsrt}

\begin{thebibliography}{10}

\bibitem{hillel2021}
Tim Hillel, Michel Bierlaire, Mohammed~Z.E.B. Elshafie, and Ying Jin.
\newblock A systematic review of machine learning classification methodologies for modelling passenger mode choice.
\newblock {\em Journal of Choice Modelling}, 38:100221, 2021.

\bibitem{Chang2019}
Ximing Chang, Jianjun Wu, Hao Liu, Xiaoyong Yan, Huijun Sun, and Yunchao Qu.
\newblock Travel mode choice: a data fusion model using machine learning methods and evidence from travel diary survey data.
\newblock {\em Transportmetrica A: Transport Science}, 15:1587--1612, 11 2019.

\bibitem{tamim2022}
Mohammad~Tamim Kashifi, Arshad Jamal, Mohammad~Samim Kashefi, Meshal Almoshaogeh, and Syed~Masiur Rahman.
\newblock Predicting the travel mode choice with interpretable machine learning techniques: A comparative study.
\newblock {\em Travel Behaviour and Society}, 29:279--296, 2022.
\newblock -ML.

\bibitem{salas2022}
Patricio Salas, Rodrigo~De la~Fuente, Sebastian Astroza, and Juan~Antonio Carrasco.
\newblock A systematic comparative evaluation of machine learning classifiers and discrete choice models for travel mode choice in the presence of response heterogeneity.
\newblock {\em Expert Systems with Applications}, 193:116253, 5 2022.

\bibitem{Grzenda2023}
Maciej Grzenda, Marcin Luckner, and \L{}ukasz Brzozowski.
\newblock Quantifying parking difficulty with transport and prediction models for travel mode choice modelling.
\newblock In {\em Computational Science – ICCS 2023: 23rd International Conference, Prague, Czech Republic, July 3–5, 2023, Proceedings, Part V}, page 505–513, Berlin, Heidelberg, 2023. Springer-Verlag.

\bibitem{cheng2019}
Long Cheng, Xuewu Chen, Jonas~De Vos, Xinjun Lai, and Frank Witlox.
\newblock Applying a random forest method approach to model travel mode choice behavior.
\newblock {\em Travel Behaviour and Society}, 14:1--10, 2019.

\bibitem{LondonDataset2018}
Tim Hillel, Mohammed Z E~B Elshafie, and Ying Jin.
\newblock Recreating passenger mode choice-sets for transport simulation: A case study of {London, UK}.
\newblock {\em Proceedings of the Institution of Civil Engineers - Smart Infrastructure and Construction}, 171(1):29--42, 2018.

\bibitem{hagenauer2017}
Julian Hagenauer and Marco Helbich.
\newblock A comparative study of machine learning classifiers for modeling travel mode choice.
\newblock {\em Expert Systems with Applications}, 78:273--282, 7 2017.

\bibitem{garcia2022}
José~Carlos García-García, Ricardo García-Ródenas, Julio~Alberto López-Gómez, and José Ángel Martín-Baos.
\newblock A comparative study of machine learning, deep neural networks and random utility maximization models for travel mode choice modelling.
\newblock {\em Transportation Research Procedia}, 62:374--382, 2022.

\bibitem{batarce2015use}
Marco Batarce, Juan~Carlos Mu{\~n}oz, Juan de~Dios~Ort{\'u}zar, Sebasti{\'a}n Raveau, Carlos Mojica, and Ramiro~Alberto R{\'\i}os.
\newblock Use of mixed stated and revealed preference data for crowding valuation on public transport in {Santiago, Chile}.
\newblock {\em Transportation Research Record}, 2535(1):73--78, 2015.

\bibitem{BROWNSTONE2000}
David Brownstone, David~S. Bunch, and Kenneth Train.
\newblock Joint mixed logit models of stated and revealed preferences for alternative-fuel vehicles.
\newblock {\em Transportation Research Part B: Methodological}, 34(5):315--338, 2000.

\bibitem{Wong1978FoundationsOP}
Stanley Wong.
\newblock Foundations of {Paul Samuelson's} revealed preference theory: A study by the method of rational reconstruction.
\newblock In {\em Foundations of Paul Samuelson's Revealed Preference Theory: A study by the method of rational reconstruction}, 1978.

\bibitem{hensher1993using}
David~A Hensher and Mark Bradley.
\newblock Using stated response choice data to enrich revealed preference discrete choice models.
\newblock {\em Marketing Letters}, 4:139--151, 1993.

\bibitem{drabicki2021modelling}
Arkadiusz Drabicki, Rafa{\l} Kucharski, Oded Cats, and Andrzej Szarata.
\newblock Modelling the effects of real-time crowding information in urban public transport systems.
\newblock {\em Transportmetrica A: Transport Science}, 17(4):675--713, 2021.

\bibitem{Chen2023}
Huanfa Chen and Yan Cheng.
\newblock Travel mode choice prediction using imbalanced machine learning.
\newblock {\em IEEE Transactions on Intelligent Transportation Systems}, 4 2023.

\bibitem{Ouallane2022}
Asma {Ait Ouallane}, Assia Bakali, Ayoub Bahnasse, Said Broumi, and Mohamed Talea.
\newblock Fusion of engineering insights and emerging trends: Intelligent urban traffic management system.
\newblock {\em Information Fusion}, 88:218--248, 2022.

\bibitem{Xie2020}
Peng Xie, Tianrui Li, Jia Liu, Shengdong Du, Xin Yang, and Junbo Zhang.
\newblock Urban flow prediction from spatiotemporal data using machine learning: A survey.
\newblock {\em Information Fusion}, 59:1--12, 7 2020.

\bibitem{Buehler2011DeterminantsOT}
Ralph Buehler.
\newblock Determinants of transport mode choice: a comparison of {Germany} and the {USA}.
\newblock {\em Journal of Transport Geography}, 19:644--657, 2011.

\bibitem{Witte2013LinkingMC}
Astrid~De Witte, Joachim Hollevoet, Fr{\'e}d{\'e}ric Dobruszkes, Michel Hubert, and Cathy Macharis.
\newblock Linking modal choice to motility: a comprehensive review.
\newblock {\em Transportation Research Part A-policy and Practice}, 49:329--341, 2013.

\bibitem{Chidambaram2021WorktripMC}
Bhuvanachithra Chidambaram and Joachim Scheiner.
\newblock Work-trip mode choice in germany – affected by individual constraints or by partner interaction?
\newblock {\em Travel Behaviour and Society}, 2021.

\bibitem{Nobis2005GenderDI}
Claudia Nobis and Barbara Lenz.
\newblock Gender differences in travel patterns: Role of employment status and household structure.
\newblock In {\em Gender Differences in Travel Patterns: Role of Employment Status and Household Structure}, 2005.

\bibitem{Hanson2010GenderAM}
Susan Hanson.
\newblock Gender and mobility: new approaches for informing sustainability.
\newblock {\em Gender, Place \& Culture}, 17:23 -- 5, 2010.

\bibitem{Best2004DivisionOL}
Henning Best and Martin Lanzendorf.
\newblock Division of labour and gender differences in metropolitan car use : an empirical study in {Cologne, Germany}.
\newblock {\em Journal of Transport Geography}, 13:109--121, 2004.

\bibitem{Cao2007CrossSectionalAQ}
Xinyu Cao, Patricia~L. Mokhtarian, and Susan~L. Handy.
\newblock Cross-sectional and quasi-panel explorations of the connection between the built environment and auto ownership.
\newblock {\em Environment and Planning A}, 39:830 -- 847, 2007.

\bibitem{Reichert2015ModeUI}
Alexander Reichert and Christian Holz-Rau.
\newblock Mode use in long-distance travel.
\newblock {\em Journal of Transport and Land Use}, 8, 2015.

\bibitem{Eisenmann2018AreCU}
Christine Eisenmann and Ralph Buehler.
\newblock Are cars used differently in {Germany} than in {California}? findings from annual car-use profiles.
\newblock {\em Journal of Transport Geography}, 69:171--180, 2018.

\bibitem{Banister2005UnsustainableTC}
David Banister.
\newblock Unsustainable transport: City transport in the new century.
\newblock In {\em Unsustainable Transport: City Transport in the New Century}, 2005.

\bibitem{Racca2003FACTORSTA}
David~P. Racca and Edward~C. Ratledge.
\newblock Factors that affect and/or can alter mode choice.
\newblock In {\em FACTORS THAT AFFECT AND/OR CAN ALTER MODE CHOICE}, 2003.

\bibitem{Aziz2018ExploringTI}
H.~M.~Abdul Aziz, Nicholas~N. Nagle, April Morton, Michael Hilliard, Devin~A. White, and Robert~N. Stewart.
\newblock Exploring the impact of walk–bike infrastructure, safety perception, and built-environment on active transportation mode choice: a random parameter model using {New York City} commuter data.
\newblock {\em Transportation}, 45:1207--1229, 2018.

\bibitem{Ding2017ExploringTI}
Chuan Ding, Donggen Wang, Chao Liu, Yi~Zhang, and Jiawen Yang.
\newblock Exploring the influence of built environment on travel mode choice considering the mediating effects of car ownership and travel distance.
\newblock {\em Transportation Research Part A-policy and Practice}, 100:65--80, 2017.

\bibitem{VanAcker2007WhenTG}
Veronique~Van Acker, Bert van Wee, and Frank Witlox.
\newblock When transport geography meets social psychology: Toward a conceptual model of travel behaviour.
\newblock {\em Transport Reviews}, 30:219 -- 240, 2007.

\bibitem{Zhang2012HowBE}
Lei Zhang, Jinhyun Hong, Arefeh~A. Nasri, and Qing Shen.
\newblock How built environment affects travel behavior: A comparative analysis of the connections between land use and vehicle miles traveled in us cities.
\newblock {\em Journal of Transport and Land Use}, 5:40--52, 2012.

\bibitem{Beiro2007UnderstandingAT}
Gabriela Beir{\~a}o and J.~A.~Sarsfield Cabral.
\newblock Understanding attitudes towards public transport and private car: A qualitative study.
\newblock {\em Transport Policy}, 14:478--489, 2007.

\bibitem{Chen2011HabitualOR}
Ching-Fu Chen and Wei~Hsiang Chao.
\newblock Habitual or reasoned? using the theory of planned behavior, technology acceptance model, and habit to examine switching intentions toward public transit.
\newblock {\em Transportation Research Part F-traffic Psychology and Behaviour}, 14:128--137, 2011.

\bibitem{Clark2015ChangesTC}
Ben. Clark, Kiron Chatterjee, and Steve. Melia.
\newblock Changes to commute mode: The role of life events, spatial context and environmental attitude.
\newblock In {\em Changes to Commute Mode: The Role of Life Events, Spatial Context and Environmental Attitude}, 2015.

\bibitem{Gardner2010GoingGM}
Benjamin Gardner and Charles Abraham.
\newblock Going green? modeling the impact of environmental concerns and perceptions of transportation alternatives on decisions to drive.
\newblock {\em Journal of Applied Social Psychology}, 40:831--849, 2010.

\bibitem{Marselle2013WalkingFW}
Melissa~R. Marselle, Katherine~N. Irvine, and Sara~L. Warber.
\newblock Walking for well-being: Are group walks in certain types of natural environments better for well-being than group walks in urban environments?
\newblock {\em International Journal of Environmental Research and Public Health}, 10:5603 -- 5628, 2013.

\bibitem{Wolf2014WalkingHA}
Isabelle~D. Wolf and Teresa Wohlfart.
\newblock Walking, hiking and running in parks: A multidisciplinary assessment of health and well-being benefits.
\newblock {\em Landscape and Urban Planning}, 130:89--103, 2014.

\bibitem{GilesCorti2003RelativeIO}
Billie Giles-Corti and Robert~J Donovan.
\newblock Relative influences of individual, social environmental, and physical environmental correlates of walking.
\newblock {\em American journal of public health}, 93 9:1583--9, 2003.

\bibitem{Booth2000SocialcognitiveAP}
M~L Booth, Neville Owen, Adrian~E Bauman, Ornella Clavisi, and E.~Leslie.
\newblock Social-cognitive and perceived environment influences associated with physical activity in older {Australians}.
\newblock {\em Preventive medicine}, 31 1:15--22, 2000.

\bibitem{Handy2002HowTB}
Susan~L. Handy, Marlon~G. Boarnet, Reid Ewing, and Richard~E. Killingsworth.
\newblock How the built environment affects physical activity: views from urban planning.
\newblock {\em American journal of preventive medicine}, 23 2 Suppl:64--73, 2002.

\bibitem{Cervero2007InfluencesOB}
Robert Cervero, Olga~Luc{\'i}a Sarmiento, Enrique~R Jacoby, Luis~Fernando G{\'o}mez, and Andrea~B. Neiman.
\newblock Influences of built environments on walking and cycling: Lessons from {Bogot}{\'a}.
\newblock {\em International Journal of Sustainable Transportation}, 3:203 -- 226, 2007.

\bibitem{Ferrer2015AQS}
Sheila Ferrer, Tom{\'a}s Ruiz, and Lid{\'o}n Mars.
\newblock A qualitative study on the role of the built environment for short walking trips.
\newblock {\em Transportation Research Part F-traffic Psychology and Behaviour}, 33:141--160, 2015.

\bibitem{Forsyth2009TheBE}
Ann Forsyth, J.~Michael Oakes, Brian Lee, and Katherine~H Schmitz.
\newblock The built environment, walking, and physical activity: Is the environment more important to some people than others?
\newblock {\em Transportation Research Part D-transport and Environment}, 14:42--49, 2009.

\bibitem{Goldsmith1992NATIONALBA}
S~A Goldsmith.
\newblock National bicycling and walking study. case study no. 1: Reasons why bicycling and walking are and are not being used more extensively as travel modes.
\newblock In {\em NATIONAL BICYCLING AND WALKING STUDY. CASE STUDY NO. 1: REASONS WHY BICYCLING AND WALKING ARE AND ARE NOT BEING USED MORE EXTENSIVELY AS TRAVEL MODES}, 1992.

\bibitem{saneinejad2012modelling}
Sheyda Saneinejad, Matthew~J Roorda, and Christopher Kennedy.
\newblock Modelling the impact of weather conditions on active transportation travel behaviour.
\newblock {\em Transportation research part D: transport and environment}, 17(2):129--137, 2012.

\bibitem{laffan2018every}
Kate Laffan.
\newblock Every breath you take, every move you make: Visits to the outdoors and physical activity help to explain the relationship between air pollution and subjective wellbeing.
\newblock {\em Ecological Economics}, 147:96--113, 2018.

\bibitem{xu2021does}
Yuquan Xu, Yuewen Liu, Xiangyu Chang, and Wei Huang.
\newblock How does air pollution affect travel behavior? a big data field study.
\newblock {\em Transportation Research Part D: Transport and Environment}, 99:103007, 2021.

\bibitem{chen2018air}
Siyu Chen, Chongshan Guo, and Xinfei Huang.
\newblock Air pollution, student health, and school absences: Evidence from {China}.
\newblock {\em Journal of Environmental Economics and Management}, 92:465--497, 2018.

\bibitem{ward2016responds}
Alison L~Sexton Ward and Timothy~KM Beatty.
\newblock Who responds to air quality alerts?
\newblock {\em Environmental and resource economics}, 65:487--511, 2016.

\bibitem{kim2023role}
Suji Kim, Yekang Ko, and Kitae Jang.
\newblock Role of income on travel behavior in polluted air.
\newblock {\em Journal of Transport \& Health}, 33:101705, 2023.

\bibitem{liu2018severe}
Haoming Liu and Alberto Salvo.
\newblock Severe air pollution and child absences when schools and parents respond.
\newblock {\em Journal of Environmental Economics and Management}, 92:300--330, 2018.

\bibitem{lovelace2017propensity}
Robin Lovelace, Anna Goodman, Rachel Aldred, Nikolai Berkoff, Ali Abbas, and James Woodcock.
\newblock The propensity to cycle tool: An open source online system for sustainable transport planning.
\newblock {\em Journal of transport and land use}, 10(1):505--528, 2017.

\bibitem{Winters2010BuiltEI}
Meghan Winters, Michael Brauer, Eleanor~M. Setton, and Kay Teschke.
\newblock Built environment influences on healthy transportation choices: Bicycling versus driving.
\newblock {\em Journal of Urban Health}, 87:969--993, 2010.

\bibitem{Sun2023}
Chao Sun and Jian Lu.
\newblock The relative roles of different land-use types in bike-sharing demand: A machine learning-based multiple interpolation fusion method.
\newblock {\em Information Fusion}, 95:384--400, 7 2023.

\bibitem{Holmgren2007MetaAnalysisOP}
Johan Holmgren.
\newblock Meta-analysis of public transport demand.
\newblock {\em Transportation Research Part A-policy and Practice}, 41:1021--1035, 2007.

\bibitem{Redman2013QualityAO}
Lauren~Fielder Redman, Margareta Friman, Tommy G{\"a}rling, and Terry Hartig.
\newblock Quality attributes of public transport that attract car users : A research review.
\newblock {\em Transport Policy}, 25:119--127, 2013.

\bibitem{Curtis2016PlanningFP}
Carey Curtis and Jan Scheurer.
\newblock Planning for public transport accessibility: An international sourcebook.
\newblock In {\em Planning for Public Transport Accessibility: An International Sourcebook}, 2016.

\bibitem{Khan2021IncreasingPT}
Jamil Khan, Robert Hrelja, and Fredrik Pettersson-L{\"o}fstedt.
\newblock Increasing public transport patronage – an analysis of planning principles and public transport governance in {Swedish} regions with the highest growth in ridership.
\newblock {\em Case studies on transport policy}, 9:260--270, 2021.

\bibitem{Dodson2011ThePO}
Jago Dodson, Paul Mees, John Stone, and Matthew~I. Burke.
\newblock The principles of public transport network planning: a review of the emerging literature with select examples.
\newblock In {\em The principles of public transport network planning: a review of the emerging literature with select examples}, 2011.

\bibitem{McLeod2017UrbanPT}
Sam McLeod, Jan Scheurer, and Carey Curtis.
\newblock Urban public transport.
\newblock {\em Journal of Planning Literature}, 32:223 -- 239, 2017.

\bibitem{Guerra2018UrbanFT}
Erick Guerra, Camilo Caudillo, Paavo Monkkonen, and J.~Montaner i~Montejano.
\newblock Urban form, transit supply, and travel behavior in {Latin America}: Evidence from {Mexico's} 100 largest urban areas.
\newblock {\em Transport Policy}, 2018.

\bibitem{imekolu2015TheRO}
{\"O}zlem Şimşekoğlu, Trond Nordfj{\ae}rn, and Torbj{\o}rn Rundmo.
\newblock The role of attitudes, transport priorities, and car use habit for travel mode use and intentions to use public transportation in an urban norwegian public.
\newblock {\em Transport Policy}, 42:113--120, 2015.

\bibitem{Oa2021PublicTU}
Juan de~O{\~n}a, Esperanza Est'evez, and Rocio de~O{\~n}a.
\newblock Public transport users versus private vehicle users: Differences about quality of service, satisfaction and attitudes toward public transport in {Madrid (Spain)}.
\newblock {\em Travel behaviour and society}, 23:76--85, 2021.

\bibitem{Walker2007PurposedrivenPT}
Jarrett Walker.
\newblock Purpose-driven public transport: creating a clear conversation about public transport goals.
\newblock In {\em Purpose-driven public transport: creating a clear conversation about public transport goals}, 2007.

\bibitem{Ko2019ExploringFA}
Joonho Ko, Sugie Lee, and Miree Byun.
\newblock Exploring factors associated with commute mode choice: An application of city-level general social survey data.
\newblock {\em Transport Policy}, 2019.

\bibitem{carse2013factors}
Andrew Carse, Anna Goodman, Roger~L Mackett, Jenna Panter, and David Ogilvie.
\newblock The factors influencing car use in a cycle-friendly city: the case of {Cambridge}.
\newblock {\em Journal of transport geography}, 28:67--74, 2013.

\bibitem{hensher2001parking}
David~A Hensher and Jenny King.
\newblock Parking demand and responsiveness to supply, pricing and location in the {Sydney} central business district.
\newblock {\em Transportation Research Part A: Policy and Practice}, 35(3):177--196, 2001.

\bibitem{Ardeshiri2019LifestylesRL}
Ali Ardeshiri and Akshay Vij.
\newblock Lifestyles, residential location, and transport mode use: A hierarchical latent class choice model.
\newblock {\em Transportation Research Part A: Policy and Practice}, 2019.

\bibitem{Kuhnimhof2012TravelTA}
Tobias Kuhnimhof, Ralph Buehler, Matthias Wirtz, and D.~Kalinowska.
\newblock Travel trends among young adults in {Germany}: increasing multimodality and declining car use for men.
\newblock {\em Journal of Transport Geography}, 24:443--450, 2012.

\bibitem{Prato2017LatentLA}
Carlo~G. Prato, Katr{\'i}n Halld{\'o}rsd{\'o}ttir, and Otto~Anker Nielsen.
\newblock Latent lifestyle and mode choice decisions when travelling short distances.
\newblock {\em Transportation}, 44:1343--1363, 2017.

\bibitem{zhao2020}
Xilei Zhao, Xiang Yan, Alan Yu, and Pascal~Van Hentenryck.
\newblock Prediction and behavioral analysis of travel mode choice: A comparison of machine learning and logit models.
\newblock {\em Travel Behaviour and Society}, 20:22--35, 2020.

\bibitem{mohd2022}
Nur Fahriza~Mohd Ali, Ahmad Farhan~Mohd Sadullah, Anwar P.P.~Abdul Majeed, Mohd Azraai~Mohd Razman, and Rabiu~Muazu Musa.
\newblock The identification of significant features towards travel mode choice and its prediction via optimised random forest classifier: An evaluation for active commuting behavior.
\newblock {\em Journal of Transport \& Health}, 25:101362, 6 2022.

\bibitem{Prado2023}
Ignacio-Iker Prado-Rujas, Emilio Serrano, Antonio García-Dopico, M.~Luisa Córdoba, and María~S. Pérez.
\newblock Combining heterogeneous data sources for spatio-temporal mobility demand forecasting.
\newblock {\em Information Fusion}, 91:1--12, 2023.

\bibitem{yang2022}
Yongjiang Yang, Kuniaki Sasaki, Long Cheng, and Xingwei Liu.
\newblock Gender differences in active travel among older adults: Non-linear built environment insights.
\newblock {\em Transportation Research Part D: Transport and Environment}, 110, 9 2022.

\bibitem{Lundberg2017}
Scott~M. Lundberg and Su~In Lee.
\newblock {A unified approach to interpreting model predictions}.
\newblock {\em Advances in Neural Information Processing Systems}, 2017-December(Section 2):4766--4775, 2017.

\bibitem{grzenda2023ECML}
Maciej Grzenda, Marcin Luckner, and Przemys{\l}aw Wrona.
\newblock Urban traveller preference miner: Modelling transport choices with survey data streams.
\newblock In Massih-Reza Amini, St{\'e}phane Canu, Asja Fischer, Tias Guns, Petra Kralj~Novak, and Grigorios Tsoumakas, editors, {\em Machine Learning and Knowledge Discovery in Databases}, pages 654--657, Cham, 2023. Springer Nature Switzerland.

\bibitem{yang2019}
Jie Yang and Jun Ma.
\newblock Compressive sensing-enhanced feature selection and its application in travel mode choice prediction.
\newblock {\em Applied Soft Computing}, 75:537--547, 2 2019.

\bibitem{Marz2015}
Nathan Marz and James Warren.
\newblock {\em Big Data: Principles and Best Practices of Scalable Realtime Data Systems}.
\newblock Cambridge University Press, Greenwich, CT, USA, 2015.

\bibitem{gtfs}
Luqmaan Dawoodjee.
\newblock Gtfs: Making public transit data universally accessible.
\newblock \url{ https://gtfs.org}.
\newblock Accessed: 2022-08-14.

\bibitem{ditzler2015}
Gregory Ditzler, Manuel Roveri, Cesare Alippi, and Robi Polikar.
\newblock Learning in nonstationary environments: A survey.
\newblock {\em IEEE Computational Intelligence Magazine}, 10(4):12--25, 2015.

\bibitem{rpart}
\&~Ripley~Brian Therneau~Terry, Atkinson~Beth.
\newblock Rpart: Recursive partitioning. r package version 4.1-3.
\newblock Technical report, CRAN.R-project, 2013.

\bibitem{ranger}
Marvin~N. Wright and Andreas Ziegler.
\newblock {ranger}: A fast implementation of random forests for high dimensional data in {C++} and {R}.
\newblock {\em Journal of Statistical Software}, 77(1):1--17, 2017.

\bibitem{Chen2016}
Tianqi Chen and Carlos Guestrin.
\newblock Xgboost: A scalable tree boosting system.
\newblock In {\em Proceedings of the 22nd ACM SIGKDD International Conference on Knowledge Discovery and Data Mining}, KDD '16, page 785–794, New York, NY, USA, 2016. Association for Computing Machinery.

\bibitem{Breiman2001}
Leo Breiman.
\newblock {Random Forests}.
\newblock {\em Machine Learning}, 45(1):5--32, 2001.

\bibitem{japkowicz2011}
Nathalie Japkowicz and Mohak Shah.
\newblock {\em Evaluating Learning Algorithms: A Classification Perspective}.
\newblock Cambridge University Press, New York, NY, USA, 2011.

\bibitem{KosteleckaSzarataJacyna2015}
Aneta Kostelecka, Andrzej Szarata, and Marianna Jacyna.
\newblock Warsaw traffic measurement 2015.
\newblock Technical report, PBS Sp. z o.o, Cracow University of Technology, Warsaw University of Technology, 2015.

\end{thebibliography}

\end{document}